\documentclass[twocolumn,prX,showpacs,nofootinbib,nonbibnotes]{revtex4}

\usepackage{graphicx}% Include figure files
\usepackage{dcolumn}% Align table columns on decimal point
\usepackage{bm}% bold math

\usepackage{epsf}
\usepackage{anysize}
\usepackage{graphicx}
\marginsize{1.8 cm}{1.5 cm}{1.0 in}{4.0 cm}

\usepackage{graphicx}% Include figure files
\usepackage{dcolumn}% Align table columns on decimal point
\usepackage{bm}% bold math
\usepackage{amsmath}
\usepackage{amssymb}
\usepackage{mathrsfs}
\usepackage[latin1]{inputenc}
\usepackage{amsmath}

\makeatletter
  \newcommand\figcaption{\def\@captype{figure}\caption}
  \newcommand\tabcaption{\def\@captype{table}\caption}
\makeatother

\newcommand{\dd}{{\rm d}}

\newcommand{\sd}{Schr\"{o}dinger }

\newcommand{\G}{{\rm{Stab}}}
\newcommand{\J}{\mathcal{J}}
\newcommand{\U}{\mathcal{U}}

\newcommand{\hil}{\mathcal{H}}

\newcommand{\F}{\mathscr{F}}

\newcommand{\R}{{\mathbb{R}}}
\newcommand{\tr}{{\rm Tr}}

\newcommand{\Sp}{{\rm Sp}}
\newcommand{\OSp}{{\rm OSp}}
\newcommand{\ISp}{{\rm ISp}}

\newtheorem{theorem}{Theorem}

\newtheorem{definition}{Definition}

\begin {document}

%\preprint{APS/123-QED}

\title{Quantum Control Landscapes}

\author{Raj Chakrabarti}
\affiliation{Department of Chemistry, Princeton University,
Princeton, New Jersey 08544, USA} \email{rajchak@princeton.edu}

\author{Herschel Rabitz}
\affiliation{Department of Chemistry, Princeton University,
Princeton, New Jersey 08544, USA}

\begin{abstract}

Numerous lines of experimental, numerical and analytical evidence
indicate that it is surprisingly easy to locate optimal controls
steering quantum dynamical systems to desired objectives.  This has
enabled the control of complex quantum systems despite the expense
of solving the Schrodinger equation in simulations and the
complicating effects of environmental decoherence in the laboratory.
Recent work indicates that this simplicity originates in universal
properties of the solution sets to quantum control problems that are
fundamentally different from their classical counterparts. Here, we
review studies that aim to systematically characterize these
properties, enabling the classification of quantum control
mechanisms and the design of globally efficient quantum control
algorithms.

\pacs{03.67.-a,03.67.Lx,03.65.Yz,02.30.Yy}

%\bigskip
%\bigskip
%\bigskip

%\textbf{Word Count}: 20379 words

%\textbf{Keywords}:  quantum control theory, quantum control
%experiments, laser chemistry, optimal control, open quantum systems

\end{abstract}

\maketitle

\tableofcontents

\section{Introduction}

\begin{figure*}%[h]
\centerline{
\includegraphics[width=6.0in,height=4.5in]{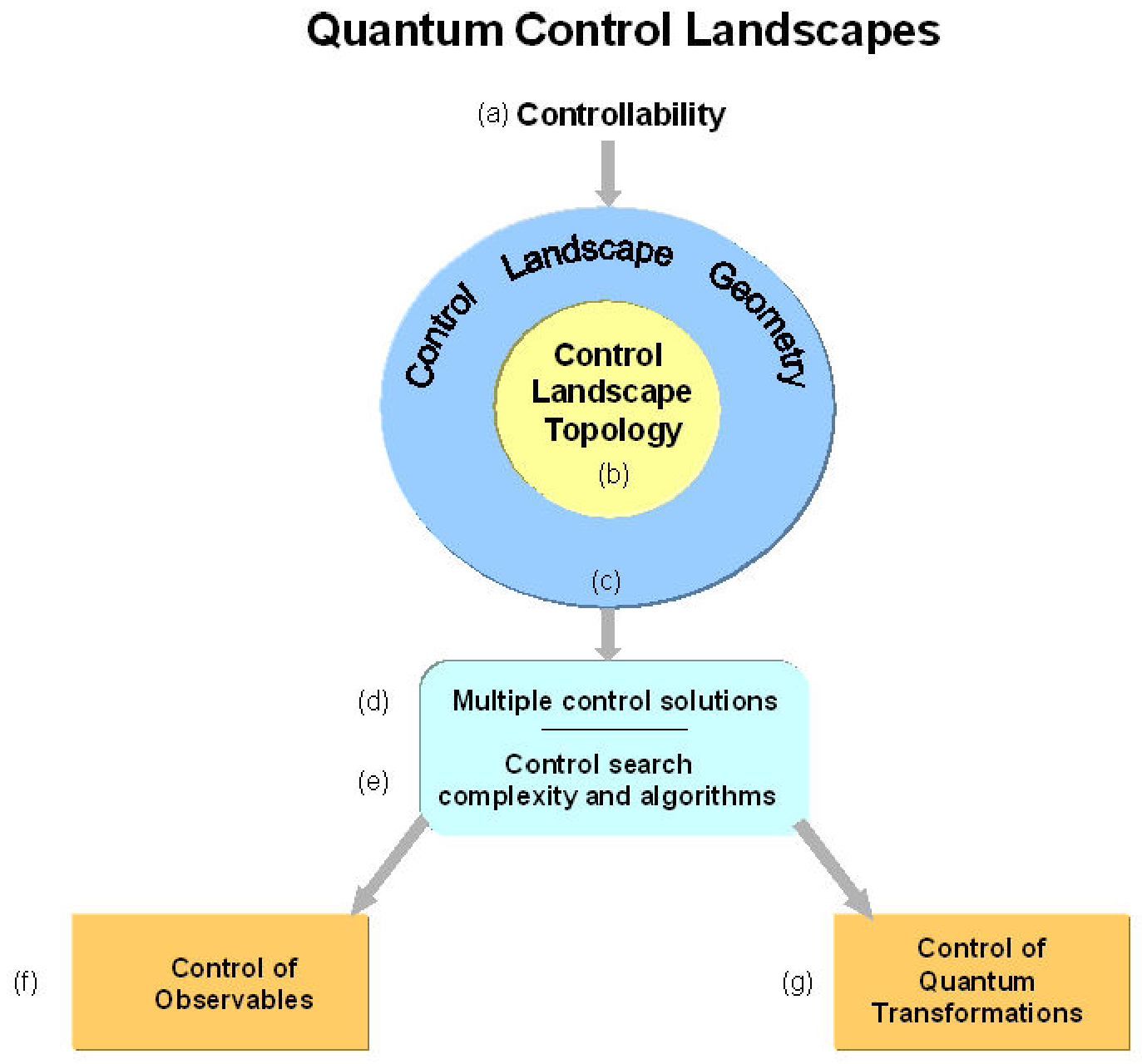}
} \caption{A control landscape is defined as the map between the
time-dependent control Hamiltonian and associated values of the
control cost functional. The entry point into their study is (a) the
controllability of the quantum system, which allows search
algorithms to freely traverse the landscape. Quantum control
landscape features can be conveniently subdivided into those
pertaining to (b) critical topology, i.e., the maxima, minima and
saddle points of the landscape, and (c) landscape geometry, namely
the characteristic local structures encountered while climbing
toward the global optimum. Study of the geometry of quantum control
landscapes reveals (d) the existence of multiple control solutions
corresponding to any given objective function value. The topology
and geometry of quantum control landscapes together determine (e)
the search complexity of the control problems, i.e. the scaling of
the effort required to locate optimal controls. An ultimate goal in
the study of quantum control landscapes is the design of global
search algorithms that attain lower bounds on this search
complexity. Such algorithms may be applied to either of the two
major classes of quantum system manipulation problems, (f) control
of quantum observables and (g) control of quantum dynamical
transformations (i.e., unitary propagators). An overarching
conclusion pertaining to both these types of landscapes is that they
contain no suboptimal traps, which has broad-scale implications for
both the experimental and computational feasibility of quantum
control.}\label{fig1}
\end{figure*}

\begin{figure*}%[h]
\centerline{
\includegraphics[width=4.5in,height=4in]{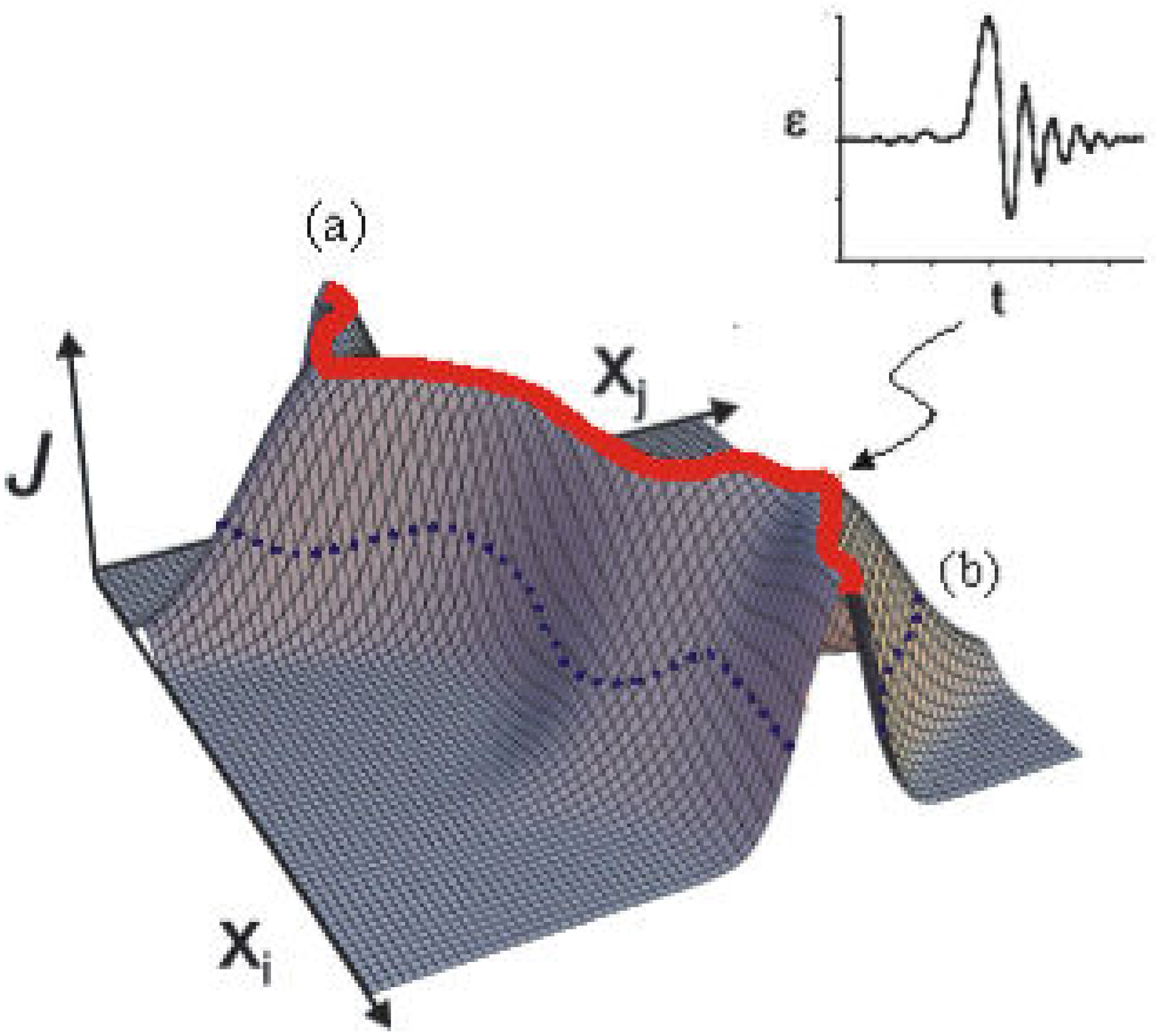}
} \caption{Schematic representation of a quantum control landscape
depicting various features discussed in the review. $x_i, x_j$
indicate two of possibly many control degrees of freedom, and $J$
denotes objective function value. A point on the landscape
corresponds to a time-dependent control field. a) Critical points of
the map correspond to locally optimal solutions to the control
problem. The number of positive, negative, and null Hessian
eigenvalues at these points determine whether they are saddle points
or local traps (Section II). In the most common case of
state-to-state population transfer, there are no local optima and
the global maximum is a continuous manifold, depicted in the picture
as the solid curve. Level sets (b) of the control landscape correspond
to control fields that produce the same objective function value at
the final dynamical time T (shown as a dotted line). Any point
on the landscape corresponds to a control field $\varepsilon(t)$,
with one on the extremal manifold indicated.
%(Note that the global maximum of the landscape is
%a continuous submanifold.)
}\label{fig2}
\end{figure*}
% of a quantum optimal control

The notion of controlling quantum systems seems inherently
problematic on several counts. First, the extreme sensitivity of
quantum systems to environmental interactions would appear to place
limits on the maximal achievable control fidelity. Second, from a
numerical perspective, given the considerable cost of propagating
the Schrodinger equation, unless the search space for optimal
controls has particularly simple properties, it would appear
impossible to locate controls for all but the smallest quantum
systems in reasonable time. However, once the methods of optimal
control began to be applied to molecular systems (thanks to
remarkable advances in laser pulse shaping technology) it rapidly
became clear that quantum control was not an ill-fated concept, but
rather, that controlling quantum systems was surprisingly easy. In
the laboratory, this conclusion was particularly apparent in the
case of so-called adaptive learning control of quantum dynamics,
wherein learning (i.e., typically genetic) algorithms are applied to
search the space of laser control parameters for the maximization of
the expectation value of a quantum observable. This search space is
high-dimensional, normally suggesting that it should be replete with
local optima  and other unfavorable features that would trap
unsophisticated search algorithms, especially in the presence of
environmental decoherence. The repeated successes of quantum optimal
control experiments and simulations indicated that the so-called
"curse of dimensionality", common in the theory of optimization in
high-dimensional spaces, was not prohibitive here.

This attractive circumstance for quantum control rests on the
mathematical underpinnings of quantum theory being surprisingly
simple, owing to the linearity of quantum mechanics and the
unitarity of the accompanying transformations. Although quantum
systems can be highly sensitive to environmental perturbations, the
rules governing their dynamics are in many ways simpler than those
governing classical dynamics. Furthermore, the presence of an
environment, rather than being an impediment, may under the right
conditions aid the control process. Recent work has aimed to
understand the precise mathematical properties of quantum mechanics
are responsible for the surprising simplicity with which quantum
phenomena can be controlled. Analytical, numerical and experimental
treatments of the problem have been explored. In order to enable the
systematic study of these features, the notion of a quantum control
landscape, defined as the map between the space of time-dependent
controls and associated values of the objective functional, was
introduced (Figs. \ref{fig1}, \ref{fig2}).

From an analytical perspective, it was recently found that for
several classes of low-dimensional problems, it is possible to
exactly solve for quantum optimal controls, without any need for
numerical search. By contrast, for the analogous classical problems,
analytical solutions do not exist. Of course, analytical solutions
are still only possible for special small systems; however, beyond
this, it has become clear that the numerical or experimental search
for optimal controls is often easier for quantum systems than for
classical systems. In this regard, the topology of the search space
is of fundamental importance. Evidence suggests that the landscapes
for both observable maximization and control of dynamical
transformations have simpler topological properties for quantum
versus classical systems, contributing to rapid convergence of
numerical or experimental searches for effective controls. Moreover,
besides the simplicity of locating quantum controls, it has been
observed that the controls themselves have remarkably simple
functional properties, in some cases enabling a direct
interpretation of the mechanisms involved in steering about the
dynamics.

The ease of locating optimal quantum controls, and the comparatively
simple structure of these controls, have pervasive implications for
a wide range of quantum technologies. The study of quantum control
landscapes is motivated by the practical goals of achieving higher
objective function yields and designing control fields with desired
properties, but in order to attain these goals, it is necessary to
embrace the mathematical framework that underlies the remarkable
properties of these landscapes. The origin of these counterintuitive
properties, and their differences with respect to classical control,
constitute the primary subject of this review.

The review is organized as follows. In section II, we examine the
topology of solution sets to quantum observable and unitary
transformation control problems. Section III reviews the analytical
solutions obtained for low-dimensional quantum control problems. In
section IV, we consider numerical studies of the solution sets to
higher dimensional quantum control problems without analytical
solution, focusing on the degeneracy in the set of controls that
reach the same objective.  Then, section V reviews experimental work
that has probed the structure of these level sets of multiple
solutions, as well as the topology of quantum control landscapes.
Section VI examines how the controllability of quantum systems,
compared to that of classical systems, impacts control landscape
properties, in particular with respect to search efficiency. In
section VII and VIII, we review approaches to the quantification of
quantum control search effort complexity and the design of global
search algorithms that aim to attain lower bounds on complexity
scaling. In Section IX, we consider the effects of quantum
decoherence on the structure of quantum control landscapes derived
in previous sections. Finally, in section X, we draw conclusions and
discuss future directions.

\section{Hamiltonian-independent properties of solution sets to quantum
control problems}\label{topology}
%optimal control problems

The most fundamental property of landscapes of solution sets to
variational problems is their critical topology, i.e., the number of
solutions, their associated functional values, and their optimality
status (Fig. 1). In the context of optimal control theory, these
critical points correspond to (possibly suboptimal) solutions of the
optimal control problem, including both global and local minimizers
of the objective functional.

An early work \cite{Demiralp1993} explicitly explored the
multiplicity of solutions to quantum optimal control problems aimed
at the maximization of the expectation value of an observable
operator at a final dynamical time $T$. Through the application of a
perturbation theory approach to the nonlinear variational equation,
it was shown that in general, a denumerably infinite number of
solutions (control fields) exist to such quantum control problems.
Moreover, multiple unitary propagators $U(T)$ are typically
associated with the various possible local and global optima of the
objective. Recent work, discussed throughout this review, has aimed
to identify how the infinite number of solutions to quantum optimal
control problems are distributed among the global and local optima
of the control landscape. The optimality status of these solutions
plays an important role in determining the performance of local
search algorithms as they traverse the landscape. A remarkable
feature of quantum optimal control landscapes is that these
properties can all be determined analytically for many problems of
interest, whereas for general variational problems outside of
quantum control, such information is exceedingly difficult to
acquire. Moreover, many of these properties are independent of the
Hamiltonian of the quantum system.

In this section we review work aimed at characterizing the critical
topologies of quantum optimal control problems. In the absence of
additional information, it is natural to expect that the control
landscape will possess multiple local maxima and minima that are
capable of trapping the search for optimal controls at suboptimal
values of the objective. We will show that in stark contrast to this
generically expected situation, the critical topologies of quantum
optimal control landscapes are surprisingly simple. In what follows
we consider a controllable (see Appendix \ref{appctrl} for a review
of the definition of controllability) quantum system of $N$ discrete
levels whose dynamics are driven by the Hamiltonian
$H=H\left[H_0,\{\varepsilon_k\}\right],$ depending on a free
Hamiltonian $H_0$ describing the uncontrolled evolution of the
system and an appropriate set $\{\varepsilon_k\}$ of control
variables (e.g., phases and amplitudes in an optimal control
experiment (OCE) pulse shaper).

A generic quantum optimal control cost functional can be written:

\begin{multline}\label{OCT}
J =\Phi(U(T), T)-\\
\textmd{Re} \left[\tr\int_{0}^T\left\{\left(\frac{\partial
U(t)}{\partial t} +
\frac{i}{\hbar}H(\varepsilon(t))U(t)\right)\beta(t)\right\}dt\right]-\\
\lambda\int_{0}^T f^0(\varepsilon(t)) \dd t
\end{multline}
where $\beta(t)$ is a Lagrange multiplier operator constraining the
quantum system dynamics to obey the \sd equation, $\varepsilon (t)$
is the time-dependent control field, and $\lambda$ weights the
importance of an auxiliary physically motivated penalty term on the
field. This latter penalty decreases the degeneracy of solutions to
the optimal control problem; a common choice for $f^0$ in
simulations is $\frac{1}{s(t)}|\varepsilon(t)|^2$, where $s(t)$ is
the pulse envelope, corresponding to a penalty on the total field
fluence. Other, even more general cost functions can easily be
generated with alternate choices for $f^0$ as well as additional
terms involving the evolving quantum state.

The two most common types of quantum optimal control problems are
the maximization of the expectation value of a Hermitian observable
and the maximization of the fidelity of a quantum unitary
transformation.  Optimizing the expectation value of a Hermitian
observable operator describes a broad variety of problems in quantum
control, such as performing selective chemical fragmentation and
rearrangement \cite{Assion1998,Baumert1997}, redirecting energy
transfer in biomolecules \cite{Herek2006}, creating ultra-fast
optical switches and tailoring high harmonic generation
\cite{Bartels2000}. This problem corresponds to the following choice
of $\Phi$:
\begin{equation}
\Phi_1(U) = \tr\left[U(T){\rho(0)}U^{\dag}(T)\Theta\right],
\end{equation}
where $\rho(0)$ is the initial density matrix of the system,
$\Theta$ is an arbitrary observable operator, and $T$ is the final
dynamical time.

The optimal control of quantum unitary transformations has recently
received increasing attention due to its applications to the field
of quantum information processing (QIP). Over the past several
years, it has become clear that the physical implementation of
logical gates in QIP, which are represented by unitary propagators,
may be facilitated by optimal control theory (OCT)
\cite{PalKos2002,Grace2006a,KhaBro2001,KhaBro2002}. However, as we
will see below, the optimal control of unitary transformations also
has applications to population transfer in atoms and molecules.
Within OCT, the problem of maximizing the fidelity of a dynamical
transformation $W$ can be framed using \cite{PalKos2002}:
\begin{eqnarray}
\Phi_2(U) &=& \sum_{i,j} |W_{ij}-U_{ij}(T)|^2\\
&=& 2N- 2\textmd{Re}\tr\left[W^{\dag}U(T)\right]
\end{eqnarray}
where $W$ is the target unitary transformation.

Solutions to these optimal control problems correspond to the
condition $\frac{\delta J}{\delta \varepsilon(t)} = 0$. In this
section, we assume $\lambda = 0$, and show that under this
%$\lambda \ll 1$
assumption many properties of the critical points of the functionals
$\Phi_1$ and $\Phi_2$ can be characterized analytically. An
infinitesimal functional change in the Hamiltonian $\delta H(t)$
produces an infinitesimal change in the dynamical propagator
$U(t,0)$ as follows:
\begin{equation}
\delta U(t,0) = - \frac {i}{\hbar} \int_0^t U(t,t') \delta H(t')
U(t',0)dt'. \end{equation} It can be shown that for the objective
functions $\Phi_1$ and $\Phi_2$ the respective changes in $\Phi$ are
$\delta \Phi_1 =-\tr \big[
\left[\Theta(T),\rho(0)\right]U^{\dag}(T,0)\delta U(T,0)\big],$
where $\Theta(T) \equiv U(T) \Theta U^{\dag}(T),$ and $\delta \Phi_2
=\tr \big[(W^{\dag}U-U^{\dag}W)U^{\dag}(T,0)\delta U(T,0)\big].$ In
the special case of the electric dipole approximation, the
Hamiltonian has the form $H(t) = H_0 - \mu \cdot \varepsilon(t)$,
where $H_0$ is the internal Hamiltonian of the system and $\mu$ is
the electric dipole operator. Then $\delta H(t) =
\triangledown_{\varepsilon} H(t) \cdot \delta \varepsilon(t)$,
$U^{\dag}(T,0)\delta U(T,0) = -\frac{i}{\hbar}\int_0^T
U^{\dag}(t,0)\triangledown_{\varepsilon}H(t)U(t,0) \cdot \delta
\varepsilon(t) \dd t,$ and the gradients of the control objective
functionals can be written:

\begin{equation}\label{grad1}
\frac{\delta \Phi_1}{\delta \varepsilon(t)} =\tr
\left(\left[\Theta(T),\rho(0)\right]B(t)\right),
\end{equation}
and \begin{equation}\label{grad2} \frac{\delta \Phi_2}{\delta
\varepsilon(t)} =\tr \left((W^{\dag}U-U^{\dag}W)B(t)\right),
\end{equation}
where $B(t) \equiv -\frac{i}{\hbar}
U^{\dag}(t,0)\triangledown_{\varepsilon}H(t)U(t,0),$ and we have
adopted the shorthand notation $U \equiv U(T)$. Within the dipole
approximation, $B(t) = -\frac{i}{\hbar}U^{\dag}(t,0)\mu U(t,0)$.

The variational problems of optimal control theory admit two types
of minimizers \footnote{In mechanics, the Lagrangian functional that
determines the equations of motion is uniquely determined by
symmetries of the system. However, in optimal control theory, the
objective functional, which determines the control law, is chosen by
the controller. This distinction lends an additional component to
the study of the topology of optimal control problems, namely the
topology of the map between dynamical propagators and associated
values of the chosen objective function.}. According to the chain
rule,
\begin{equation}\frac{\delta J}{\delta \varepsilon(t)} = \frac{\dd J}{\dd U}
\cdot \frac{\delta U}{\delta \varepsilon(t)}.
\end{equation}
The first type of minimizer corresponds to those control Hamiltonians that are
critical points of the control objective functional, but are not
critical points of the map between control fields and associated
dynamical propagators (i.e., points at which $\frac{\dd J}{\dd
U}=0$, while the Frechet derivative mapping from the control
variation $\delta \varepsilon(t)$ to $\delta U(T)$ at $t=T$ is
surjective).
%$\frac{\delta U}{\delta \varepsilon(t)} \neq 0, \forall t$
The second type corresponds to critical points of the latter map
(i.e., points at which the mapping from $\delta \varepsilon(t)$ to
$\delta U(T)$ is not locally surjective)
%$\frac{\delta U}{\delta \varepsilon(t)} = 0, \forall t$)
\cite{Wu2007}. In this section, we consider the first type of
critical point, which are referred to as kinematic critical points
or normal extremal controls. The second type, which are called
abnormal extremal controls, are comparatively rare in quantum
control problems, and we defer their study to section VII.
%$$\delta J = \int_0^T \frac{\dd J}{\dd U}
%\cdot \frac{\delta U}{\delta \varepsilon(t)}\delta \varepsilon \dd t.$$

The class of normal extremal controls is Hamiltonian-independent and
captures the most universal topological features of general quantum
control landscapes. Because the map $\varepsilon(t) \rightarrow
U(T)$ is locally surjective at these points, each of the matrix
elements $U_{pq}$ must be uniquely
%assumption that the class of singular extremal controls vanishes,
%Our assumption that the quantum system is fully controllable implies
%that the Jacobian $\frac{\dd J}{\dd U}$ is locally surjective at
%$\varepsilon(t)$ on the set of realizable gates at the time $T$.
addressable by the control field $\varepsilon(t)$ for all $p$ and
$q$ values in keeping with $U$ being unitary, i.e., the set of $N^2$
functions $\frac{\delta U_{pq}}{\delta \varepsilon(t)}$ should be
linearly independent. Hence, the critical condition for normal
extremal controls is equivalent to $\frac {\dd J}{\dd U} = 0$ for
arbitrary $\frac{\delta U}{\delta \varepsilon(t)}$. Moreover, it can
be shown (see Appendix \ref{appmap}) that the optimality status
(i.e., minimum, maximum or saddle point) of a critical control field
$\varepsilon(t)$ will be equivalent to that of the resulting
propagator $U(T)$ on the unitary group $U(N)$. The local
surjectivity of $\varepsilon(t) \rightarrow U(T)$ has important
connections to the controllability of the quantum system, discussed
in sections VI and VII. In what follows, we provide expressions for
the gradient and Hessian of the above objective functions on both
the domain of control fields and unitary propagators, and use these
results to characterize their
critical topologies. %$\frac{\dd J}{\dd U}$

\subsection{Observable maximization}

In 1937, John von Neumann first addressed the critical topology of a
problem that has direct implications for optimizing quantum
observables \cite{VonNeumann1937a, VonNeumann1937b}. Although it is
unclear whether von Neumann anticipated the applications of his work
to quantum control, this paper may be considered the first work in
the theory of quantum optimal control landscapes. This work was
recently extended by several authors \cite{RabMik2004,Koch1998}.

Within the electric dipole approximation, the gradient (\ref{grad1})
can be explicitly written \cite{HoRab2006a}:
%\begin{widetext}
%\begin{eqnarray*}
\begin{multline}\label{obsgrad}
\frac{\delta \Phi_1}{\delta \varepsilon(t)} =
-\frac{i}{\hbar}\tr\left[\left[\Theta(T),\rho(0)\right]\mu(t)\right]=\\\frac{i}{\hbar}\sum_i
\rho(0)\langle i|\Theta(T)\mu(t)-\mu(t)\Theta(T)|i\rangle = \\
\frac {i}{\hbar} \sum_{i=1}^n p_i \sum_{j=1}^N \Big[\langle
i|\Theta(T)|j\rangle \langle j|\mu(t)|i\rangle -\langle
i|\mu(t)|j\rangle \langle j|\Theta(T)|i\rangle \Big]
\end{multline}
%\end{eqnarray*}
%\end{widetext}
where the initial density matrix is given as $\rho(0)=\sum_{i=1}^n
p_i |i\rangle\langle i|, p_1 > ...> p_n > 0, \quad \sum_{i=1}^n p_i
= 1,$ and $\mu(t)\equiv U^{\dag}(t)\mu U(t)$. The local surjectivity
of $\varepsilon(t) \rightarrow U(T)$ at normal extremal controls
implies that the $N^2$ functions of time $\langle i|\mu(t)|j\rangle
$ are linearly independent. As discussed above, under this
assumption the critical condition is equivalent to that on the
domain of unitary propagators, $\frac {\dd \Phi_1}{\dd U} = 0$.
Because the gradient $\frac{\delta \Phi_1}{\delta \varepsilon(t)}$
depends on the eigenvalue spectra of $\rho(0)$ and $\Theta$, it is
convenient to simplify the analysis by investigating the critical
topology on the domain $U(N)$. Expanding the argument of the
objective function by $U \rightarrow U\exp{(iAs)},$ where $s$
parametrizes an arbitrary curve in the Lie algebra of $U(N)$ in the
direction $A$, the critical condition $\frac{\dd \Phi_1}{\dd U}=0$
can be expressed as

\begin{equation}
\frac{\dd \Phi_1}{\dd s} = i \tr\left(A\left[U^{\dag}\Theta
U,\rho(0)\right]\right)=0.
\end{equation}

The maximal subset of $U(N)$ which satisfies this condition is
composed of the matrices of the form

\begin{equation}
\hat U_l = QP_lR^{\dag}
\end{equation}

where $P_l,\quad l=1,\cdots, N!$ is an $N-$fold permutation matrix
whose nonzero entries are complex numbers
$e^{i\phi_1},\cdots,e^{i\phi_N}$ of unit modulus, and
$\rho(0)=Q^{\dag}\varepsilon Q$ and $\Theta=R^{\dag}\lambda R$.
$\varepsilon_1,\varepsilon_2,...,\varepsilon_N$ and $\lambda_1,
\lambda_2,...,\lambda_N$ are the eigenvalues of $\rho(0)$ and
$\Theta$ with associated unitary diagonalization transformations $Q$
and $R$, respectively. The critical set is the union of $N-$torii
$\bigcup_l T_l^N$  with each torus corresponding to a distinct
permutation $P_l$.

The number of critical submanifolds corresponding to suboptimal
landscape values scales factorially with system dimension for fully
nondegenerate $\rho(0)$ and $\Theta$.  If $\rho(0)$ and $\Theta$
have arbitrary numbers of degeneracies $D_1,...D_m$, and
$E_1,...,E_n$, respectively, it can be shown \cite{WuMike2007}  that
the critical submanifold dimension on the domain $U(N)$ for a
particular critical manifold $M_k$ is
\begin{equation}
d(M_k)=\sum_{s=1}^mD_s^2+\sum_{t=1}^nE_t^2-\sum_{l=1}^ro_l^2.
\end{equation}
%$$d(M_k)=N+\sum_{s=1}^mD_s^2+\sum_{t=1}^nE_t^2-\sum_{l=1}^ro_l^2.$$
where the $o_l$'s are the overlap numbers (numbers of overlapping
elements) between the degenerate blocks $D_1,...,D_m$ and
$E_1,...,E_n$, for the permutation matrix corresponding to that
manifold. These degeneracies cause neighboring submanifolds to
merge, and give rise to subspaces that are invariant to eigenvalue
permutations. If $\rho(0)$ or $\Theta$ is a pure state projector,
the number of critical manifolds scales linearly with Hilbert space
dimension $N$ \cite{Mike2006a}. In the limiting case where both
$\rho(0)$ and $\Theta$ are pure state projectors, $\Phi$ at the
extrema only has the values zero and unity, corresponding
respectively to no control or perfect control (i.e., the landscape
is convex).

In the electric dipole approximation, the Hessian of the objective
function can be written

\begin{eqnarray}\label{obshess}
\hil^{\varepsilon}(t,t') =
-\frac{1}{\hbar^2}\tr\big(\big[\left[\Theta(T),\mu(t)\right],\mu(t')\big]\rho(0)\big)
\end{eqnarray}
The Hessian quadratic form (HQF), defined as
\begin{equation}
\langle \omega|\hil|\omega \rangle =  \int_0^T \int_0^T \omega(t)
\frac{\delta^2 \Phi_1}{\delta \varepsilon(t)\delta
\varepsilon(t')}\omega(t') \dd t \dd t'
\end{equation}

where $\omega(t)$ is an arbitrary real function, is a polynomial
representation of the Hessian that facilitates the identification of
increasing, null and decreasing directions at each critical point.
The explicit representation of the HQF for the general case of
nondegenerate $\rho$ and $\Theta$ is complicated and is reviewed in
Appendix A.1.b.
%\ref{apphess}
Based on this representation, it can be shown that all of the
suboptimal critical submanifolds corresponding to $\Phi$ values less
than the global maximum are saddle regions, and thus will not act as
traps for optimal control searches.

\begin{figure*}%[h]
\centerline{
\includegraphics[width=5in,height=3in]{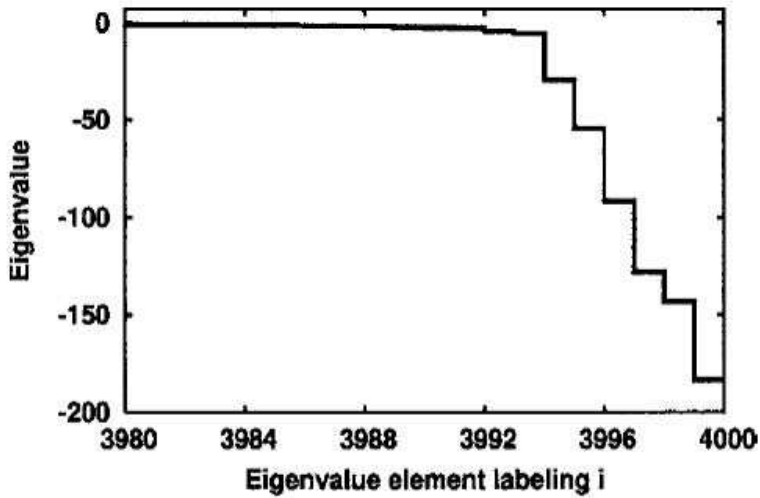}
} \caption{ %(1)
The dominant eigenvalues of the Hessian matrix for state-to-state
population transfer in a simple four level system. Only the last 20
eigenvalues are shown, and the remaining ones are essentially zero.
For this system with dimension N=4, it is evident that the 2N-2 rule
is satisfied with six nonzero eigenvalues being present. (From ref
\cite{Shen2006}.)
 }\label{shenfig}
\end{figure*}

As we will show in Section \ref{analytic} below, the dimension of
the global optimum on $U(N)$ is useful for exploring the degeneracy
of solutions to quantum observable control problems.  This number
ranges from $N$, for fully nondegenerate $\rho(0)$ and $\Theta$, to
$N^2-(2N-2)$, when $\rho(0)$ and $\Theta$ are both pure state
projectors.  In the former case, the number of decreasing directions
at the global maximum is the greatest ($N^2-N$), whereas in the
latter case it is the smallest ($2N-2$). Note that the landscape
mapping analysis in Appendix \ref{appmap} indicates that the number
of positive and negative principal axis directions of the Hessian
matrix will be preserved upon passage from the domain $U(N)$ to the
domain $\varepsilon(t)$, with the remainder of the directions on
$\varepsilon(t)$ being flat. A recent numerical analysis
\cite{Shen2006} confirmed these predictions, by considering the
problem of optimizing the expectation value of a pure state
projector over a $N=4$ quantum system, initialized in a pure state,
with the goal of $|0\rangle \rightarrow |3\rangle$ population
transfer. In this case, the Hessian can be expanded on a basis of
$2N-2$ linearly independent functions $\beta_l$, as

%CONVEX OPTIMIZATION
\begin{eqnarray}
\hil_{k'k}&=&\frac{\partial^2 \Phi_1}{\partial \eta_{k'}\partial
\eta_k}\\&=&\int_0^T\int_0^T\dd t \dd t' \frac{\delta
\varepsilon(t')}{\partial \eta_{k'}}\frac{\delta^2\Phi_1}{\delta
\varepsilon(t')\delta \varepsilon(t)}\frac{\partial
\varepsilon(t)}{\partial
\eta_k}\\&=&-\sum_{l=1}^{2N-2}\beta_l(k')\beta(k),
\end{eqnarray}
where $\eta$ is a vector of appropriate control parameters.
Diagonalization of the Hessian evaluated at the landscape maximum
revealed $10 = 4^2 - 2 \times 4 +2$ zero-valued eigenvalues, as
expected (Fig. \ref{shenfig}). Importantly, when the dipole coupling
strengths were reduced to negligible values for all but two of the
levels, the number of zero-values Hessian eigenvalues dropped to
$2=2^2-2 \times 2 + 2$, revealing the reduction of the 4-level to a
quasi 2-level system. Thus, the robustness of observable
maximization control solutions depends not only on the actual
Hilbert space dimension, but also on the effective number of states
that are accessible to the dynamics.

A related work recently studied the critical topology of observable
maximization on the domain $SU(N)$ (instead of $U(N)$)
\cite{Helmke2006}. It was shown that the set of maxima decomposes
into two connected components, and an explicit description of both
components was derived. Separately, Glaser et al. \cite{Glaser1998}
examined the topology of observable maximization for non-Hermitian
observables, such as those operators that arise in quadrature
detection. These works, as well as others
\cite{KhaBro2001,KhaBro2002}, were concerned with the problem of
identifying optimal controls for quantum observables in two steps:
first, solving numerically for the set of unitary matrices $U$ that
maximize the expectation value $\tr(U\rho(0)U^{\dag}\Theta)$ of the
observable $\Theta$, and then, finding a control field
$\varepsilon(t)$ that produces the quantum gate $U$ at time $t=T$.
We discuss analytical solutions to the latter problem in section III
below.

\subsection{Quantum gate control}

In the electric dipole approximation, the critical point condition
corresponding to the gradient (\ref{grad2}) can be explicitly
written \cite{RabMik2005,HoRab2006b}
\begin{equation}
\frac{\delta \Phi_2(U)}{\delta \varepsilon(t)} =
-\frac{i}{\hbar}\sum_i \sum_j (W^{\dag}U-U^{\dag}W)_{ij}\langle
j|\mu(t)|i \rangle = 0.
\end{equation}
The critical topology of the gate fidelity cost function on $U(N)$
was first studied by Frankel \cite{Frankel1965}. Assuming local
surjectivity of $\varepsilon(t) \rightarrow U(T)$, a necessary and
sufficient condition for the critical points is
$W^{\dag}U=U^{\dag}W$ or
\begin{equation}
(W^{\dag}U)^2=I.
\end{equation}

The solutions to this equation are the roots of $I$, i.e. $W^{\dag}U
= \textmd{diag}(\lambda_1,...,\lambda_n, \quad \lambda_i =
(-1)^{n_i}), n_i = 0,1$. These solutions fall into equivalence
classes indexed by the number of eigenvalues $\lambda_i = 1$. Thus,
there are a total of $N+1$ critical manifolds, taking on $\Phi$
values of $0,4,...,4N$. The number of suboptimal critical regions
hence grows only linearly with respect to the system Hilbert space
dimension $N$, a slower scaling than that of the landscape for
observable maximization when both $\rho$ and $\Theta$ are
nondegenerate full rank matrices.

The Hessian at the critical points is
\begin{equation}
\hil(t,t') = \frac{1}{\hbar^2}\tr
\{W^{\dag}U(\mu(t)\mu(t')+\mu(t')\mu(t))\}
\end{equation}
\bigskip
which can be expanded as \cite{HoRab2006b}

\begin{widetext}
\begin{multline} \hil(t,t') = \frac{2}{\hbar^2} \sum_i(-1)^{n_i}\langle
i|\mu(t)|i\rangle \langle i| \mu(t') |i \rangle + \frac{2}{\hbar^2}
\sum_i \sum_{j>i} \left[(-1)^{n_i}+(-1)^{n_j}\right] \\
\times \left[\textmd{Re}\left(\langle i |\mu(t)|j\rangle \right)
\times \textmd{Re}\left(\langle i|\mu(t')|j\rangle
\right)+\textmd{Im}\left(\langle i|\mu(t)|j\rangle \right) \times
\textmd{Im}\left(\langle i | \mu(t') | j \rangle \right)\right]
\end{multline}
\end{widetext}
The number of positive, negative and null directions at the critical
points can be determined by simple inspection of the Hessian
quadratic form. The expression for the HQF is
%(NULL SHOULD BE THE SAME?)
\begin{widetext}
\begin{multline}
\langle \omega | \hil | \omega \rangle = \frac{2}{\hbar^2} \sum_i
(-1)^{n_i} \left(\int_0^T \langle i | \mu(t) |i \rangle \omega(t)
\dd t \right)^2 + \frac{2}{\hbar^2} \sum_i
\sum_{j>i}\left[(-1)^{n_i}+(-1)^{n_j}\right] \\
\times \left[ \left(\int_0^T \textmd{Re}\left(\langle i | \mu(t)|
j \rangle \right) \omega(t) \dd t\right)^2 +
\textmd{Im}\left(\langle i | \mu(t)| j \rangle \omega(t) \dd t
\right)^2 \right]
\end{multline}
\end{widetext}

At the suboptimal critical points, there are $N-m$ even and $m$ odd
integers $n_i$. It can be shown that the number of positive and
negative Hessian eigenvalues equals the number of odd and even
$n_i$, respectively, and that the remaining eigenvalues are zero
\cite{RabMik2005}. The numbers of positive and negative directions
at a critical point $m$ are thus
\begin{equation}
h_+ = m^2; h_-=(N-m)^2,
\end{equation}
whereas all the remaining principal axis directions are flat. In
particular, all the local suboptima are saddle manifolds, and we see
again that there are no local traps in the quantum control
landscape. In contrast to the multiplicity of unitary matrices that
solve the observable maximization problem, the kinematic critical
regions of the landscape corresponding to global optima are isolated
unitary matrices \cite{RabMik2005}, although an infinite number of
controls may steer the system to those matrices.

\subsection{Continuous variable quantum control}

The kinematic landscape critical topology for controlling continuous
variable quantum dynamical transformations for systems with
quadratic Hamiltonians was recently studied \cite{WuRaj2007}. These
systems are relevant for the implementation of continuous variable
quantum information processing\cite{LloBra1998}. Continuous variable
transformations can be realized by harmonic oscillators, molecular
rotors, or coupled modes of the electromagnetic field. Dynamical
transformations for such systems can be represented by symplectic
propagators (Appendix \ref{symp}).

The critical topology of such landscapes offer insight into the
differences between control landscapes for discrete and continuous
quantum systems. Since the (quantum) symplectic gate $U$ is a
faithful unitary representation of a symplectic matrix $S$, it is
reasonable to define the gate fidelity analogously to that for
discrete gates as

\begin{multline}\label{control landscape}
\J[\varepsilon(t)]=\tr(S-W)^T(S-W)+(s-w)^T(s-w),\\
\quad S_s\in \ISp(2N,\R),
\end{multline}
where $s$ and $w$ denote phase space displacements. Importantly, the
symmetries of this objective functional again permit an analytical
characterization of the critical topology, although this topology is
more complex than that of the control landscapes for discrete
quantum systems. If we write the singular value decomposition of $W$
as $W=UEV$, the critical submanifolds can be expressed as
\cite{WuRaj2007}:
\begin{equation}
S^*=R^TDR,\quad R\in \G(E),
\end{equation}
where $R$ is an arbitrary orthogonal symplectic matrix in the
stabilizer of $E$ in $\OSp(2N,\R)$:
\begin{eqnarray}
\G(E)&=&\{R\in \OSp(2N,\R)|R^TER=E\}\\
&=&\OSp(2n_0)\times O(n_1)\times \cdots O(n_s).
\end{eqnarray}
The characteristic matrix $D$ consists of different operations on
the separate modes represented by diagonal blocks of three different
types, depending on the singular values of the target gate $W$. In
particular, $D$, and hence the critical topology, differs depending
on the degeneracy of the singular values. Again, all suboptimal
critical points are found to be saddle manifolds, indicating that
the control landscape for these infinite-dimensional quantum gates
is devoid of local traps.
%(The
%properties of these types of critical points, including their
%optimality status, are summarized in Appendix .)

\begin{figure*}
\centerline{
\includegraphics[width=4.5in,height=3in]{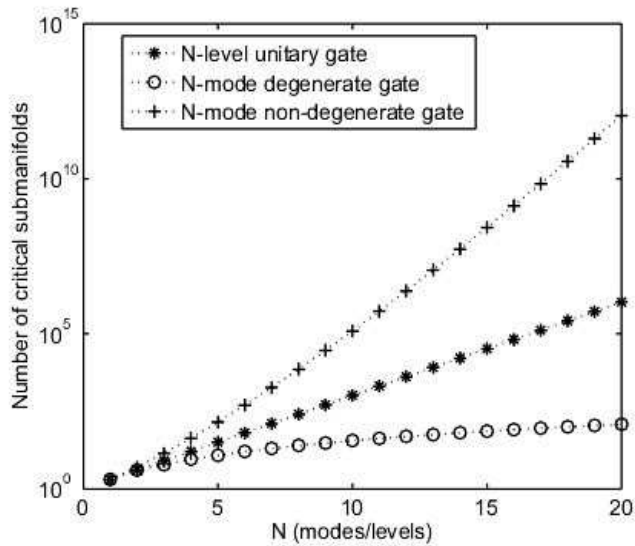}
} \caption{The scaling of the numbers of critical submanifolds for
discrete quantum (unitary) and continuous quantum (symplectic) gate
control landscapes with system size\cite{WuRaj2007}.
Degenerate/non-degenerate refer to the singular values of the
symplectic matrix representing the continuous quantum propagator.
Landscapes for the control of discrete quantum propagators all have
identical critical topologies, whereas those for continuous quantum
propagators are target-dependent. See the text for definitions of
quantum gate terminology. } \label{numbers}
\end{figure*}

Although this landscape is devoid of traps, it can be shown
\cite{WuRaj2007b} that the lower symmetry of the continuous variable
fidelity function, compared to that of the discrete quantum fidelity
function, can result in a more rapid scaling of the number of
critical manifolds. Moreover, the critical topology is dependent on
the target gate (Fig. \ref{numbers}).

The dynamical transformations of classical systems with quadratic
Hamiltonians can also be represented by symplectic propagators.
However, unlike quantum observables, classical observables may be
expressed as arbitrary smooth functions on phase space (and hence
the space of symplectic propagators); thus, unlike quantum control
landscapes, classical observable control landscapes have no
universally characterizable critical topology, either for quadratic
or more general classes of Hamiltonians.

For each of the control landscapes discussed above, an important
issue is the size of the attracting regions of these critical
manifolds and the behavior of the gradient flows of the objective
function around them. Unlike the critical topology, these gradient
flows (which represent a geometric property of the landscape) are
Hamiltonian-dependent, and we will revisit them, including their
connection to topology, in Section VII.

Although the above analytical results were derived under the
assumption that the fluence penalty coefficient $\lambda = 0$, they
remain valid in many cases in the presence of a significant cost on
the field fluence. In particular, in the case of observable
maximization where $\rho(0)$ and $\Theta$ are pure state projectors,
it was shown for the model system described in section IIA
\cite{Shen2006} that the Hessian retains $2N-2$ nonzero eigenvalues
in the presence of substantial fluence costs. In the next section,
we discuss how the imposition of such constraints facilitates the
identification of analytical solutions for certain (low-dimensional)
observable and gate control problems in quantum mechanics.

%\section{Analytical solutions to quantum optimal control problems}
\section{Analytical features of quantum control landscape
geometry}\label{analytic}

\subsection{The role of analysis in exploring control landscape geometry}

The previous sections showed that the critical point
\textit{topology} of the most common quantum optimal control
problems can be established analytically, and display properties
favorable for optimal search. In this section, we show that
analytical statements can also be made regarding the
\textit{geometry} of quantum control landscapes. The geometry of a
control landscape can be broken down into two components: 1) the
relationship among controls producing the same objective function
value (level sets), and 2) the search trajectory followed in
locating the optimal objective function value.

The geometry of quantum control landscapes is Hamiltonian-dependent.
The universal monotonicity of quantum control landscapes ensures the
convergence of local algorithms, but does not provide a direct
measure of the search effort involved in finding optimal solutions.
A further reduction in the search effort involved in locating
optimal controls could be aided by analytical landscape geometry.
Although these analytical results may not fully identify the
solution set to an optimal control problem, they may nonetheless
restrict its structure.

In the case that no constraints are placed directly on the controls
or on the time required for reaching the objective, an infinite
number of solutions exist to quantum control problems. As such,
analytical results pertaining to the geometry of the landscape are
restricted to statements regarding generic features of the controls.
When constraints or auxiliary costs are imposed in the objective
function, it is possible in the case of many low-dimensional
problems to explicitly solve for the optimal controls. In these
cases, multiple solutions may still exist, but they are distributed
among distinct unitary propagators.

Studying analytical solutions to quantum control problems provides
insight into the geometry of control landscapes in several ways.
First, there can exist quantum symmetries that reduce the
dimensionality of the domain of control fields $\varepsilon(t)$ over
which the control search must be carried out. From either a
computational or experimental perspective, this means that simpler
parametrizations of the control fields can be used in optimizations,
thereby reducing search effort.

Second, several important low-dimensional quantum optimal control
problems are analytically soluble (integrable) once auxiliary
constraints are imposed on the objective function. In particular,
the problem of realizing a quantum unitary transformation in minimal
time or with minimal fluence can be solved for Hilbert spaces of
several different dimensions. As shown in Section II, a continuous
submanifold of unitary matrices is associated with the maximization
of any observable expectation value. The set of points on this
submanifold can be identified (either analytically or numerically)
at minimal cost. Therefore, a submanifold of the set of all
solutions (one corresponding to each unitary matrix) to any
observable control problem for such systems can be obtained through
a combination of analytical and numerical methods with substantially
reduced search effort, due to the existence of analytical solutions
to gate control problems \cite{KhaBro2001,KhaBro2002}. This
submanifold consists of all solutions that minimize the auxiliary
cost (e.g., field fluence). Thus, analytical solutions to gate
control problems provide a means of further delineating the level
set geometry of observable control landscapes.

For these reasons, we review in this section analytical results
pertaining to the solution of quantum control problems. We discuss
1) analytical results pertaining to control mechanisms in arbitrary
Hilbert space dimension; 2) the integrability  of low dimensional
problems (which in some cases relies on 1), and 3) the reasons that
integrability breaks down in higher dimensions, whereas mechanistic
simplicity is retained.

The basic theorem of optimal control theory used for solving
problems of this type is the Pontryagin maximum principle, reviewed
in Appendix \ref{max}. Consider the control problem of minimizing a
cost associated with steering the system

\begin{equation}
\dot x = f(x,\varepsilon), \quad x \in \R^n, \quad \varepsilon \in
\Omega \subset R^k
\end{equation}
from some initial state $x(0)=x_0$ to some final state $x_1$. For
quantum gate and state control, we are dealing with the
right-invariant control systems $\dot U = f(U,\varepsilon) =
-\frac{i}{\hbar}\left[H_d+\mu\varepsilon(t)\right]U$  and $\dot \psi
= f(\psi,\varepsilon) = -\frac{i}{\hbar}
\left[H_d+\mu\varepsilon(t)\right]\psi$, respectively (see Appendix
\ref{appctrl} for a definition of right-invariance). Note that these
equations can be generalized to the case of $m$-independent
controls, which could take the form of, e.g., components of the
time-dependent electromagnetic field coupled to independent Pauli
spin operators in an NMR control experiment. When analytically
solving for optimal controls satisfying Pontryagin's maximum
principle (PMP), the maximization of the function $\Phi$ in equation
(\ref{OCT}) is often imposed as a fixed constraint in addition to
the Schrodinger equation, and the "cost" to be minimized takes the
form of the last term in this equation. Such problems are framed
most conveniently in Hamiltonian form. If we denote the cost as
$\int_0^T f^0(x,\varepsilon) \dd t$, then the PMP-Hamiltonian
function is defined as:
\begin{equation}
h(x,\lambda,\varepsilon) = \langle \lambda, f(x,\varepsilon) \rangle
+ \lambda_0f^0(x,\varepsilon)
\end{equation}
where the first term on the RHS is either a matrix or vector inner
product, depending on whether the problem is defined on the space of
state vectors or dynamical propagators, and the $\lambda$ play the
role of PMP-conjugate momenta (Appendix \ref{max}). The
PMP-Hamiltonian function takes on the following form for gate and
state control, respectively:

\begin{equation}
h(M,\lambda_0, u)\equiv
\tr\left[M\left(U_0^{\dag}(t)(H_d+\mu\varepsilon(t))U_0(t)\right)\right]+
\lambda_0f^0(t)
\end{equation}

\begin{equation}\label{hamstate}
h(P,\lambda_0, u)\equiv\langle
P,(H_d+\mu\varepsilon(t))\psi(t)\rangle + \lambda_0f^0(t)
\end{equation}
where $M$ is the conjugate PMP-momentum for gate control, $P$ is the
conjugate PMP-momentum for state control, and where we have
considered only the case of pure state population transfer in the
latter case, for simplicity. The Pontryagin maximum principle
(Appendix \ref{max}) then specifies the PMP-Hamiltonian equations of
"motion" for the control system; the solutions to these equations of
motion correspond to the solutions to the control problem, i.e.,
trajectory/control couples $(x(t),\varepsilon(t))$.

The auxiliary cost $\int_0^Tf^0 \dd t$ can take on several canonical
forms. The most common are (assuming $m$ independent controls): 1)
The field fluence for fixed transfer time $T$,
$E=\int_{0}^{T}\sum_{i=1}^m \varepsilon_i^2\dd t$, and 2) The total
transfer time, with fluences either unconstrained or subject to the
constraint $\int_0^T\sum_{i=1}^m \varepsilon_i^2 \leq C$, with $C$
an arbitrary positive constant, on the field amplitudes. Note that
in the case of 2), the final time in $\Phi(T)$ can be taken to be a
variable rather than a fixed parameter. An important distinguishing
feature between solutions corresponding to these different auxiliary
costs is the time-dependent structure of the corresponding optimal
control fields. When controls are bounded, the optimal fields are
typically resonant with the system transition frequencies, as
discussed further below. By contrast, when controls are unbounded,
the optimal fields are often singular (i.e., short sequences of hard
pulses) \cite{KhaBro2001}.

Since landscape geometry is Hamiltonian-dependent, it is important
to specify the class of Hamiltonians when studying analytical
solutions. We restrict the analysis primarily to state and gate
control problems defined on the special unitary group $SU(N)$.
Consider the following right-invariant control system on $U(N)$ with
$m$ controls:

\begin{equation}
\dot U(t) = -\frac{i}{\hbar} \left[H_d + \sum_{i=1}^m \mu_i
\varepsilon_i(t)\right]U(t)
\end{equation}
The matrices $H_d,\mu_i, i=1,...,m$ are skew-Hermitian matrices. If
we write $\bar H_d=D_{H_d}+H_d$ and $\bar \mu_i = D_{\mu_k}+\mu_k,
i=1,...,m$, with $D_{H_d}=\textmd{diag}(\frac{1}{2}\tr(\bar
H_d),\frac{1}{2}\tr(\bar H_d))$,
$D_{\mu_i}=\textmd{diag}\frac{1}{2}\tr(\bar
\mu_k),\frac{1}{2}\tr(\bar \mu_i)),i=1,...,m$, the matrices
$D_{H_d},D_{\mu_i}$ give a pure phase contribution to the solution.
These terms do not contribute to the relative phases of the
components of the state vector and therefore can be neglected since
states that differ only by a phase are physically indistinguishable.
Thus, we consider the physically equivalent problem on the domain of
traceless skew-Hermitian matrices, the Lie group $SU(N)$.

Just as analytical solutions to the Schrodinger equation exist only
for the simplest quantum systems, formal analytical solutions to the
PMP for quantum control problems involving arbitrary Hamiltonians
are very scarce. However, for certain classes of Hamiltonians, or
under certain physically reasonable approximations, control problems
of real practical interest may be integrable. In what follows, we
assume that the rotating wave approximation (RWA), reviewed in
Appendix \ref{rwa}, holds for the dynamics under consideration. It
is important to note that this latter condition is often not
satisfied, necessitating the use of numerical methods to solve the
most general class of control problems (Section \ref{level}).
However, in several of the most commonly encountered quantum control
problems, the RWA does hold to a reasonable approximation.

The methods of geometric control theory and sub-Riemannian geometry
\cite{Jurdjevic1997,KhaBro2001} provide a means of obtaining, in
certain specific cases, analytical solutions for the optimal control
fields reaching a given objective. Although geometric control theory
was originally developed in the context of classical control, it has
recently been shown that the Pontryagin maximum principle in a
geometric framework can be used to obtain analytical solutions for
optimal control fields for such low-dimensional quantum control
problems. We examine these specific solutions after briefly
reviewing analytical results pertaining to the control mechanisms
for fluence-minimizing state controls.

%\subsection{Analytic reduction of control dimensionality}
\subsection{Analytical solutions to state control
problems}\label{statesoln}

For discrete quantum control problems, the existence of symmetries
on the Hilbert space of states often allows a significant reduction
in the dimensionality of the problem and the parameterization of the
controls. This feature extends beyond the limited subset of
low-dimensional problems with analytical solution to discrete
quantum control problems in arbitrarily high dimensions. Let
$V(t)=\sum_{i=1}^m \varepsilon_i(t) \mu_i$ denote the total time
dependent control Hamiltonian. A problem of particular interest for
chemical applications is where control laser fields couple only
neighboring energy levels of the system, i.e., $V_{j,k} = V_{k,j} =
0$ if $j \neq k \pm 1$. This is a common scenario in strong field
control experiments. The optimal controls for these problems are
often in resonance with the transition frequencies of the
uncontrolled system. It can be shown that for auxiliary cost 1 above
(i.e., fluence minimization) with Hamiltonians of this form, the
controls will always satisfy a more general condition of "weak"
resonance. In either case, the search space of the problem is then
reduced from the Hilbert sphere $S^{2N-1}$ to $S^{N}$.

\begin{definition} (Resonance, weak resonance of optimal controls) A control $V_{j,k}(t)$ is said to be resonant with respect to the
uncontrolled system with state function $\psi$ if it has the
following form:

\begin{equation}V_{j,k}(t)=A_{j,k}(t)\exp
\left(i\left[(E_j-E_k)t+\pi/2+\phi_{j,k}\right]\right)
\end{equation}
where $A_{j,k}(.): \left[0,T\right] \rightarrow \R,
~A_{j,k}=-A_{k,j}$, $\phi_{j,k}\equiv arg(\psi_j(0))-arg(\psi_k(0))
\in \left[-\pi,\pi\right].$ Physically, this means that the lasers
oscillate with frequency $(E_j-E_k)/2\pi$; $A_{j,k}$ describes the
field amplitudes. A control $V_{j,k}$ is weakly-resonant if it is
resonant in each interval of time in which the states that it is
coupling (i.e., $\psi_j$ and $\psi_k$) are different from zero.
\cite{Boscain2002}.
\end{definition}
Let us denote these latter intervals $I_{j,k,l}$, where $j,k$
indexes the matrix elements of the Hamiltonian and $l$ indexes the
time interval where $\psi_j,~\psi_k \neq 0$. Then according to the
terminology above, $V_{j,k}$ is weakly resonant if
\begin{multline}
H_{j,k}(t)\mid_{I_{j,k,l}}=A_{j,k,l}(t)\exp i\phi_{j,k,l}\\
A_{j,k,l}(.): I_{j,k,l} \rightarrow \R, \quad A_{j,k,l}=-A_{k,j,l}.
\end{multline}

The phenomenon of resonance may be viewed geometrically as
originating from a rotational symmetry in Hilbert space, under the
transformation $Rot_{\alpha}: (\psi_1,...,\psi_n) \rightarrow
(e^{i\alpha_1}\psi_1,...,e^{i\alpha_n}\psi_n).$ The two admissible
curves $\psi(.) = (\psi_1(.),...,\psi_n(.))$ and
$Rot_{\alpha}(\psi(.))$ on $\left[0,T\right]$ have the same cost. In
particular, any point within the set $T_{\psi^2}$ generated by the
action of any element of $Rot_{\alpha}(\psi(.))$ on the state vector
$\psi^2$ can be reached at the same cost from any point within the
set $T_{\psi^1}$, defined analogously. Let us represent the controls
as
\begin{equation}
V_{j,k}^{l}(t) \equiv
V_{j,k}(t)\mid_{l}\equiv\left(u_{j,k}^{l}(t)+i\nu_{j,k}^{l}(t)\right)\exp(i\beta_{j,k}^{l}(t)),
\end{equation}
decomposing them into non-resonant and resonant time-dependent
parts. If $\psi(.):\left[0,T\right] \rightarrow S^{2n-1}$ is a
minimizing trajectory between sets $M_{\psi^1},M_{\psi^2}$, then the
transversality condition of the maximum principle (Appendix
\ref{max}) implies that $\langle P(t), TM_{\psi(t)} \rangle = 0$. We
write $\dot \psi_j
=\sum_k\left(u_{j,k}^{\l}F_{j,k}^{l}(\psi)+v_{j,k}^{l}G_{j,k}^{l}(\psi)\right)$,
%decomposing the time-dependent control Hamiltonians $u^{l}(t)$ and
%$\nu^{l}(t)$ into real and complex time-independent matrices $F$ and
%$G$  respectively, and corresponding scalar time-dependent
%coefficients $u_{j,k},~v_{j,k}$.
where $F(\psi)$ and $G(\psi)$ are vector fields subsuming the action
of the resonant contribution to the (real, imaginary) control
Hamiltonians on the state vector $\psi$. It can then be shown
\cite{Boscain2007} that $G_{j,k}^{l}(\psi)$ is always tangent to a
submanifold of $S^{2n-1}$ whose points are reached with the same
cost, i.e. $G_{j,k}^{l}(\psi(t)) \in TM_{\psi(t)}, ~ \forall t$ .

Therefore, in the maximum principle, $\langle
P(t),G_{j,k}^{l}(\psi(t))\rangle = 0$, and the maximality condition
of the maximum principle implies that $v_{j,k}(t)=0$. It follows
that it is possible to join any two eigenstates $\psi_j,~\psi_k$ by
a trajectory that is in resonance. For states that are not
eigenstates, the weak resonance condition holds.

State control and gate control for discrete quantum systems can both
be framed in terms of the identification of geodesic trajectories
under suitable metrics. The existence of a set of symmetries is
important for the identification of analytical solutions to these
problems. In particular, for state control problems, the reduction
in control dimensionality following from resonance is essential for
obtaining analytical solutions in low Hilbert space dimensions.

Analytical solutions to problems of the general class described
above can be obtained for population transfer in two- and
three-level quantum systems, for off-diagonal control Hamiltonians
whose Lie algebra spans the entire dynamical group. For instance,
for population transfer in three-level systems using fluence as the
cost with two controls, the PMP-Hamiltonian (\ref{hamstate}) becomes
\begin{multline}
h(P,\lambda_0, u)=\langle
P,(H_d+\mu_1\varepsilon_1(t)+\mu_2\varepsilon_2(t))\psi(t)\rangle
+\\ \frac{1}{2}\lambda_0 (\varepsilon_1^2(t)+\varepsilon_2^2(t)).
\end{multline}
Since our primary focus here is the relationship between multiple
quantum control solutions, we relegate a summary of this problem to
Appendix \ref{appsoln}. For our present purpose, its most important
features are that the assumption of resonant controls permits the
(sub-Riemannian) problem to be mapped from $S^5$ to $S^3$, and that
the resulting reduced Hamiltonian system is integrable. In four
dimensions, the corresponding state control problem resides on
$S^7$. It appears that the Hamiltonian system given by the maximum
principle is not integrable in this case, but the resonance
condition still holds, and simplifies numerical search in this and
higher dimensions.

Although the the class of systems described above - where controls
couple only two neighboring levels - is practically important, for
more general systems the optimal controls may not be resonant or
weakly resonant. Control mechanisms for such systems have been
studied using numerical methods described in Section \ref{level}.

\subsection{Analytical solutions to gate control
problems}\label{gatesoln}

As discussed above, quantum gate control solutions can be used to
significantly reduce the search effort required to obtain solutions
to the corresponding observable control problems. Analytical
solutions have been found for gate control problems in dimensions
2,4 and 8 for important classes of Hamiltonians, under the
assumption of unbounded controls \cite{KhaBro2001,KhaBro2002}. These
problems can be framed as so-called adjoint control problems, a type
of sub-Riemannian control problem where the optimal control
minimizes length on a geometric space under constraints on the
possible paths. An essential prerequisite for their analytical
solution is that the Riemannian space display certain symmetries,
which in this case are endowed by the geometry of the special
unitary group.

Instead of seeking fluence minimizing controls (cost 1) we consider
here the minimization of the  transfer time as the auxiliary cost
(cost 2 above), with unbounded controls. This problem is
particularly important for minimizing the time required for
coherence transfer in nuclear magnetic resonance (NMR) experiments
with radiofrequency pulses; the resultant time optimal pulses
outperform those typically used in NMR by a significant margin
\cite{KhaBro2001, KhaBro2002}. In the previous section we focused
attention on systems whose control Hamiltonians span the entire
dynamical group. Here, we examine the more common case where the
control Hamiltonians span only a subgroup of the dynamical group; if
controls are unbounded, the latter assumption is required for a
lower bound on the evolution time to exist, since unbounded controls
can attain the target in arbitrarily small time. The PMP-Hamiltonian
is then
\begin{equation}
h(M,\lambda_0, u)\equiv\tr\left[M
\left(U_0^{\dag}(t)(H_d+\mu\varepsilon(t))U_0(t)\right)\right] +
\frac{1}{2}\lambda_0.
\end{equation}
Note (Appendix A2) that the maximum principle for time optimal
control differs slightly from that for the problem of minimal field
fluence with fixed final time. In fact, for time-optimal control
problems where the controls do not span the dynamical group, framing
the problem in terms of an adjoint control system (and an associated
"adjoint maximum principle") facilitates solution, as shown below.

To understand this approach, let $G$ denote the special unitary
group $SU(N)$.  Under the assumption of full controllability of the
system, the algebra $s$ generated by the entire control system
$\{H_d,\mu_1,...,\mu_m\}$ is equal to the Lie subalgebra $su(N)$,
and the corresponding group $S$ is equal to $G$. We call the
subalgebra generated by the controls $\{\mu_1,...,\mu_m\}$ $l$, and
the corresponding subgroup $K$. Since the controls are unbounded,
any element of $K$ can be reached in arbitrarily small time.
Consider the problem of driving the evolution from $U_1$ to $U_2$ in
the shortest possible time, and let the coset $KU_1 = \{kU_1|k \in K
\}.$  We then need to find the fastest way to move from $KU_1$ to
$KU_2$, since the time required to travel anywhere within a coset is
negligible.

If we decompose $G=p \oplus l$ such that $p$ is orthogonal to $l$,
then $p$ represents all possible directions to move in $G/K$. All
directions in this set can be generated by using the control
Hamiltonians to place the system at appropriate starting points $k
\in K$, from which the drift Hamiltonian moves the system in the
directions given by $k_1^{\dag}H_dk_1$. However, we cannot access
all these possible directions directly. All motion in $G/K$ is
generated by the drift Hamiltonian $H_d$. These directions are
represented by

\begin{equation}
Ad_K(H_d) = \{Ad_{k_1}(H_d)=k_1^{\dag}H_dk_1| k_1 \in K\} \in
p,
\end{equation}
called the adjoint orbit of $H_d$ under the action of the subgroup
$K$. This form of direction control has been defined as an adjoint
control system, reviewed in Appendix \ref{max}.
%The set
%$Ad_K(-iH_d)$ is called the adjoint orbit of $-iH_d$ under the
%action of the subgroup $K$.

The goal is to find the shortest path between two points in $G/K$
under the constraint that the tangent direction must
always be in the adjoint orbit (Fig. \ref{khanfig}).%This is the content
%of the time-optimal torus theorem (Appendix \ref{torus}).
The shortest paths between points on a manifold subject to the
constraint that the tangent to the path always belongs to a subset
of all permissible directions are called sub-Riemannian geodesics.
The problem of finding time optimal control laws then reduces to
finding sub-Riemannian geodesics in the space $G/K$, where the set
of accessible directions is the set $Ad_K(-iH_d)$. This problem can
be framed in terms of an adjoint cost function,
$f(P)=\tr(\lambda^{\dag}\hil P)$, with $P\lambda^{\dag} \in p$. The
adjoint-PMP Hamiltonian is
$h(P(t),\lambda(t),\hil(t))=\tr(\lambda^{\dag}(t)\hil(t)P(t))$.
Solutions to the control problem follow from the adjoint maximum
principle (Appendix A2).

%PROB NEED THIS TO SUMMARIZE BELOW
For one and two spin systems, $G/K$ is a Riemannian symmetric space.
In this case, the decomposition
\begin{equation}
g= p \oplus l, \quad p = l^{\perp}
\end{equation}
satisfies the commutation relations $\left[l,l\right] \subset l,
\quad \left[p,l\right] = p, \quad \left[p,p\right] \subset l$. This
property implies that the tangent vectors to the path through $G/K$
must commute, since if they do not, a component of the path must lie
within $K$, and hence the path cannot be time optimal. Let $h
\subset p$ denote a subspace of maximally commuting directions or
generators in $G/K$ space.  Any unitary propagator $U_F$ can then be
written $U_F = k_2\exp(Y)k_1$, where $Y \in h$. According to the
time-optimal torus theorem, the fastest way to reach $U_F$ is on the
shortest path between the identity and the propagator $\exp(Y)$ such
that all tangent directions commute. If we express $Y$ as
\begin{equation}
Y=\sum_{i=1}^p \alpha_iAd_{k_i}(H_d), \quad \alpha_i> 0,
\end{equation}
it can be shown that the shortest path corresponds to the choice of
$\alpha_i$ with the smallest value of $\sum_{i=1}^p \alpha_i$. For
two-dimensional systems, $G/K$ is of rank 1, so $Y=\alpha Ad_k(H_d),
\quad \alpha > 0$ for some $k \in K$, and the time optimal path to
coset $\exp(Y)$ is to flow along $Ad_k(H_d)$ for time $\alpha$.
Qualitatively, the optimal controls are pulse-drift-pulse sequences,
i.e. hard pulses followed by evolution under drift and then some
hard pulses again. For four-dimensional systems, it is necessary to
pulse the controls intermittently to generate new $k$s, to create a
chained pulse-drift-pulse sequence.

Appendix \ref{appsoln} reviews the explicit construction of the
geodesic trajectories and associated minimal times for the problems
in dimensions 4 and 8, based on Pontryagin's maximum principle and
the time-optimal torus theorem. By first numerically determining the
set of unitary propagators mapping to a particular observable
expectation value, and then applying these analytical results to
identify the optimal field producing that propagator in minimal
time, we can obtain analytical insight into the relationship among
time-minimizing controls on any given level set of a quantum
observable control landscape.

In higher dimensions, $G/K$ is no longer a Riemannian symmetric
space, and it is necessary to move back and forth in noncommuting
directions to obtain the optimal path through $G/K$. In these cases,
the approach of representing the invariant control system on a Lie
group as an adjoint control system is still applicable, but analytic
solutions have not yet been found. Nonetheless, the application of
sub-Riemannian geometry to problems of quantum gate control is
currently a topic of intense interest, and promises to afford
additional analytical insights into the geometry of quantum control
landscape level sets.

Further analytical studies on the control of unitary transformations
(in two-level quantum systems) were carried out by D'Alessandro and
Dahleh \cite{DAlessandro2001a}. These authors studied the related
problem of fluence minimizing controls driving a two-level quantum
system to a target unitary propagator at fixed final time $T$. In
two dimensions, this problem also has analytical solutions, for both
single and multi-input control systems. The resulting optimal
controls have a more complicated temporal structure; it was shown
that the optimal  fields are always Jacobi elliptic functions
\cite{DAlessandro2001a}. Thus, the fluence minimizing solution does
not have the simple singular behavior of the unbounded
time-minimizing solutions, consistent with the phenomenon of
resonance discussed in the previous section.

\begin{figure*}
\centerline{
\includegraphics[width=3in,height=3in]{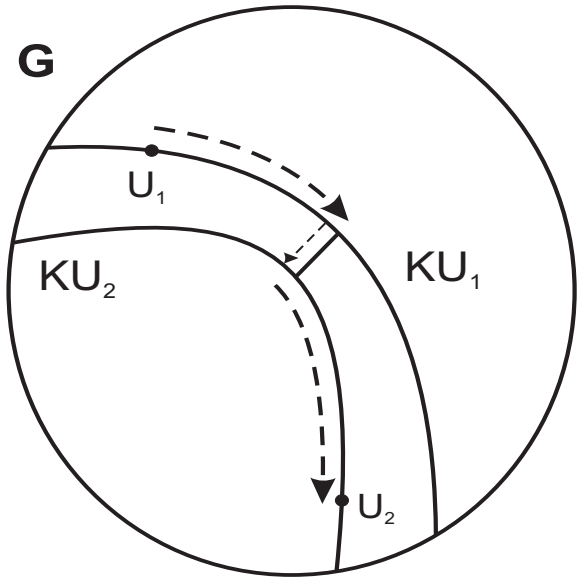}
} \caption{Time optimal path (dashed arrows) between elements $U_1$
and $U_2$ belonging to $G$. The long arrows depict the fast portion
of the path involving movement within coset $KU$ and correspond to
the pulse; the short arrow depicts the slow portion of the path
connecting different cosets and corresponds to evolution of the
couplings.}\label{khaneja2}
\end{figure*}

\section{Numerical exploration of quantum control landscape level sets}\label{level}
%Landscape level set properties

\subsection{Algorithms for level set exploration}
%Numerical exploration of level sets

We have seen that for certain classes of low-dimensional quantum
optimal control problems, analytical solutions for the control
fields exist.  However, as the system dimension increases,
analytical solutions become increasingly difficult to obtain.
Moreover, we have seen that the optimal controls for these problems
display particularly simple mechanistic properties, such as
resonance with the transition frequencies of the system.

For the problems studied above, auxiliary costs were imposed on the
controls, such as minimal time or minimal fluence. In the absence of
these constraints, quantum optimal control problems generally
possess an infinite number of solutions \cite{Demiralp1993}. A
quantum control level set (Fig. 2) consists of the collection of all
fields that produce a particular value for the target observable,
regardless of the intervening temporal dynamics (i.e., control
mechanism). A natural question concerns the relationship between
these degenerate solutions, and whether their associated control
mechanisms retain the simplicity of those for low-dimensional
systems. From a practical standpoint, a high degeneracy of solutions
will enable control fields to be tailored for specialized
applications in quantum technology, through the imposition of
auxiliary costs. Level set degeneracies also improve the robustness
of control solutions. There will inevitably be physical inaccuracies
in the experimental implementation (owing to the presence of noise,
decoherence) of a particular solution, and one would like the
nonideal fields to also produce dynamics that reach the objective.

Conventional algorithms for optimal control, being designed for
identification of the optima of the objective function, are not
well-suited to exploring the level sets of quantum control
landscapes. For this purpose, Rothman et al.
\cite{Rothman2005,Rothman2006a,Rothman2006b} developed a
diffeomorphic homotopy procedure for systematically exploring
diverse control fields on a landscape level set, referred to as
diffeomorphic modulation under observable-response-preserving
homotopy (D-MORPH), which we summarize here. The procedure can
selectively explore the control fields on a level set that display
desired properties. For example, the algorithm allows one to
numerically explore the control fields producing the various unitary
propagators on a level set with minimal fluence or in minimal time,
for control systems that are analytically intractable.

It is convenient to parametrize the field and its variation by the
exploration variable $s$:

\begin{equation}
\varepsilon(t)\Rightarrow \varepsilon(s,t)
\end{equation}
\begin{equation}
\dd \varepsilon(t) \Rightarrow \dd \varepsilon(s,t)
\end{equation}
where $0 \leq s \leq 1$. Since the goal is to explore the set of
control fields that are compatible with a given observable
expectation value, the solutions $\varepsilon(s,t)$ satisfy the
nonlinear equation

\begin{eqnarray}
F(s)&=&\langle \Theta(s) \rangle_T - C_T \\&=& \langle
\Theta(\left[\varepsilon(s,t),H_d(s),\mu(s)\right],T)\rangle -
C_T=0,
\end{eqnarray}
as a function of $s$, where $C_T$ is the desired observable
expectation value.

The maintenance of in $\langle \Theta \rangle$ over an infinitesimal
step $\dd s$ through the level set can be written
\begin{equation}\label{beqn}
\frac{\dd}{\dd s} \langle \Theta \rangle = \int_0^T \frac{\delta \langle \Theta \rangle}{\delta \varepsilon(s,t)}\frac{\partial\varepsilon(s,t)}{\partial s}\dd t = 0
\end{equation}
The neglected higher-order terms only become relevant near an
extremum, where $\frac {\delta \langle \Theta \rangle}{\delta
\varepsilon(t)}=0$. The relationship in equation (\ref{beqn}) is
highly underspecified for determining $\varepsilon(s,t)$ as $s$
traverses a level set. As shown in Appendix \ref{appmorph}, the
integral equation may be expressed as an equivalent initial value
problem

\begin{equation}
\frac{\partial \varepsilon(s,t)}{\partial s}= S(t)
\{f(s,t)-\frac{\gamma(s)}{\Gamma(s)}a_0(s,t,T)\}, \quad s \geq 0
\end{equation}
where \begin{multline} a_0(s,t,T)=\frac{\delta \langle \Theta
\rangle}{\delta \varepsilon(s,t)} = \\
-\frac{1}{i\hbar}\langle \psi_0|\left[U^{\dag}(T,0)\Theta
U(T,0),U^{\dag}(t,0)\mu U(t,0)\right]|\psi_0\rangle,
\end{multline}
Here $S(t)$ is an arbitrary weight function (e.g., it can bias the
control field towards a short pulse that approaches zero at the
endpoints of the time interval), $\gamma(s)= \int_0^T
S(t)f(s,t)a_0(s,t,T) \dd t$ and $\Gamma(s)=\int_0^T
S(t)\left[a_0(s,t,T)\right]^2 \dd t.$ The ability to freely choose
the function $f(s,t)$ permits exploration of the multiplicity of
solutions to the original integral equation. Regardless of the
choice of $f(s,t)$, $\langle \Theta(s) \rangle$ will remain
invariant over the $s \geq 0$ trajectory.
%A particular choice of $f(s,t)$ corresponds to a choice of auxiliary cost $f_a$
%in the Hamiltonian for the maximum principle, and will result in
%$\varepsilon(s,t)$ following a particular trajectory through the
%level set.

\subsection{Quantum control mechanisms and robustness}

Perhaps the most compelling reason to explore quantum control level
sets is the insight they offer into control mechanisms.  The ability
to transform one successful control into another, and therefore one
control mechanism into another, must be considered when seeking to
establish the mechanism of any particular quantum control problem.
Indeed, before making any definitive statements about mechanisms, it
is necessary to understand the diversity of controls on a level set.
How diverse can the solutions be, given the relatively simple
structure of the optimal controls for integrable problems?

\begin{figure*}
\centerline{
\includegraphics[width=3.5in,height=4in]{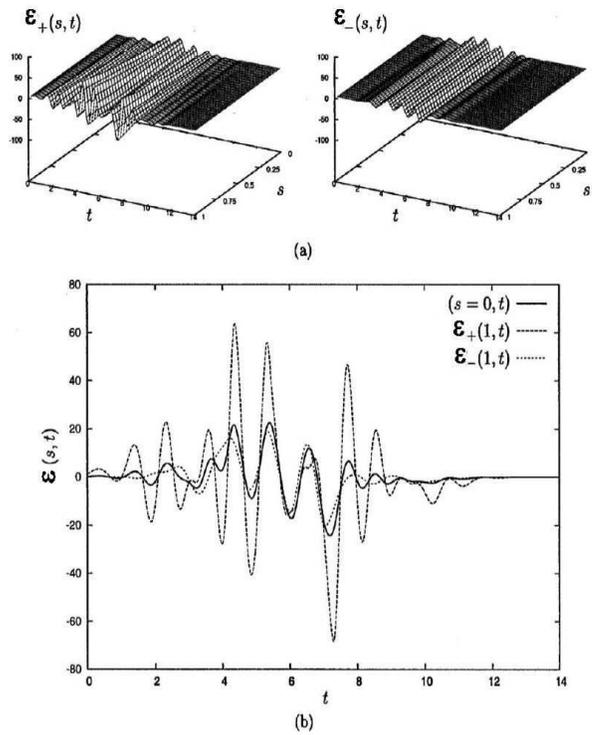}
} \caption{Starting from an initial control field
$\varepsilon(s=0,t)$, the control fields $\varepsilon_{\pm}(s,t)$
are evolved on the interval $s \in \left[0,1\right]$ subject to the
functions $f_{\pm}$, where $\pm$ refer to fluence maximization and
minimization, respectively. a) Under fluence maximization the
control field grows in amplitude and incorporates complex structure;
under fluence minimization the progression of fields decreases in
amplitude. b) Cross sections of the fields in a) are plotted. The
field at $s=0$ is the same for both $\varepsilon_{\pm}(s,t)$, but
different free functions $f(s,t)$ cause the fields to evolve in
dramatically different ways with s. (From ref
\cite{Rothman2006b}.)}\label{levelf}
\end{figure*}

In order to investigate this question, Rothman et al.
\cite{Rothman2006b} applied the D-MORPH technique to an eight-level
Hamiltonian with nondegenerate energy levels and with couplings only
between adjacent, next-nearest and next-next nearest states. The
control objective was state-state population transfer $|1\rangle
\rightarrow |8\rangle$; level sets of both high and low yield were
explored (Fig. \ref{levelf}.) In the case of a choice of $f(s,t)$
corresponding to fluence minimization, the control field asymptotes
as $s \rightarrow \infty$ towards a field of minimal fluence,
although a different asymptotic field is produced for each initial
field $\varepsilon(0,t)$.  By contrast, for a fluence maximizing
function $f(s,t)$, the distance between $\varepsilon(0,t)$ and
$\varepsilon(s,t)$ increases without bound. In the latter case, the
use of multiple transition pathways over the $s$ interval suggests
that the level sets are rich with fields producing vastly different
dynamics. The observation that controls of minimal fluence often
involve simpler mechanisms is consistent with the mechanisms
apparent in the fluence and time-minimizing analytical control
solutions discussed in the previous section. In order to assess the
robustness of control fields to noise along the level set, the
Hessian (\ref{obshess})  was evaluated at various points along the
trajectory. In the case of fluence maximization, the trace of the
Hessian was not preserved, indicating varying degrees of robustness
of $\langle \Theta(T) \rangle$ to noise in the control field
$\varepsilon(s,t)$.

\begin{figure*}
\centerline{
\includegraphics[width=6.5in,height=3in]{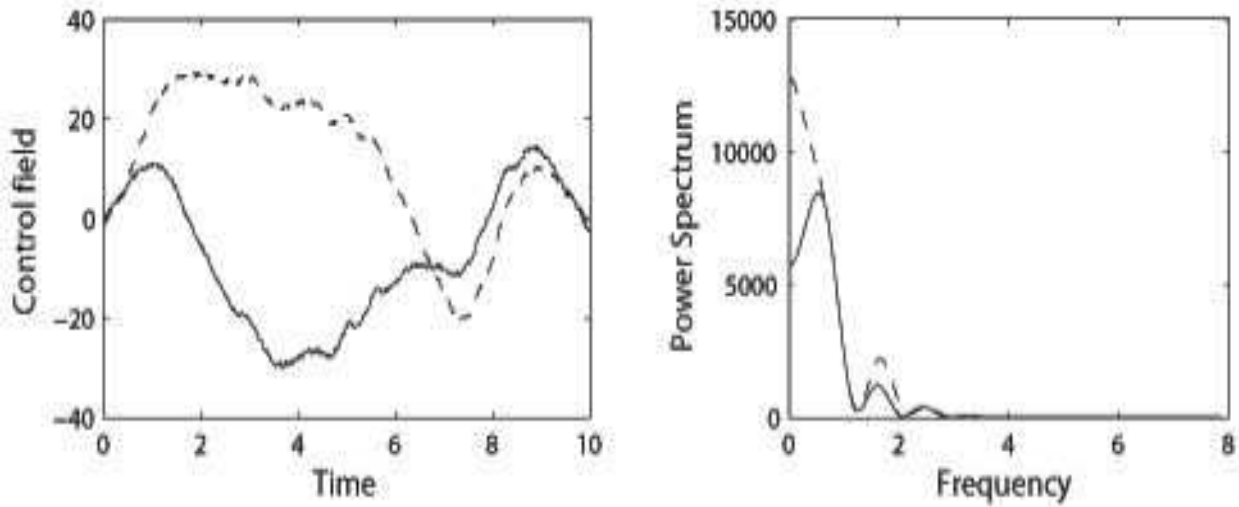}
} \caption{The optimal control fields and corresponding Fourier
power spectra for continuous quantum SUM gate control in a
controllable system, using two distinct control Hamiltonians. The
solid and dashed lines depict the associated control fields in
arbitrary units. Compare to the characteristic beat field structure
in Fig. \ref{levelf} representative of optimal controls for discrete
quantum systems. (From ref \cite{WuRaj2007}.)}\label{sumf}
\end{figure*}

In the weak-field regime, optimal fields for control of discrete
quantum systems are typically in resonance with the transition
frequencies of the system. By contrast, for continuous quantum
systems, the fields are usually not simply related to natural
resonant frequencies.  Wu, Chakrabarti and Rabitz compared the
mechanisms of optimal control for discrete and continuous variable
quantum gates \cite{WuRaj2007}. Figure \ref{sumf} depicts the
optimal control fields for achievement of the SUM logic gate (a
4-dimensional symplectic matrix) for a model two-mode continuous
variable system. The resulting optimal control fields typically
display complicated Fourier spectra that suggest a richer variety of
possible control mechanisms. The larger diversity of mechanisms at
work in continuous variable quantum control may have implications
for the comparative effort of locating continuous variable versus
discrete quantum control, since in the latter case the dimension of
the search space cannot be reduced by requiring that the optimal
controls adopt a canonical shape.

%strongly

\subsection{Hamiltonian-dependence of landscape level set geometry}
%structure

An important quantum control goal is to discern the distinct
controls that can achieve the same objective in each member of a set
of similar quantum systems. A common outcome might be, for example,
breaking the same type of bond in a set of molecules or creating an
analogous excited state in a family of related systems. The notion
of families of reactants in chemistry can be given a rigorous
meaning in terms of the similarity of optimal control fields driving
systems with related internal Hamiltonians to the same final state.

\begin{figure*}
\centerline{
\includegraphics[width=3.3in,height=3in]{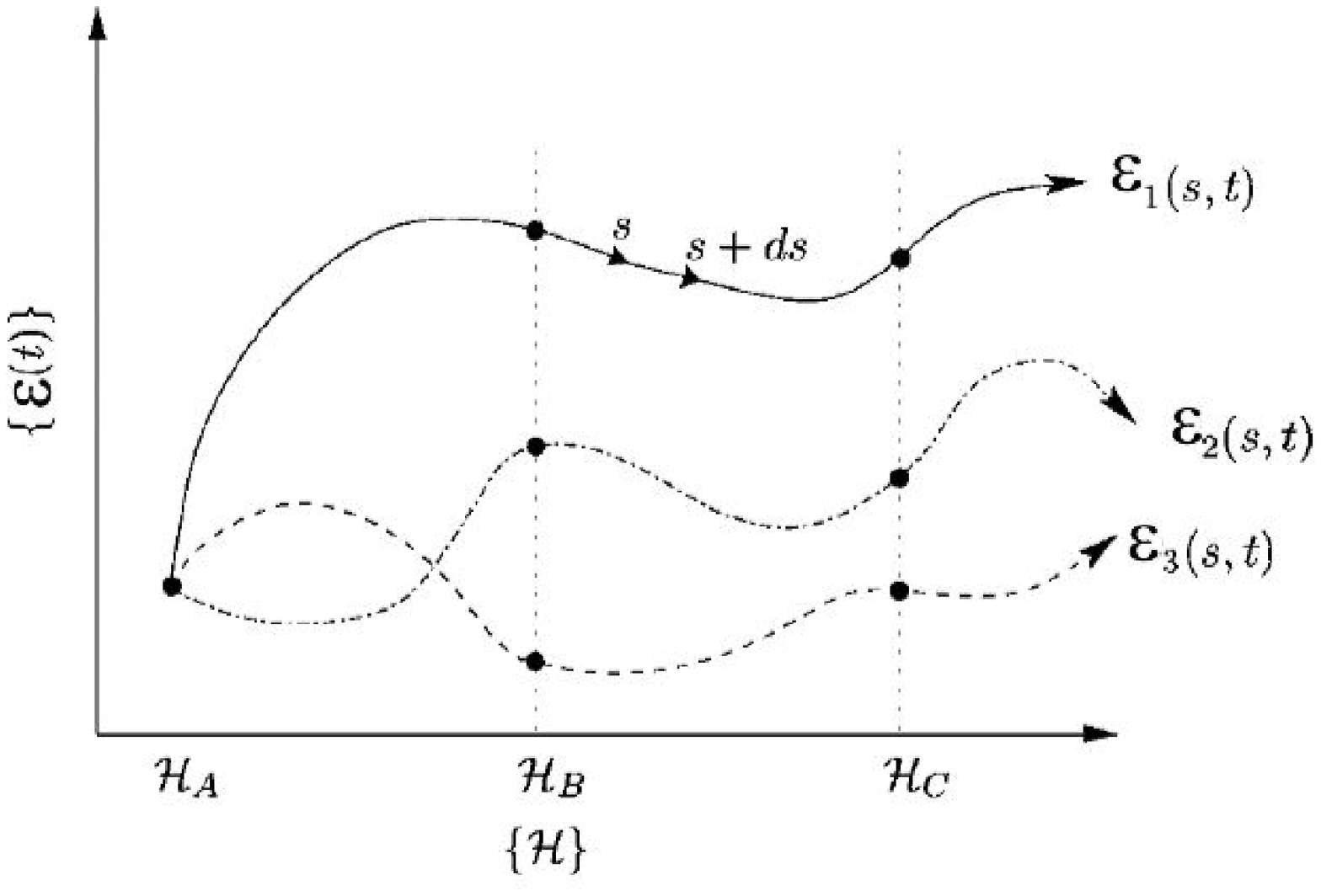}
\includegraphics[width=3.3in,height=3in]{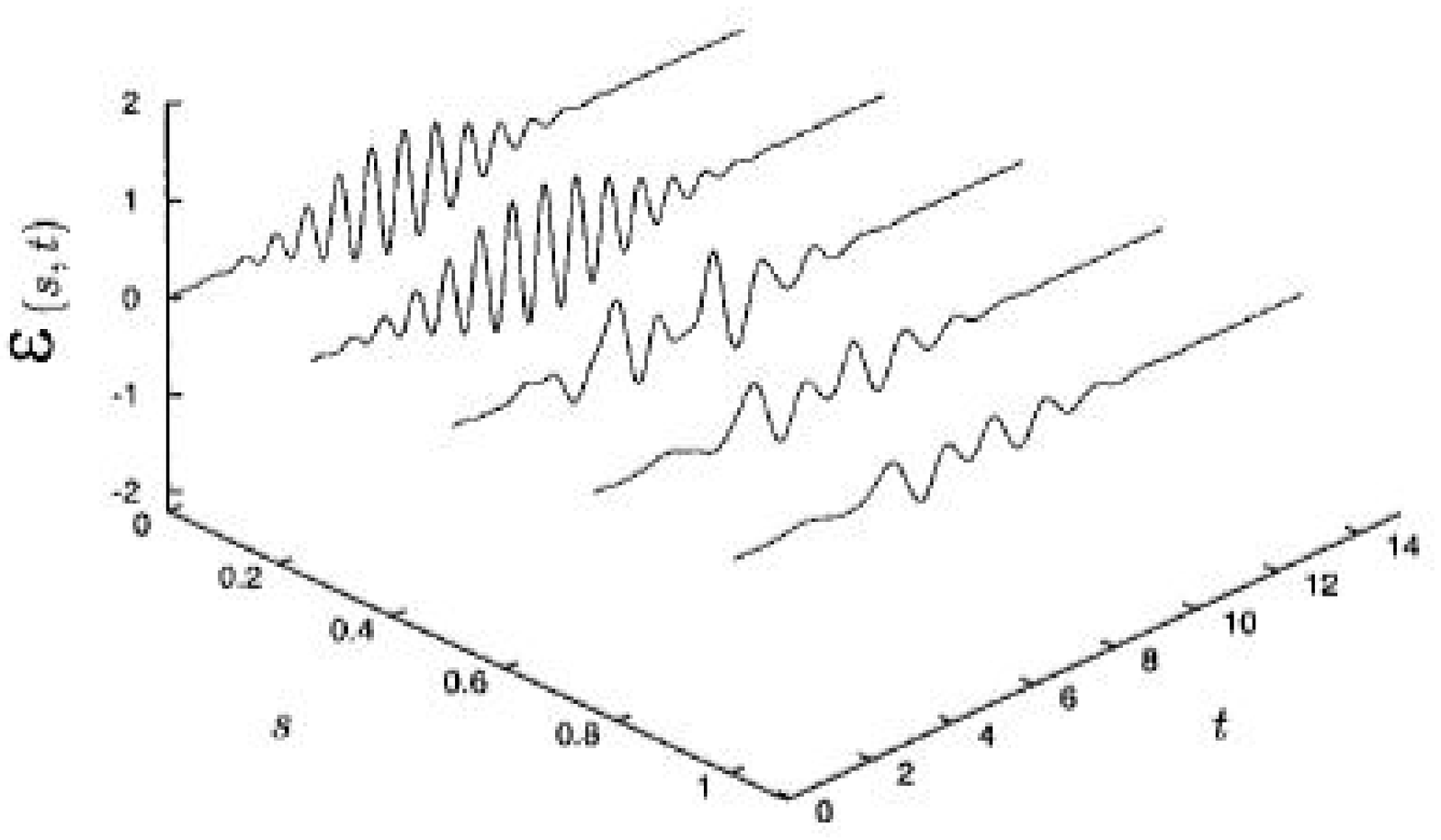}
} \caption{ (a) Starting from quantum system A with Hamiltonian
$H_A$, three paths are shown passing through systems B and C with
their associated Hamiltonians $H_B$ and $H_C$. Three particular
control fields $\varepsilon_i(s,t), i=1,2,3$ characterize the
pathways along which the common observable $\langle \Theta(T)
\rangle$ is preserved; these distinct observable-preserving controls
are specified by the auxiliary functions $f_i(s,t), i=1,2,3$. (b)
Control fields as functions of s and t for the combined internal
Hamiltonian/dipole diffeomorphism example discussed in the text
(with fluence minimization). Target observable population in state
$\mid 3 \rangle$ is preserved at $t=T$ for all $s$. (From ref
\cite{Rothman2005}.)}\label{hmorph}
\end{figure*}

The diffeomorphic homotopy approach described above for level set
exploration can be extended to study the relationship among controls
producing the same expectation value for homologous quantum systems
\cite{Rothman2005}. Diffeomorphic changes in the system Hamiltonian
are introduced by scanning over a homotopy parameter s and then
monitoring the control field response needed to maintain the value
of a  specified target observable. The time-dependent Hamiltonian is
written as a function of the homotopy parameter $s$ as
\begin{equation}
H(s,t)=H_d(s)-\mu(s)\varepsilon(s,t).
\end{equation}
Figure \ref{hmorph} schematically displays the concept of a
trajectory through Hamiltonian space.

%DONT USE O
It is possible to derive a differential equation for
$\frac {\partial \varepsilon(s,t)}{\partial s}$ for remaining on the
level set under s-dependent changes in the system Hamiltonian. The
following two terms

\begin{widetext}
%\begin{multline*}
\begin{eqnarray}
a_1(s,t,T)&=& -\frac{1}{i\hbar}\langle \psi_0 |
\big[U^{\dag}(T,0)\Theta U(T,0),U^{\dag}(t,0)\frac{\dd \mu(s)}{\dd
s}U(t,0)\big] | \psi_0\rangle\\
%\end{multline*}
%\begin{multline*}
a_2(s,t,T)&=& -\frac{1}{i\hbar}\langle \psi_0 |
\big[U^{\dag}(T,0)\Theta U(T,0),U^{\dag}(t,0)\frac{\dd H_d(s)}{\dd
s}U(t,0)\big]|\psi_0 \rangle
%\end{multline*}
\end{eqnarray}
\end{widetext}

analogous to $a_0$ above account for changes in the internal
Hamiltonian and dipole operator, respectively, along the trajectory.
As summarized in Appendix \ref{appmorph}, we obtain the explicit
initial value problem:

\begin{equation}
\frac{\partial \varepsilon(s,t)}{\partial s} =  f(s,t)+\frac{\left(b(s,T)-\gamma(s)\right)a_0(s,t,T)}{\Gamma(s)},\\
s \geq 0.
\end{equation}

Rothman et al. carried out numerical D-MORPH simulations across a
family of related three-level model systems forming a homologous
set. In these studies, transfer of pure state population was
considered, although the D-MORPH methodology is applicable to
arbitrary observable maximization problems originating from
arbitrary mixed states. The dipole moment operator and internal
Hamiltonian were varied both independently and in unison.

In the case of independent dipole diffeomorphism, the goal was to
move the population from state $|1\rangle$ to state $|3\rangle$. The
direct transition was initially allowed, by setting $\mu_{13}(0)
\neq 0,$ but finally forbidden, $\mu_{13}(1)=0$, while along the
alternative dynamical route the opposite situation exists, i.e.
$\mu_{12}(0)=\mu_{23}(0)=0$ and $\mu_{12}(1) \neq 0$ and
$\mu_{23}(1)\neq 0$.  Therefore, the population transfer occurs by
two different mechanisms at $s=0.0$ and $s=1.0$.  Two different
trajectories were followed through Hamiltonian space, one along a
straight path between the two dipole operators, and one along a
curved path. Figure \ref{hmorph} depicts the variations in the
control field required to preserve the observable across these
homologous quantum systems.

It is possible to redefine the problem of level set exploration for
families of related quantum systems, by identifying the set of
dynamically homologous quantum systems that produce the same
expectation value of a quantum observable when subjected to a fixed
time-dependent control field \cite{Beltrani2007}. Rather than
tracking over an arbitrary path in Hamiltonian space and determining
the change in the field that preserves the observable expectation
value, in this case the path in Hamiltonian space is determined by
the constraint that the control field does not change. An infinite
number of such paths are possible, just as in the former problem.
Topologically connected and disconnected families of homologous
Hamiltonians have been shown to exist under various conditions.
Numerical calculation of the Hessian of the associated cost
functional indicates that the critical topology of this landscapes
displays remarkably similar features to that for observable
expectation value control; in particular, the critical points appear
to be saddles rather than local traps, and the rank of the Hessian
at the critical points displays the same behavior with respect to
the degeneracies in the matrices $\rho$, $\Theta$.

\section{Experimental exploration of quantum control
landscapes}\label{exptl}

\subsection{Level sets}

Experimental methods are currently being developed for exploring
control landscape level sets. Because the domain of control fields
is infinite dimensional, the experimental investigation of quantum
control landscapes requires careful choice of parametrization of the
field such that the landscape can be sampled sufficiently.

\begin{figure*}
\centerline{
\includegraphics[width=6.5in,height=3in]{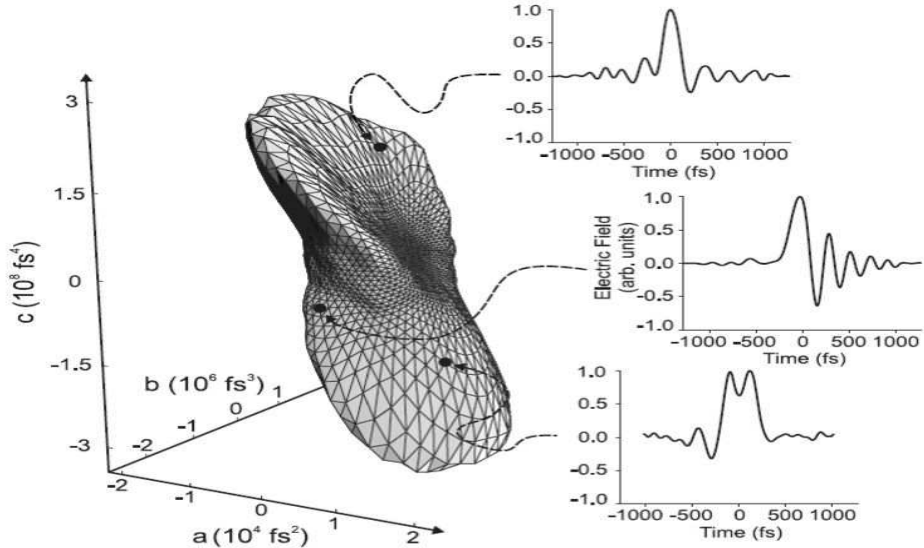}
} \caption{Experimental second harmonic generation (SHG) level set
surface for a yield of $50\%$. Included are three experimentally
retrieved control fields located on the surface. (From ref
\cite{Roslund2006}.)}\label{explevel}
\end{figure*}

\begin{figure*}
\centerline{
\includegraphics[width=6in,height=3in]{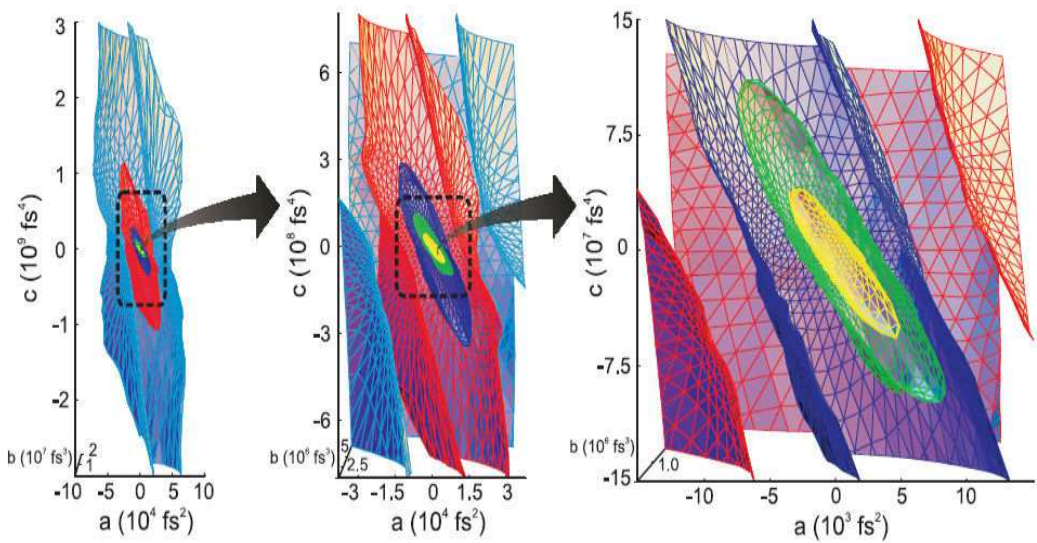}
} \caption{Experimental second harmonic generation (SHG) level set
surfaces with increasing magnification moving to the right for
$\alpha=0.10$ (light blue), $0.25$ (red), $0.50$ (dark blue), $0.75$
(green) and $0.90$ (gold). Each surface is sliced along its a-c
plane at $b=0$ so that the interior is visible. (From ref
\cite{Roslund2006}.)}\label{nested}
\end{figure*}

Roslund and Rabitz \cite{Roslund2006} explored the level set
surfaces for second harmonic generation and related nonresonant
two-photon absorption. The second harmonic spectral field is given
by

\begin{equation}
E_2(\Omega_2) \sim \int_{-\infty}^{\infty}
E_1(\Omega')E_1(\Omega_2-\Omega')\dd \Omega'
\end{equation}
where $E_2(\Omega_2)$ and $E_1(\Omega)$ are the complex spectral
envelopes of the second harmonic and control pulses, respectively,
and the frequencies of these envelopes are relative to their
spectral center, i.e. $\Omega = \omega-\omega_0$ and $\Omega_2 =
\omega-2\omega_0$. The time integrated signal, given by
\begin{equation}
S \propto \int_{-\infty}^{\infty} |E_1(t)|^4 \dd t = \int_{-\infty}^{\infty}
|E_2(\Omega_2)|^2 \dd \Omega_2
\end{equation}
is measured. The spectral phase
$\phi(\Omega)$ serving as the control is a truncated Taylor
expansion around the center frequency $\omega_0$,

\begin{equation}
\phi(\Omega) = \frac{a}{2} \Omega^2 +
\frac{b}{6}\Omega^3+\frac{c}{24}\Omega^4
\end{equation}
where the zeroth and first-order terms are discarded because they
simply correspond to an arbitrary constant phase and shift in the
time origin of the pulse, respectively. The level sets were
expressed on the domain of phase parameters $a,b,c$; multiple level
sets were identified, with one surface at $50\%$ yield shown in Fig.
\ref{explevel}. The level sets of different yield were nested in the
phase parameter space (Fig. \ref{nested}). Each of the continuously
varying control fields over a given level set preserves the
observable value by its own distinct manipulation of constructive
and destructive quantum interferences. Thus, the richness of quantum
control fields meeting a particular observable value is accompanied
by an equally diverse family of control mechanisms.

For these systems, the level sets were shown to be closed surfaces
in the control parameter space. In order to explore the origin of
phenomenon, the Hessian of the cost functional was computed
numerically at several points progressively farther away from the
global optimum. Based on the positive-definiteness of the Hessian,
it was shown that the level sets are predicted to be ellipsoids (see
below), consistent with the experimental observation. In general,
however, level sets can be unbounded in extent.

\subsection{Landscape topology}

In section II, we demonstrated that formally, no traps exist in
quantum control landscapes in the absence of direct costs or
constraints placed on the control field \footnote{Again, this
statement holds rigorously in the absence of abnormal extremal
controls. See section VII for a detailed discussion.}. Roslund and
Rabitz \cite{Roslund2007} made the first explicit experimental
demonstration of the trap-free, monotonic behavior of unconstrained
quantum control landscapes in the case of two systems, unfiltered
and filtered second harmonic generation (SHG). These landscapes were
randomly sampled and interpolated up to landscape level of data
noise. In order to explore the topology of the landscapes, 1500
trajectories originating at random points were propagated along the
gradient flow of the objective functional, according to the equation

\begin{equation}
\textbf{r}(s) = \textbf{r}(0) + \int_0^s \triangledown
S_t\left[\textbf{r}(s')\right] \dd s'
\end{equation}
where the gradient in the integrand was determined from the
laboratory SHG landscape data. $3\%$ (48) of these trajectories did
not converge to the global optimum, and ended up distributed among
two other local maxima, but these latter maxima were demonstrated to
be artifacts due to noise in the control apparatus.

\begin{figure*}
\centerline{
\includegraphics[width=4.5in,height=3in]{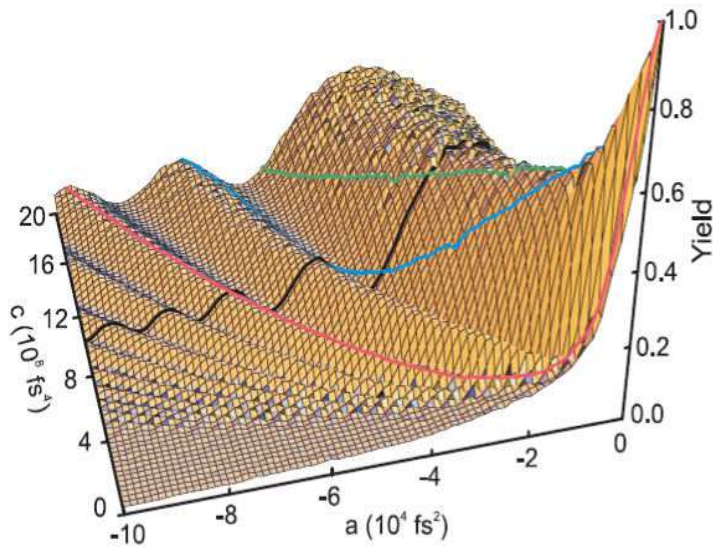}
} \caption{Experimental, unsmoothed quantum control landscape for
filtered second harmonic generation (SHG). Four trajectories are
shown; the landscape possesses a single global optimum that may be
reached monotonically by the curvilinear channels that slice through
the landscape. (From ref \cite{Roslund2007}.)}\label{filtered}
\end{figure*}

Figure \ref{filtered} depicts several possible search trajectories
along the filtered SHG landscape. (Filtering the SHG signal to be
evaluated at $\Omega_2=2\omega_0$ removes the dependence of the
signal on the cubic Taylor coefficient in the polynomial basis.)
Several of these paths correspond to simple parameterizations of the
control field. As can be seen, restricted parameterization of the
control field will generally produce artificial structure by forcing
projections of the original full infinite dimensional control space.

Although linear trajectories in the polynomial phase representation
considered above are incapable of following the gradient flow
trajectory, they do display the favorable property of preserving the
intrinsic topology of the landscape.  This can be verified by
perturbation analysis near the global maximum of the filtered SHG
landscape. Expanding the exponential phase to second order around
the optimal solution $\phi(\omega)=0$, it may be shown
\cite{Roslund2007} that the perturbative signal $\delta
S_f(\Omega)=S_f^*-S_f(\Omega)$, where $S_f^*$ is the transform
limited signal, indicates that the level sets are ellipsoidal. In
addition, assuming a spectral amplitude of the form
$A(\Omega)=\exp(\frac{-\Omega^2}{2\Delta^2})$, the normalized signal
variation under a Taylor expansion of the phase functional preserves
this landscape topology. In general, however, the appropriate choice
of control parameters or variables that preserve landscape topology
may not be apparent \textit{a priori}. Methodologies exist
\cite{Cardoza2005} for transforming to an optimal local basis set of
laboratory control parameters. A physically convenient
parameterization is first chosen prior to the outset of the
experiment, followed by a Hessian analysis to determine a locally
separable representation. This methodology has been successfully
illustrated using the example of molecular fragmentation of
$\textmd{CH}_2\textmd{BrI}$.

A possible cause for the appearance of local traps in quantum
control landscapes is the existence of costs or constraints on the
controls, which in some cases may be impossible to avoid. Numerical
results \cite{Shen2006} indicate that landscape topology should be
reasonably preserved even in the presence of small penalties on the
field fluence. These results suggest that in weak-field experiments,
where the control field constraints may not be particularly
limiting, the topology of experimental quantum control landscapes
should remain monotonic.

In the strong field regime, however, local traps may appear more
readily. For example, Wells et al. \cite{Wells2005} applied adaptive
search algorithms to the fragmentation of a complex molecule,
octahedral sulfur ($S_8$), investigating several different control
field parametrizations. The structure of the optimal pulses obtained
using these various parameterizations were considerably different,
although they produced comparable signal enhancements. The fluence
of the control fields were constrained considerably, corresponding
to large values of $\lambda$ in equation (\ref{OCT}). Although the
sampling was not exhaustive enough to rigorously establish level set
structure or landscape topology, the results suggested that local
maxima existed in the $S_8$ fragmentation landscape for the
parameterizations employed. However, even in the strong field
regime, the fundamental monotonicity of the landscape need not be
compromised. In related work \cite{Wollenhaupt2005}, the control
landscape for the strong-field ionization of potassium atoms was
sampled by phase modulated pulses. The intensity of the
Autler-Townes components in the photoelectron spectra were
controlled by a sinusoidal phase modulation function.  The use of a
two-dimensional parameter space enabled constrained, but effective
sampling of the control landscape. The maxima and minima of the
landscape were identified, with clear level sets, and no evidence
for local traps was found.

In most cases of practical interest in the weak-field regime,
unconstrained quantum control landscapes possess no suboptimal
traps. Thus, local experimental search algorithms should be
effective in locating optimal controls; moreover, since
gradient-based algorithms can take advantage of landscape structure,
they may perform better than "blind" algorithms (e.g. genetic
algorithms) under suitable conditions. In the aforementioned work of
Roslund and Rabitz, only the total SHG yield was measured; the
gradient of the objective was determined based on radial basis
function interpolation.  More recent work \cite{Roslund2007b} has
addressed the question of how to experimentally measure and follow
the gradient flow of the observable maximization objective function,
given that the basis must change at each step along the curvilinear
path. The gradient of the filtered SHG objective was measured using
a moment-based method with only 30 observable measurements on a
128-dimensional parameter space. Despite the statistical uncertainty
in the measurements and the presence of noise, following this flow
resulted in convergence to $> 90\%$ achievement in half the number
of steps that were required for a GA. Note that accurate estimation
of the gradients of observable expectation values with respect to
control field parameters is properly a subject of quantum
statistical inference \cite{Malley1993}. Quantitative assessment of
the statistical uncertainties associated with these estimates, as a
function of the number of measurements made, is essential for
determining the speedup that can be achieved by using gradient
versus adaptive search algorithms.

Following the gradient flow of the objective function exploits the
favorable topology of quantum control landscapes, but does not
explicitly exploit their geometry. In particular, the best path to
the global maximum in Fig. \ref{filtered} is not necessarily the
gradient flow path. A further possibility is to use local algorithms
that make use of local gradient information, but do not follow the
gradient flow directly, instead tracking alternative observable
expectation value paths. In this regard, an important question is
whether certain observable paths are expected to display more rapid
convergence to the global optimum than others. Section \ref{search}
discusses how geometric features of quantum control landscapes
indicate that certain observable paths may in fact be globally more
efficient than the gradient flow. Because they do not follow the
path of steepest ascent, these algorithms may require more accurate
estimation of the gradient. Such methodologies could also be applied
to explicit experimental tracking of predetermined level set
trajectories \cite{RajWu2007}.

In summary, improved methods for the measurement of the gradient in
the presence of noise, and the accessibility of control field
parameterizations capable of tracking the gradient are the central
challenges for implementing experimental control algorithms that
exploit the favorable topological and geometric structure of quantum
control landscapes. Note that the analytical results of Section III
pertaining to dominant quantum control mechanisms can also be used
to help choose appropriate control field parameterizations that
simplify experimental landscape search.

\section{Quantum system controllability and landscape structure}\label{control}

As discussed in section II, the full controllability of a quantum
system is a necessary condition for the nonexistence of local traps
in the control landscape. If the target transformation is reachable
at the final time T, but intermediate dynamical propagators are not,
the path to the target may be plagued with local traps. It is
therefore important to review the conditions for controllability of
quantum systems at a fixed dynamical time T. Moreover, since the
choice of T is to some extent arbitrary, it is important to identify
the possible choices of T that lead to full controllability.

The conditions for the controllability of finite-dimensional quantum
systems were established \cite{Ramakrishna1995} based upon earlier
work on the controllability of systems on compact Lie groups
\cite{JurdSuss1972}. In \cite{Ramakrishna1995}, an easily
implementable algorithm for ascertaining the controllability of a
finite-level quantum system was provided. Consider the
right-invariant system described by Definition \ref{invar} (Appendix
\ref{appctrl}); let $S$ be the subgroup of the dynamical group $G$
generated by the internal and control Hamiltonians (with
corresponding Lie algebra $s$), and let $L$ be the subgroup
generated by the control Hamiltonians alone (with corresponding Lie
algebra $l$). The following theorem establishes sufficient
conditions for the full controllability of such systems.

\begin{theorem}
Theorem (Controllability of right-invariant systems on Lie groups).
The reachable set from the identity matrix in $G$ is contained in
$S$. If $S$ is compact then the reachable set from the identity
matrix equals $S$. In particular, if the dimension of the Lie
algebra $l$ equals the dimension of the ambient Lie group $G$, and
the Lie group is compact, then the control system is controllable.
Furthermore, in this case it is possible to reach any matrix with an
admissible control which is bounded in amplitude.
\end{theorem}
%by an arbitrary finite set of constants.

Controllability can be checked using the rank condition, which
states that if the dimension is $N^2$ for the Lie algebra spanned by
$H_d$, $\mu_i$ and their commutators such as $[H_d,\mu_i]$,
$[H_d,[H_d,\mu_i]]$, $[\mu_i,[H_d,\mu_i]]$, etc., then the system is
controllable. This is equivalent to requiring that $l$ be the Lie
algebra of all $N \times N$ skew-Hermitian matrices, which in turn
is equivalent to requiring that the dimension of $l$ as a vector
space over the real numbers is precisely $N^2$. This criterion for
controllability of quantum dynamical propagators extends to that of
states; all coherent superpositions of states can be achieved if $S$
equals $U(N)$. The fact that the controllability of a linear control
system can be checked via a simple rank criterion which, in
addition, does not vary from point to point, is an important
property which is generally not valid for a nonlinear control
system. Usually this condition guarantees only accessibility
\cite{Ramakrishna1995}.

\subsection{Exact-time controllability of discrete quantum systems}

The above theorem establishes the necessary conditions for the
existence of a time $T$ at which the system is controllable, but
does not constructively define $T$. Landscape search would be
simplest if the quantum control system satisfied the conditions for
strong controllability.

\begin{definition} A control system $F=(A,B_i,u_i(t))$ is strongly controllable
over a subgroup $M$ if for any $T > 0$ any point of $M$ is reachable
from any other point by $F$ in $T$ or fewer units of time.  A
control system is said to be strongly controllable if the property
of strong controllability holds for the entire dynamical group $G$.
A control system is exact time controllable at time $T$ if any point
of $M$ is reachable from any other point by $F$ in exactly $T$ units
of time.
\end{definition}

Just as there exist analytical solutions to the Pontryagin maximum
principle for finite-dimensional quantum systems (i.e.,
right-invariant systems on a compact Lie group), there are also
powerful general theorems for establishing exact-time and strong
controllability of these systems. Also by analogy, there are
differences in the conditions establishing exact-time and strong
controllability of dynamical transformations versus that of quantum
states.

For finite-dimensional quantum systems, strong controllability can
be guaranteed if two controls are used, and these controls span the
whole Lie algebra of the dynamical group. In the more common case of
one control, strong controllability cannot be guaranteed, but
exact-time controllability across a wide range of times $T$ can be
straightforwardly established.

D'Alessandro and Dahleh \cite{DAlessandro2000} have studied the
exact-time controllability of two-qubit gates. They showed that if a
two qubit gate is controllable at time $T_1$ (called the critical
time), then it is also controllable at any time $T_2 > T_1$.  We
summarize their proof here because of its possible extensions to
higher-dimensional quantum systems. Let $R(T)$ be the reachable set
from $I$, i.e., the set of possible values for $X(T)$ obtained by
varying the controls $u_1,\cdots, u_m$ within the set of continuous
functions defined on $\left[0,T\right]$. We also define the sets

\begin{equation}
\R(\leq T) = \bigcup_{0\leq t\leq T}R(t)
\end{equation}
\begin{equation}
\R = \bigcup_{0 \leq t < \infty} R(t)
\end{equation}
Because of right-invariance, $R(I,T)S = R(S,T)$ for every $S \in
SU(2)$ and every $T$. Therefore, it suffices to consider the
(exact-time) controllability properties of the set reachable from
the identity. For quantum systems of arbitrary dimension, the
following theorem establishes the existence of a critical time
beyond which the reachable set $\R(T)$ is equal to the entire
unitary group $U(N)$ of dynamical propagators.

\begin{theorem}\label{jurd} \cite{JurdSuss1972}. Let $S$ denote the subalgebra generated
by the controls $A, B_1,\cdots,B_m$.  If $(A,B_i,u_i(t))$ is a
right-invariant control system on a Lie group $G$ and $S$ is
compact, (i) $\R = S$; (ii) There exists a $T > 0$ such that
$\R(\leq T) = \R$.
\end{theorem}

Powerful exact-time controllability results may be proven in
dimension 2 because a Lie algebra isomorphism $\gamma$ exists
between $su(2)$ and $so(3)$. It follows from Lie's third theorem
\cite{DAlessandro2000}, that $\rho$ induces a homomorphism $\gamma'$
mapping $SU(2)$ onto $SO(3)$. It can be shown that if a system is
controllable on $SO(3)$, any element of the subgroup $SO(2)$ can be
reached in arbitrarily small time (i.e., the system is small-time
controllable on $SO(2)$). It is then straightforward to demonstrate
that on either $SO(3)$ or $SU(2)$, if $T_1 \leq T_2$, then $R(T_1)
\subseteq R(T_2)$ and therefore $\R(\leq T) = R(T)$ for each $T \geq
0$ \cite{DAlessandro2000}. In other words, if $T_1 < T_2$, all the
points reachable at $T_1$ are reachable at $T_2$. Combining this
result with Theorem \ref{jurd}, we obtain the the following
exact-time controllability result in dimension 2:

\begin{theorem} There exists a time $T_c$, such that $R(T) = SU(2)$ for every
$T>T_c$. The critical time $T_c$ is the least time such that for
every $T > T_c$, it is possible to drive the system from the
identity to an arbitrary matrix in $SU(2)$.
\end{theorem}

An important question is whether this property can be extended to
finite quantum systems of arbitrary dimension $N$, i.e., whether the
above equivalence between the sets $R$ and $\R$ holds in general. If
this is the case, quantum control landscapes for larger systems will
display a homogeneous structure for all $T$ above a critical time,
such that simulations and experiments need not sample extensively
over the time $T$ in order to obtain a landscape with simple
topology.

\subsection{Controllability of classical and continuous variable
quantum systems}

For the noncompact Lie groups describing the evolution of classical
or continuous variable quantum systems, strong controllability is
not established by the above Lie algebra rank condition. This
implies that the critical topology of control landscapes for such
systems may change considerably as the final dynamical time $T$ is
varied, underscoring the comparative simplicity of discrete quantum
control landscapes.

For continuous variable (infinite-dimensional) quantum systems, the
rank condition is also sufficient for establishing controllability
on noncompact symplectic groups (i.e., those with quadratic
Hamiltonians) in the common case where $H_0$ is compact, but there
is no guarantee of exact-time controllability, i.e., some particular
gates may only be reachable after an extremely long time. However,
exact-time controllability can be achieved at arbitrary positive
times (i.e., $R(T) = \R(\leq T) = \R = U(N)$  for every $T>0$) if we
can employ two control Hamiltonians that span the whole Lie algebra
of the group of dynamical propagators. Thus, the topology of control
landscapes for the subset of infinite dimensional quantum gates
described in section II will be largely insensitive to the final
time T if two independent controls are used.

The strong controllability of general infinite-dimensional quantum
systems (i.e., those with nonquadratic Hamiltonians) was studied by
Wu and Tarn \cite{WuTarn2006}. Such systems were shown to be
associated with dynamical symmetries represented by noncompact Lie
groups with infinite-dimensional unitary representations. A
criterion for approximate strong controllability, called smooth
controllability, was given, showing that such systems, which possess
an uncountable number of levels, can be well manipulated using a
finite number of control fields.

\begin{figure*}
\centerline{
\includegraphics[width=3.5in,height=3in]{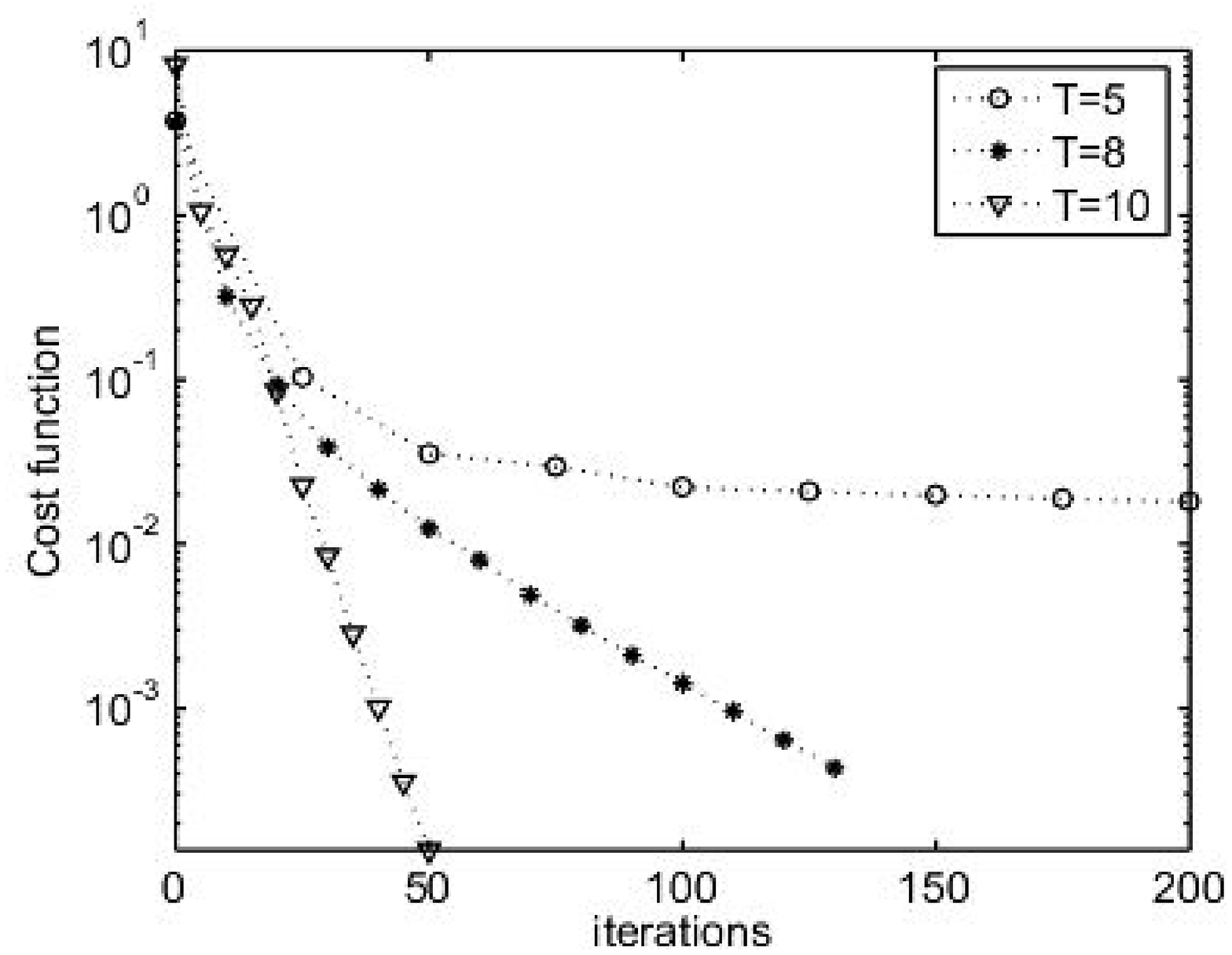}
\includegraphics[width=3.5in,height=3in]{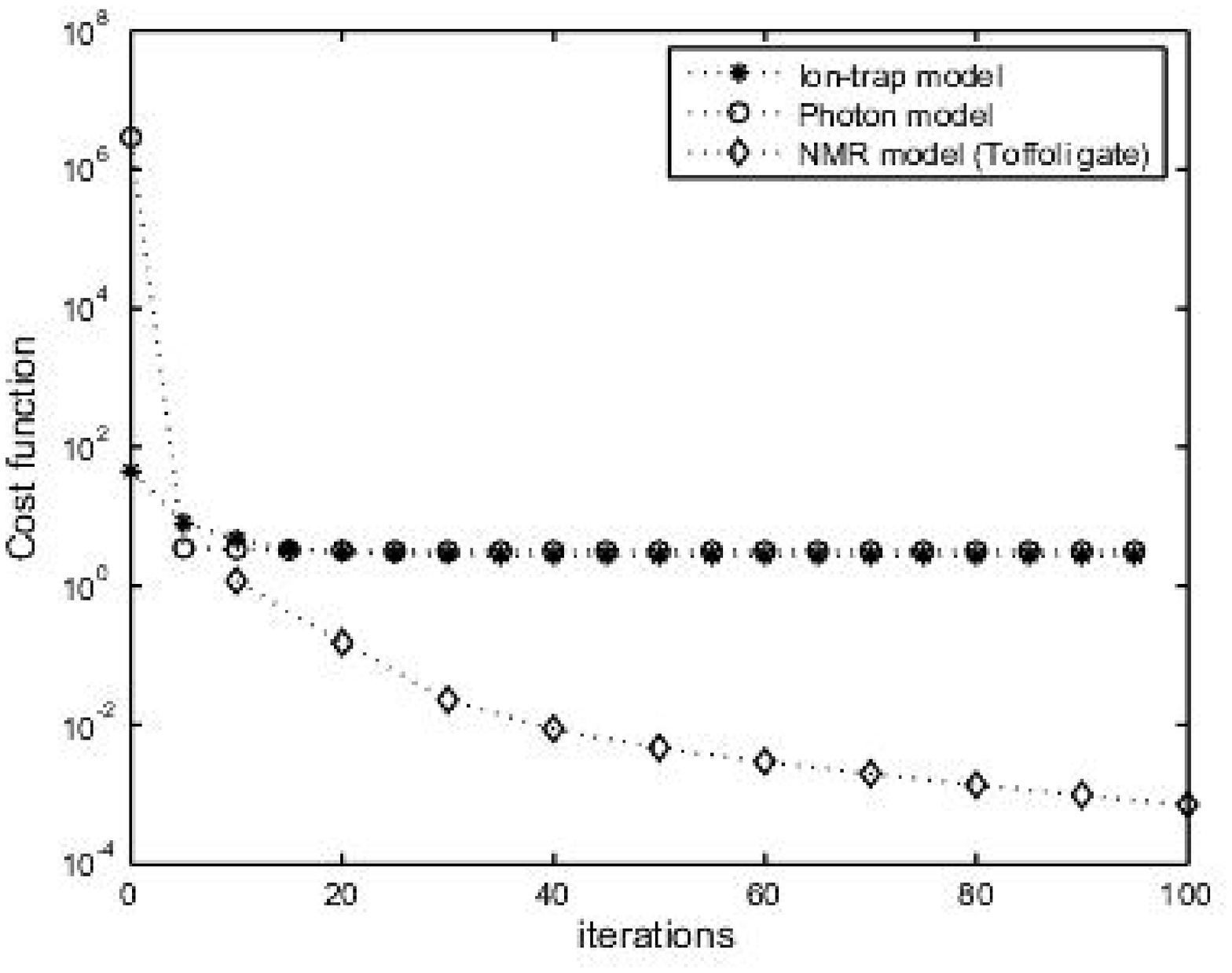}
} \caption{Effects of exact-time controllability on the optimal
control fidelity of discrete vs. continuous quantum systems. (a) The
convergence of control field searches for the 2-qunit continuous
quantum SUM gate, with a weakly controllable continuous variable
system, using conjugate gradient algorithms and different final
times. (b) The convergence of optimal searches for a 3-qunit SUM
gate with a weakly controllable system (photon model), 3-qunit SUM
gate with an uncontrollable system (ion-trap model) and 3-qubit
discrete quantum Controlled-CNOT gate with a standard NMR spin
coupling model. See text for definitions of quantum gate
terminology.(From ref \cite{WuRaj2007}.)}\label{controllable}
\end{figure*}

\begin{figure*}
\centerline{
\includegraphics[width=5in,height=4in]{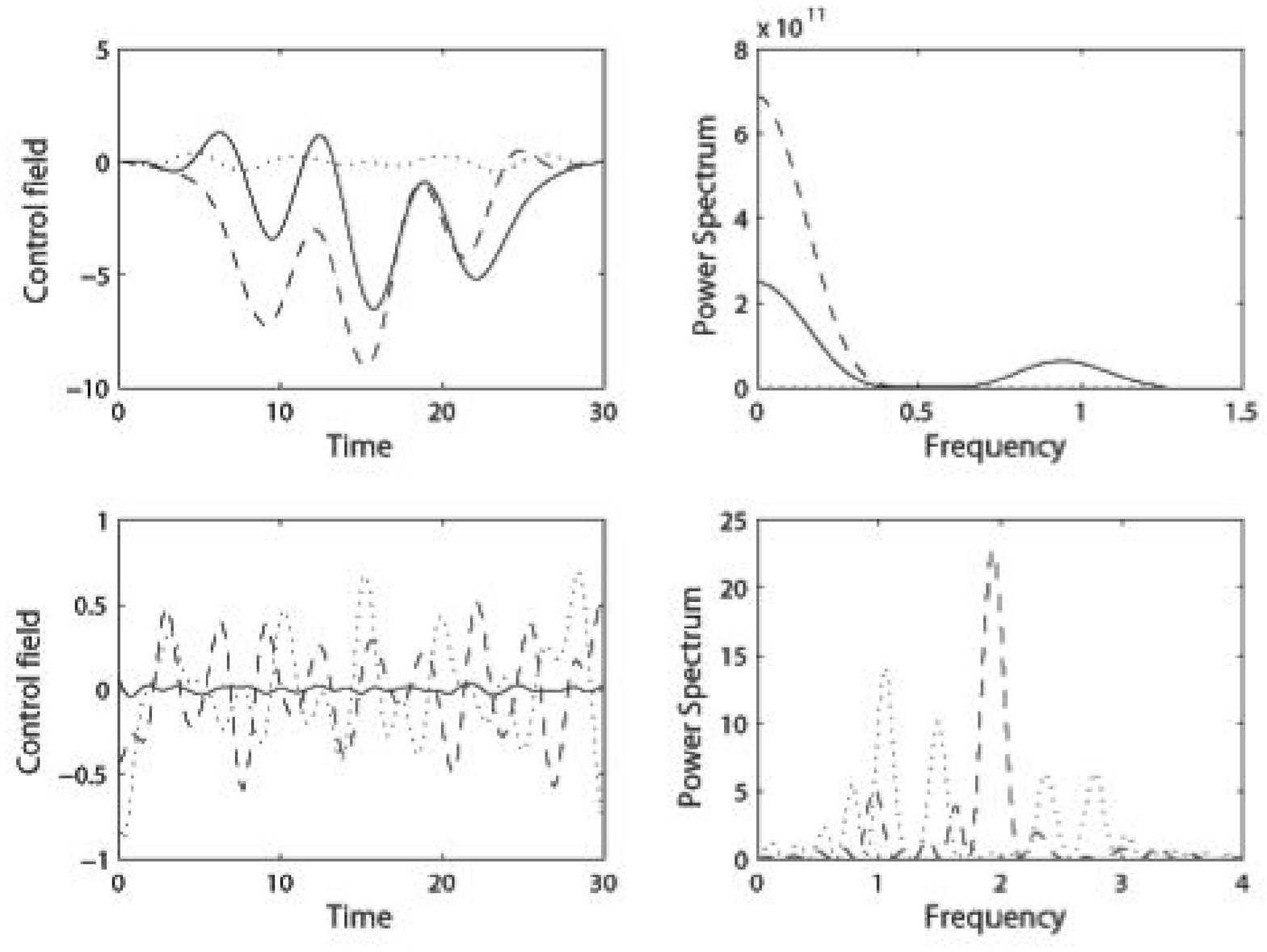}
} \caption{Comparison of the optimal control fields for discrete and
continuous variable quantum gates depicted in Fig.
\ref{controllable}b. Three independent controls were employed,
represented by solid, dashed, and dotted lines, respectively. (a)
Optimal fields for 3-qunit continuous quantum SUM gate control in a
weakly controllable system; (b) Optimal fields for 3-qubit discrete
quantum Controlled-CNOT gate control in a standard NMR spin coupling
model. Control field amplitudes are arbitrary. (From ref
\cite{WuRaj2007}.)}\label{controllable2}
\end{figure*}

The effects of exact-time controllability of discrete versus
continuous variable quantum systems on the simplicity of control
field search were studied by Wu, Chakrabarti and Rabitz
\cite{WuRaj2007}. Fig. \ref{controllable} compares the convergence
of optimal searches for achieving the CV SUM logic gate (symplectic
propagator) versus the computationally equivalent discrete CNOT gate
(unitary propagator), using identical gradient-based algorithms. The
CV gates were implemented using several models with varying degrees
of controllability, including a weakly controllable system
(employing ion trap interactions) and an uncontrollable system
(employing photon-atomic spin interactions). As can be seen, the
maximal achievable fidelity is highly sensitive to the choice of
final time for uncontrollable or weakly controllable systems.
Moreover, even when the gate was reachable, the search effort
required for convergence was found to increase with decreasing
controllability. By contrast, a randomly chosen final time was
sufficient for achieving near perfect fidelity in the discrete
quantum system. The weaker controllability of continuous systems can
also result in control fields with more complicated Fourier power
spectra, as shown in Fig. \ref{controllable2}.

\section{Computational complexity of quantum control
landscapes}\label{complexity}

The scaling of the expense of quantum simulation with system
dimension is of fundamental importance in quantum chemistry.
Similarly, the scaling of the expense for quantum control search -
either OCT or OCE - lies at the heart of the applicability of
quantum control to the large molecules of practical interest in many
applications. Whether the search is carried out numerically or
experimentally, this scaling is referred to as the problem's
computational complexity. Compared to other central optimization
problems in quantum technology and quantum information, such as
state or process reconstruction \cite{Hradil2003}, the complexity of
quantum control problems is more difficult to assess since the
optimization is carried out over an infinite-dimensional parameter
space.

We have seen that analytical solutions to quantum optimal control
problems appear to exist only for low-dimensional systems, and in
some cases only for restricted Hamiltonians. Nonetheless, the
existence of analytical solutions is rare in classical OCT, and begs
the question of how the difficulty of finding solutions to
nonintegrable quantum control problems increases with system
dimension. Of course, this complexity is algorithm-dependent. It is
also a function of the system Hamiltonian, but may display
homogeneous features across families of Hamiltonians. Recall that in
Section III, we showed that for certain classes Hamiltonians,
analytical results pertaining to control mechanisms permits
reduction in the dimension of the search space.

The search effort required in OCT calculations for observable
maximization is remarkably insensitive to the dimension of the
quantum system. A multitude of such calculations have been reported
in the literature \cite{Maday2003,Balint-Kurti2005,Amstrup1995}, on
systems ranging in dimension from 2 up to more than $10^2$. Even for
the largest systems studied, the number of iterations required for
convergence seldom exceeds $10^3$, with most calculations requiring
between $10^2$ and $10^3$ iterations\cite{Brixner2003,Levis2001}.
Much larger systems have been subjected to OCE studies; in the
common case of selective bond dissociation, a continuum of states is
accessed. Nonetheless, the search effort reported in most OCE
studies is of the same order of magnitude as in OCT.

Since the majority of quantum OCT algorithms are based on the
gradient of the objective function, and the gradient is relatively
straightforward to implement in OCE, it is natural to ask about the
complexity of optimal control when the search is carried out using
these algorithms.

\subsection{Gradient flows and search complexity}

In Section II, we showed how the critical topology of the most
common objective functions in quantum optimal control can be
determined analytically and display features favorable optimal
search. In Section III, we established several important analytical
results pertaining to the geometry of control landscapes, in
particular their level sets, finding that in certain cases, control
mechanisms can be exploited to reduce the dimensionality of the
control search space. Here we examine another feature of the
geometry of control landscapes that permits analytical
investigation, namely the kinematic gradient flows of the objective
function $\Phi$. On the domain $\U(N)$, these gradient flows
themselves represent integrable dynamical systems. As a result of
this feature, it is possible to identify the system-independent
contribution to the scaling with system dimension of the search
effort for locating quantum optimal controls.

The gradient flow is the trajectory followed by the search algorithm
when the algorithmic step is defined according to the differential
equation

\begin{equation}
\frac{\dd U}{\dd s} = -\triangledown_U J(U)
\end{equation}

The unitary gradient flow equations for observable maximization and
gate optimization are then, respectively,

\begin{eqnarray}
  \left(\frac{dU}{ds}\right)_1 &=& -U\left[\rho(0),U^{\dag}\Theta
U\right], \\
  \left(\frac{dU}{ds}\right)_2 &=& W - UW^{\dag}U
\end{eqnarray}
In the case that $\rho(0)$ has only one nonzero eigenvalue,
corresponding to an initial pure state, it was shown that under the
change of variables $\rho(T,s) = |\psi(s)\rangle \langle
\psi(s)| $, $|\psi(s)\rangle  = (c_1(s),\cdots,c_N(s))$, $x(s)
\equiv (|c_1(s)|^2,\cdots,|c_N(s)|^2)$, the gradient flow of
$\Phi_1$ can be explicitly integrated to give \cite{RajWu2007}:
\begin{eqnarray}
x(s) &=& \frac{e^{2s\Theta}\cdot(|c_1(0)|^2,\cdots,|c_N(0)|^2}{\sum_{i=1}^N|c_i(0)|^2e^{2s\lambda_i}} )\\
&=&
\frac{e^{2s\lambda_1}|c_1(0)|^2,\cdots,e^{2s\lambda_N}|c_N(0)|^2}{\sum_{i=1}^N|c_i(0)|^2e^{2s\lambda_i}}
\end{eqnarray}
where $\lambda_1,\cdots, \lambda_N$ denote the eigenvalues of
$\Theta$. The explicit solution for the gradient trajectory of
objective functional $\Phi_2$ was shown to be
\begin{multline}
W^{\dag}U(s) = (\sinh (s) + \cosh (s) W^{\dag}U_0)\cdot\\
\cdot(\cosh(s) + \sinh (s) W^{\dag}U_0)^{-1}
\end{multline} where the initial
condition is $U_0 = U(0)$ \cite{RajWu2007}.

Chakrabarti et al.\cite{RajWu2007b} have calculated upper bounds of
the convergence times of these unitary gradient flows into a ball of
radius $\epsilon$ around the solution. For the class of observable
maximization problems above, this bound was found to be

\begin{equation}
t_{c,1}(H) = \textmd{max} \leq
\frac{1}{2\mu}\Big[\ln\Big(\frac{2Nk}{\epsilon^2}\Big)+2\ln\frac{(N-k-2)\lambda_{k+1}}{k(\lambda_{(1)}-\lambda_{k+1})}\Big].
\end{equation}
where $N$ is the Hilbert space dimension, $k$ is the degeneracy of
the largest eigenvalue of the observable operator $\theta$ and $\mu$
is the absolute value of the difference of the two largest
eigenvalues of $\Theta$. The upper bound on the convergence time for
the unitary gradient flow of the gate fidelity function was shown to
be
\begin{equation}
t_{c,2}(H) = t_{c,2}(\epsilon) \leq \frac{1}{2}\ln
\Big(\frac{4N}{a^2\epsilon}\Big)
\end{equation}
where $a = \frac{\sin{\theta_0}}{1-\cos{\theta_0}}$, for small
$\epsilon$. Since both of these critical times scale logarithmically
with the Hilbert space dimension, the problem of optimizing the
objective functions on the domain of unitary propagators (via
gradient algorithms) belongs to a logarithmic analog complexity
class, referred to as CLOG within the analog complexity literature
\cite{Fishman1999}.

\begin{figure*}
\centerline{
\includegraphics[width=6in,height=2in]{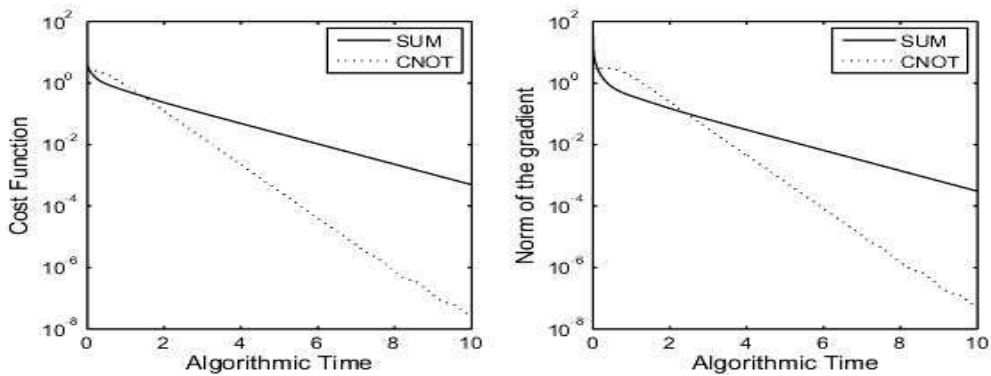}
} \caption{The convergence of the kinematic gradient flows for
optimal search of the SUM gate on $\Sp(4,\R)$ and CNOT gate on
$\U(4)$. (From ref \cite{WuRaj2007}.) }\label{flowcomp}
\end{figure*}

By contrast, the kinematic gradient flows for optimization of the
corresponding objective functions for classical or continuous
variable systems are not integrable. The kinematic flows for
optimization of the continuous variable SUM gate and the discrete
variable CNOT gate are shown in Fig. \ref{flowcomp}
\cite{WuRaj2007}, revealing that the kinematic contribution to
control optimization is less favorable for continuous variable
systems.

The integrated flow derived above for the problem of observable
maximization applies only to the case where $\rho(0)$ is a pure
state. For the more general problem of a mixed initial state,
analytic solutions are more difficult to obtain \cite{RajWu2007}.
However, insight into the scaling of the search effort for these
problems may be obtained from numerical simulations, as well as
consideration of the dimension of the manifold on which the gradient
flow evolves.

As such, numerical optimizations \cite{GregRab2007} of the
observable expectation value function on the domain of unitary
propagators were carried out for $\rho(0)$ and $\Theta$ operators of
various possible ranks and degeneracies, for Hilbert space
dimensions ranging from 2 to 40. From a kinematic perspective, the
optimization problem is symmetric with respect to these two
operators. It was found that the scaling was completely determined
by the number of nondegenerate eigenvalues of $\rho(0)$; in
particular, if $\rho(0)$ is a pure state, the spectrum of $\Theta$
does not alter the complexity class (or vice versa). In the latter
case, the scaling of effort was observed to be roughly logarithmic
in the Hilbert space dimension, consistent with the analytical
result above. In the limiting case where $\rho(0)$ is a full rank
matrix with nondegenerate eigenvalues, the effort was found to scale
linearly with system dimension.

The origin of the observed numerical scaling has been probed
\cite{HoRab2006b} by examining the dimension of the subspace on
which the gradient evolves. It is found that this dimension is
identical on the domain of unitary propagators and control fields,
although in the latter case the basis functions change continuously
along the optimization trajectory. The gradient can be expanded in
terms of at most $N(N-1)$ linearly independent functions of the
time-dependent dipole operator $\mu(t)$. Let $\rho$ consist or $r$
subsets of degenerate eigenvalues $p_1,\cdots, p_r$ with
multiplicities $n_1,...,n_r$, and write $\rho(0)=\sum_{i=1}^r p_i
|i\rangle \langle i|$. The expression (\ref{obsgrad}) for the
gradient from section II can be expanded to give

\begin{widetext}
\begin{multline}
\frac{\delta \Phi_1}{\delta \varepsilon(t)}=\frac{i}{\hbar}
\sum_{k=1}^r p_k \sum_{i=s_k+1}^{s_{k+1}}\Big[\sum_{j=1}^{s_k} +
\sum_{j=s_{k+1}+1}^N \Big] \{\langle i|\Theta(T)|j\rangle \langle
j|\mu(t)|i\rangle -
\langle i|\mu(t)|j \rangle \langle j|\Theta(T)|i \rangle \} \\
+\frac{i}{\hbar}\sum_{k=1}^rp_k\sum_{i=s_k+1}^{s_{k+1}}\sum_{j=s_k+1}^{s_k+1}
\{\langle i|\Theta(T)|j\rangle \langle j|\mu(t)|i\rangle - \langle
i|\mu(t)|j \rangle \langle j|\Theta(T)|i \rangle \}
\end{multline}
\end{widetext}
where $n_i$ are the degeneracies of the eigenvalues $p_i$ of
$\rho(0)$, $s_1=0, s_k=\sum_{i=1}^{k-1} n_i,
k=2,\cdots,r+1,s_{r+1}=n$. The terms in the second summation, of
which there are $\sum_i n_i^2$,  add to zero, from which it can be
shown that the dimension of the subspace of skew-Hermitian matrices
upon which the gradient flow evolves is \cite{HoRab2006b}

\begin{equation}
D=N^2-(N-n)^2-\sum_{i=1}^r n_i^2 = n(2N-n)-\sum_{i=1}^r n_i^2.
\end{equation}

Therefore, when $\rho(0)$ is full rank and nondegenerate, the
gradient can be expressed in terms of a linear combination of
$N(N-1)$ basis functions, irrespective of the spectrum of $\Theta$.
Increasing degeneracy in the spectrum of $\rho(0)$ reduces the
dimension of the subspace on which the gradient evolves, such that
when $\rho(0)$ has $n$ degenerate eigenvalues, this dimension is
equal to $2n(N-n)$. Although this result does not establish the
kinematic contribution to the scaling of search effort with Hilbert
space dimension, it reveals that the dimension of the subspace on
which the gradient evolves for generic observable maximization
problems scales less favorably with $N$ when $\rho$ and $\Theta$
have more nondegenerate eigenvalues.

The existence of a low-dimensional basis set of functions upon which
the gradient can be expanded is especially useful given the
difficulty of implementing effective high-dimensional control field
parameterizations in the experimental setting (section \ref{exptl}).
Although this basis set varies from point to point along the
landscape, recent work \cite{MooreRab2007} suggests that in many
cases, it can be remarkably homogeneous, providing a rational means
of estimating the minimal control parameterization dimensionality
needed to effectively climb the landscape.

\subsection{Relation between dynamic and kinematic gradient flows}

The integrated $U$-gradient flows of the observable maximization and
gate fidelity cost functions identify Hamiltonian-independent
contributions to the scaling of quantum control search effort when
using gradient algorithms. In this section, we derive the
Hamiltonian-dependent relationship between these $U$-gradient flows
and the $\varepsilon$-gradient flows that are followed by OCT and
OCE algorithms.

The $\varepsilon$-gradient flows are the solutions to the
differential equations
\begin{equation}\label{Egrad}%
\frac{\dd \varepsilon(s,t)}{\dd s}= \triangledown
\Phi(\varepsilon(t)) =\alpha \frac {\delta \Phi(s,T) }{\delta
\varepsilon(s,t)}
\end{equation}
where $s$ is a continuous variable parametrizing the algorithmic
time evolution of the search trajectory, and $\alpha$ is an
arbitrary scalar that we will set to 1. The gradient on
$\varepsilon(t)$ is related to the gradient on $\U(N)$ through
\begin{equation}\label{chain}%
\frac{\delta \Phi}{\delta \varepsilon(t)}=\sum_{i,j}\frac{\delta
U_{ij}}{\delta \varepsilon(t)}\frac{\dd \Phi}{\dd U_{ij}}.
\end{equation}
Now suppose that we have the gradient flow of $\varepsilon(s,t)$
that follows (\ref{Egrad}) and let $U(s)$, the system propagator at
time $T$ driven by $\varepsilon(s,t)$, be the projected trajectory
on the unitary group $\U(N)$. The (algorithmic) time derivative of
$U(s)$ is then
\begin{equation}\label{Us}
  \frac{\dd U_{ij}(s)}{\dd s}=  \int_0^T \frac{\delta U_{ij}(s)}{\delta \varepsilon(s,t)}\frac{\partial
\varepsilon(s,t)}{\partial s} \dd t
\end{equation}
which, combined with (\ref{Egrad}) and (\ref{chain}), gives
\begin{equation}\label{dot Us}
     \frac{\dd U_{ij}(s)}{\dd s}=\int_0^T \frac{\delta U_{ij}(s)}{\delta \varepsilon(s,t)}\sum_{p,q}\frac{\delta U_{pq}(s)}{\delta \varepsilon(s,t)}\frac{\dd \Phi}{\dd U_{pq}} \dd t.
\end{equation}
It is convenient to write this equation in vector form, replacing
the $N \times N$ matrix $U(s)$ with the $N^2$ dimensional vector
$\textbf{u}(s)$:
\begin{multline}\label{vecu}
\frac{\dd \textbf{u}(s)}{ds}=\left[\int_0^T
\frac{\delta\textbf{u}(s)}{\delta
\varepsilon(s,t)}\frac{\delta\textbf{u}^T(s)}{\delta
\varepsilon(s,t)}\dd t\right]\triangledown
\Phi\left[\textbf{u}(s)\right]\\\equiv
\textmd{G}[\varepsilon(s,t)]\triangledown
\Phi\left[\textbf{u}(s)\right]
\end{multline} where the superscript $T$ denotes the transpose. This
relation implies that the variation of the propagator in $\U(N)$
caused by the natural gradient flow in the space of control field is
Hamiltonian-dependent, where the influence of the Hamiltonian is
contained in the $N^2$-dimensional symmetric matrix
$\textmd{G}[\varepsilon(s,t)]$.

Thus, although the convergence times for the $U$-gradient flows
above scale favorably with system size, the $\varepsilon$-gradient
flows do not generally follow the same paths, and
Hamiltonian-dependent effects may dominate the scaling when
following the gradient on the domain of control fields. OCT
calculations suggest a difference in the scaling of observable
maximization and gate control search effort when using local
gradient-based algorithms \cite{PalKos2002}, although the
$U$-gradient flow scalings are similar for these two problems.
Systematic dynamical OCT studies have been carried out for unitary
gate optimization on systems of dimension ranging from 2 to 32,
using iterative algorithms. The computational effort was found to
scale exponentially in the Hilbert space dimension.

A natural question is whether the Hamiltonian-dependent unfavorable
scaling of local OCT or OCE algorithms can be mitigated by employing
global algorithms whose optimization trajectories are less sensitive
to the system Hamiltonian. In the next section, we discuss such
global search algorithms and the properties of quantum control
landscapes that render these algorithms effective.

\section{Global search algorithms for quantum
control} \label{search}

\subsection{Scalar and matrix tracking algorithms}

\begin{figure*}
\centerline{
\includegraphics[width=4.5in,height=5in]{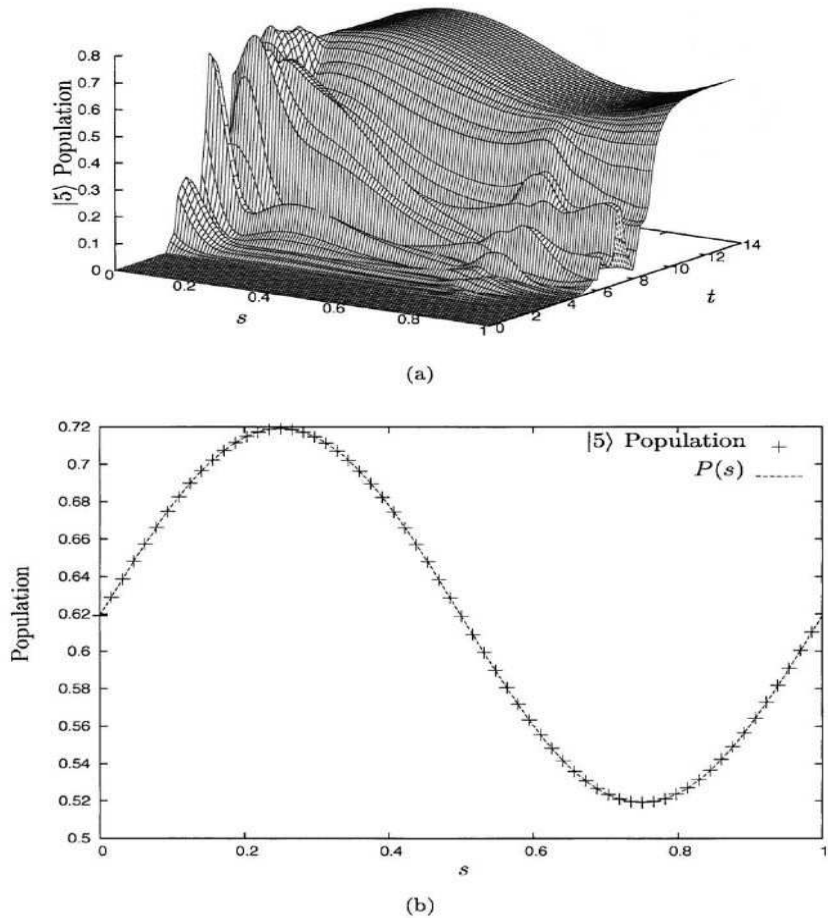}
} \caption{(a) Population surface of state $\mid 5\rangle$ as a
function of s and t for the tracking example described in the text
($P(s) = \mid \langle 5 \mid \psi(s=0,T)\rangle \mid^2+\sin(2\pi
s)$). The elements $\mu_{34}$ and $\mu_{14}$ of the dipole operator
were simultaneously modulated according to $\mu_{34}(s)=s,
\mu_{14}(s)=0.5(1-s)$, with other elements kept fixed. At $s=0$, the
only available direct transition pathway to $\mid 5 \rangle$ is
$\mid 1 \rangle \rightarrow \mid 4 \rangle \rightarrow \mid 5
\rangle$, whereas at $s=1$ the only possible pathway is ladder
climbing through all states. The change of the temporal dynamics
with s reflects the use of different dynamical pathways throughout
the s interval. (b) Cross section of a) at time $t=T$. The
calculated observable population follows the imposed track very
well. (From ref \cite{Rothman2006a}.)}\label{track}
\end{figure*}

As shown above, the convergence time of gradient OCT or OCE
algorithms is dependent on the system Hamiltonian. By contrast, it
is possible to employ algorithms that follow a predetermined track
of observable expectation values, independent of the system
Hamiltonian. An efficient algorithm for following a predetermined
track

\begin{equation}
P(s) = \langle \psi(s,T)|\Theta|\psi(s,T) \rangle, \quad 0 \leq s
\leq 1 \end{equation}
for the expectation values of a target
observable operator $\Theta$ at final time $T$ may be derived from
the diffeomorphic homotopy formalism described in section IV,
originally developed for level set exploration. By making the
substitution $\int_0^T a_0(s,t,T)\frac{\partial
\varepsilon(s,t)}{\partial s} \dd t = b(s,T) + \frac {\dd \langle
\Theta(s,T) \rangle}{\dd s},$ in equation (\ref{beqn}), we obtain
the following expression for the algorithmic step for the control
field (Appendix \ref{appmorph}):

\begin{equation}
\frac{\partial \varepsilon(s,t)}{\partial s} = f(s,t) + \frac
{\left(b(s,T)+\frac{\dd P}{\dd s}
-\gamma(s)\right)a_0(s,t,T)}{\Gamma(s)}
\end{equation}
If the Hamiltonian is kept fixed during tracking, $b(s,T) = 0$. As
mentioned in section IV, such algorithms may also be implemented
experimentally, but require a precise measurement of the gradient.
Analogous equations can be derived for following a predetermined
track of gate fidelity function values.

At each step in this approach, the observable expectation value is
specified, but the unitary propagator is not; many unitary
propagators will map to the same expectation value. The tracking
errors that occur in this approach will be system-specific,
depending on the system dimension and Hamiltonian. The method was
applied to a five-level quantum system initially in state $\mid 1
\rangle$, with the goal of transferring population to state $\mid 5
\rangle$. Figure \ref{track} depicts the changes in the control
field along the track for $P(s) = \mid\langle 5
\mid\psi(s=0,T)\rangle \mid^2+\sin(2\pi s)$. In this case, the
dipole operator was also morphed along the track, in order to
demonstrate the feasibility of simultaneous Hamiltonian variation.

It is natural to consider the prospects of tracking paths in the
space of dynamical propagators, rather than observable expectation
values. Doing so may permit a more system-independent definition of
quantum control complexity. Although this may be difficult to
achieve in OCE using current technology, it can be easily
implemented in simulations. Indeed, diffeomorphic homotopy provides
a natural means of tracking paths $U(s,T), \quad 0 \leq s \leq 1$ in
the group of quantum dynamical propagators. Consider the problem of
tracking the unitary gradient flow for observable expectation value
maximization, integrated in section VII.

In order for the projected flow from $\varepsilon(t)$ onto $U(T)$ to
match the integrated gradient flow on $U(T)$, the quantity
$\frac{\partial{\varepsilon(s,t)}}{\partial s}$ that corresponds to
movement in each step must satisfy a generalized differential
equation:
\begin{equation} \frac{\dd U(s)}{ds} = \int_0^T
\frac{\delta U(s)}{\delta
\varepsilon(s,t)}\frac{\partial{\varepsilon(s,t)}}{\partial s}\dd
t=\triangledown \Phi\left[U(s)\right].
\end{equation}
In the dipole approximation, this relation becomes the following
matrix integral equation:
\begin{equation}
\int_0^T \mu (s,t)\frac{\partial {\varepsilon(s,t)}}{\partial s}\dd
t = U^{\dag}(s)\triangledown \Phi \left[U(s)\right],
\end{equation}
where $\mu(s,t) \equiv U^{\dag}(s,t)\mu U(s,t).$ When $\Phi$ is the
observable expectation value objective function, we have

\begin{equation}
\int_0^T \mu(s,t) \frac{\partial \varepsilon(s,t)}{\partial s}\dd t
= -\left[\rho(0),U^{\dag}(s)\Theta U(s)\right].
\end{equation}
On the basis of eigenstates, the matrix integral equation is written
\begin{equation}\label{matint}
\int_0^T \mu_{ij}(s,t)\frac {\partial \varepsilon(s,t)} {\partial s}
\dd t = i\hbar\langle i|U^{\dag}(s,T)\triangledown \Phi
\left[U(s,T)\right]|j\rangle.
\end{equation} To solve this equation, we first
note that the flexibility in the choice of the representation of the
variation in $\varepsilon(s,t)$ allows us to expand it on the basis
of functions $\mu_{ij}(s,t)$, as
\begin{equation}\frac {\partial
\varepsilon(s,t)}{\partial s} = \sum_{i,j} x_{ij} \mu_{ij}(s,t).
\end{equation}
Inserting this expansion into the above equation produces
\begin{multline}
\sum_{p,q} x_{pq}(s) \int_0^T \mu_{ij}(s,t) \mu_{pq}(s,t) \dd t =\\
i\hbar \langle i|U^{\dag}(s,T)\triangledown
\Phi\left[U(s,T)\right]|j\rangle
\end{multline}
If we denote the correlation matrix $\textmd{G}(s)$ as
\begin{eqnarray}
\textmd{G}_{ij,pq}(s) &=& \int_0^T \mu_{ij}(s,t)\mu_{pq}(s,t) \dd t\\
&=& \int_0^T \langle i |\mu(s,t)|j\rangle \langle
p|\mu(s,t)|q\rangle \dd t,
\end{eqnarray}
(as in eqn (\ref{vecu}) above, but now specifically in the case of
the dipole approximation) and define
\begin{equation}
\Delta_{ij}(s) \equiv i\hbar \langle i|U^{\dag}(s,T)\triangledown
\Phi\left[U(s,T)\right]|j\rangle,
\end{equation}
it can be shown \cite{Dominy2007,RajWu2007} that the matrix
integral equation (\ref{matint}) can be converted into the following
$N^2$-dimensional algebraic (nonsingular) differential equation
(through a procedure analogous to that described in Appendix
\ref{appmorph} for scalar tracking):
\begin{equation}
\frac{\partial \varepsilon}{\partial s} =  f_s + \Big(v(\Delta) -
\alpha \Big)^T\textmd{G}^{-1}v({\mu(t)})
\end{equation}
where $f_s = f_s(t)$ is a "free" function resulting from the
solution of the homogeneous differential equation (analogous to
$f_s$ in Appendix \ref{appmorph}), the operator $v$ vectorizes its
matrix argument (as in eqn (\ref{vecu})) and $\alpha\equiv\int_0^T
v(\mu(t)) f_s \dd t$. This equation is similar to equation
\ref{finalmorph} in Appendix \ref{appmorph}, with the notable
distinctions that the scalar functions $a_0(s,T), a_1(s,T),
a_2(s,T)$, $\gamma$ and $\Gamma$ are now replaced by
$N^2$-dimensional vector and $N^4$-dimensional matrix counterparts,
respectively.

The computational overhead required for implementing such unitary
matrix tracking, compared to scalar tracking, scales (roughly) as
$N^4$, the expense of inverting the matrix $\textmd{G}$. However, if
this track can faithfully be followed, the scaling of the path
length and convergence time to the optimal control solution could
mirror that of the gradient flow of the objective function on the
domain of unitary propagators.  In the case of observable
maximization, the analysis in the previous section suggests that for
mixed initial states, the optimization trajectory followed by the
unitary gradient flow will scale unfavorably for highly
nondegenerate $\rho(0)$. Importantly, it is possible to choose a
global path that is even more favorable than the integrated unitary
gradient flow of the gate objective function. Indeed, unitary matrix
tracking algorithms have recently been developed \cite{Dominy2007}
that follow the shortest path between the initial guess and the
target matrix, namely the geodesic between these points in the
unitary group. Across a wide variety of target gates and system
dimensions, these algorithms were capable of tracking geodesic paths
in $U(N)$ with almost negligible error. As the geodesic is the
globally optimal path in $U(N)$, this indicates that it may be
possible to define a system-independent search complexity for
quantum optimal control problems in terms of the scaling of the
length of the near-geodesic path with Hilbert space dimension.

The ability to track globally optimal paths in $U(N)$, originating
in the favorable properties of the input-state map for discrete
quantum systems (below), may also be useful for design of more
efficient experimental quantum control algorithms. In particular, it
is possible to specify the observable expectation value path that
corresponds to the geodesic path in $U(N)$ for a given $\rho(0)$ and
$\Theta$. Tracking this path experimentally may cause the control
system to follow a path in $U(N)$ that is globally more optimal, and
more system-independent, than the projected path of the
$\varepsilon$-gradient, permitting a more system-invariant
definition of landscape complexity for OCE.

In this approach \cite{RajWu2007}, the unitary propagator $W$ that
maximizes the observable expectation value is first determined
numerically. This can be achieved at minimal computational cost if
$\rho(0)$ and $\Theta$ are known. The observable track that
corresponds to the geodesic $U(s) = \exp{(i\log{(W^{\dag}U_0)}s)}$
that connects $U_0$ and $W$, i.e. $\langle \Theta(s) \rangle$ =
$\tr\left(U(s)\rho(0)U(s)^{\dag} \Theta \right),$ is compatible with
an infinite number of paths $V(s)$ in $\U(N)$, and the set of these
paths can be written
\begin{multline}
M_T = \{V(s) \mid \tr\big(V(s)\rho(0)V(s)^{\dag} \Theta\big) =\\
\tr\big(U(s)\rho(0) U(s)^{\dag} \Theta \big)=\langle \Theta(s)
\rangle\}.
\end{multline}
%It can be shown
%that the volume of $M_T$ varies as a function of the degeneracies of
%the matrices $\rho(0)$ and $\Theta$ \cite{RajWu2007}.
In order to implement a global OCE search along the desired
observable track, computational overhead must be paid in order to
reconstruct the initial density matrix of the system. Research in
quantum statistical inference \cite{Malley1993} has demonstrated
that this reconstruction can generally be achieved with high
statistical certainty at comparatively low cost. As in the case of
gradient measurements, $n$ repeated observations are made on
identically prepared quantum systems. Each quantum measurement is
described by a positive operator-valued measure (POVM)
\cite{Rehacek2007}.
%FURTHER DETAILS NEEDED? REF?
Denoting by $\F_i$ the POVM corresponding to the $i$-th observation,
the likelihood functional
\begin{equation}
L(\rho(t)) = \prod_{i=1}^n \tr(\rho(t) F_i)
\end{equation}
describes the probability of obtaining the set of observed outcomes
for a given density matrix $\rho(t)$. The method of maximal
likelihood estimation (MLE) maximizes this function over the set of
density matrices \cite{Rehacek2007}. An effective parameterization
of $\rho(t)$ is $\rho(t)= T^{\dag} T$, which guarantees positivity
and Hermiticity, and the condition of unit trace is imposed via a
Lagrange multiplier $\lambda$, to give
\begin{equation}
L(T) = \sum_{i=1}^n \ln \tr (T^{\dag} T F_i) - \lambda
\tr(T^{\dag}T).
\end{equation}
%$T$ is a complex lower triangular matrix with real diagonal elements
Standard numerical techniques, such as Newton-Raphson or downhill
simplex algorithms, may be used to search for the maximum over the
$N^2$ parameters of the matrix $T$. Note that other methods for
state reconstruction, such as the maximum entropy method or Bayesian
quantum state identification \cite{Malley1993}, can alternatively be
used.

It is possible (and in many cases imperative for good results) to
constrain the optimization trajectory to follow still more precise
paths in $\U(N)$, by specifying observable tracks $\langle
\Theta_i(s) \rangle$ corresponding to the desired unitary track
$U(s)$ for multiple observable operators $\Theta_i$. In the limit
that the set $\left\{\Theta_i\right\}$ constitutes a complete
tomographic set of observables ($N^2-1$ orthogonal observables
$\Theta_i$), and $\rho(0)$ is nondegenerate, the observable tracking algorithm becomes effectively identical to the unitary matrix tracking algorithm described above. For $1 \leq i
\leq N^2-1$ observables, the geodesic track $U(s)$ is approximated
to progressively greater accuracy, for only incrementally greater
cost. Moreover, it can be shown \cite{RajWu2007} that the $k$
gradients $\frac{\delta \langle \Theta_i(T)\rangle}{ \delta
\varepsilon(t)}$ are closely related, such that the overhead
required to statistically sample $\frac{\delta \langle
\Theta_1(T)\rangle}{ \delta \varepsilon(t)}$ assists in the
determination of $\frac{\delta \langle \Theta_2(T)\rangle}{ \delta
\varepsilon(t)},...,\frac{\delta \langle \Theta_k(T)\rangle}{ \delta
\varepsilon(t)}$.

The accuracy with which globally optimal paths in $U(N)$ can be
tracked depends on the properties of the input-state map $M:
\varepsilon(t) \rightarrow U(T)$. In particular, if the matrix $G$
is singular or close-to-singular, the tracking errors will be
greater and the performance of these globally optimal algorithms
will be compromised. It can be shown that the requirement that $G$
is nonsingular amounts to a more demanding form of controllability
than full controllability, i.e., from any given unitary propagator
$U(T)$, it must be possible to track to any neighboring $U'(T)$ to
first order in $\varepsilon(t)$. An important advantage of
(orthogonal) observable tracking compared to unitary matrix tracking
is that these algorithms place less stringent demands on
controllability while reaping many of the benefits of of following a
globally direct path to the solution \cite{RajWu2007}.

In the next section, we review properties of discrete quantum
control systems that render them less likely to encounter tracking
problems, compared to classical control systems.

\subsection{Extremals of the input-state map}\label{input}

In section II, we saw that the critical points of quantum optimal
control variational problems that are not critical points of the map
between control fields and unitary propagators can be identified
analytically. These critical points are called normal (or regular)
extremals in the terminology of control theory. In this section, we
examine the properties of the so-called input-state map
$\varepsilon(t) \rightarrow U(T)$ between control fields and
associated dynamical propagators, comparing to the classical case.
The critical points of this map are called abnormal (or singular)
extremals of the control problem (see Appendix \ref{max} for a
formal definition).

Given the simple critical topology of normal extremals for quantum
optimal control problems, and the integrability of the gradient
flows of the objective functions on the domain of dynamical
propagators, the properties of the input-state map play a dominant
role in determining the effort required for locating optimal
controls. As such, the behavior (and design) of advanced search
algorithms for quantum optimal control are largely governed by the
properties of this map.

It can be shown that abnormal extremals correspond to control fields
$\varepsilon(t)$ which produce a singular matrix $G$
\cite{Dominy2007}. Because singular extremals correspond to places
where algorithms that track paths in the space of dynamical
propagators break down, their abundance plays a central role in
determining the maximum possible efficiency of optimal control
search.

For general problems of sub-Riemannian geometry \cite{Agrachev1995},
abnormal extremals exist in abundance. In particular, a dynamical
system can be strongly controllable and still possess abnormal
extremals. For discrete quantum systems, however, it is possible to
prove certain analytical results pertaining to the existence of
abnormal extremals that suggest that these extremals are
particularly rare.

D'Alessandro and Dahleh \cite{DAlessandro2001a} gave a complete
classification of normal and abnormal extremals for the
(single-input) optimal gate control problem on $SU(2)$, with field
fluence as the cost. It was shown that the only abnormal extremal in
this case is the control $\varepsilon(t)=-\tr\left( H_d \mu \right)
/ \tr\left(\mu\mu\right)$. For two control fields, if we write
$A:=a_1\mu_1+a_2\mu_2+a_3\left[\mu_1,\mu_2\right]$, the only
abnormal extremal is $\varepsilon_1=-a_1$, $\varepsilon_2=-a_2$. For
three controls, there are no abnormal extremals.

We sketch the proof of this result for a single control field.
Recall that in this case, the PMP Hamiltonian function for the
maximum principle can be written
\begin{equation}
h(M,\lambda_0, \varepsilon)=\tr\left[M
\left(U_0^*(t)(H_d+\mu\varepsilon(t))U_0(t)\right)\right] +
\frac{1}{2}\lambda_0\varepsilon^2(t)
\end{equation}
as above, where $M$ again plays the role of the conjugate momentum
corresponding to the dynamical propagator state variable. We first
demonstrate that all optimal controls except
$\varepsilon(t)=-\tr\left( H_d \mu \right) / \tr\left(\mu\mu\right)$
are normal. It is useful to define
$\left[\left[H_d,\mu\right],\mu\right]=c_1H_d+c_2\mu$
$\left[\left[H_d,\mu\right],H_d\right]=d_1H_d+d_2\mu$ where
$c_2=-d_1,\quad d_1/c_1 = \tr\left(H_d\mu\right)/\tr\left(
\mu\mu\right)$.

Now, if $\lambda_0 = 0$ (i.e., if the control is an abnormal
extremal), the maximizer of the PMP-Hamiltonian must satisfy
$\tr\left(M, U_0(t)\mu U_0(t)\right) = 0$. Differentiating twice, we
obtain
%\begin{multline*}
\begin{widetext}
\begin{equation}
c_1\varepsilon(t)\tr(MU_0^*(t)H_dU_0(t))+
x_2\varepsilon(t)\tr\left(MU_0^*(t)\mu U_0(t)\right)+
d_1\tr\left(MU_0^*(t)U_0(t)\right) + d_2\tr\left( MU_0^*(t)\mu
U_0(t) \right) = 0
\end{equation}
%\end{multline*}
\end{widetext}
which implies
\begin{equation}
(c_1\varepsilon(t) + d_1)\tr \left( MU_0^*(t)H_dU_0(t)\right)=0.
\end{equation}
It can be shown that $\varepsilon(t) \neq \tr\left(H_d\mu\right) /
\tr\left(\mu\mu \right)$ implies $c_1+\varepsilon(t)d_1\neq 0$.
Thus, $\tr\left( M U_0^*(t)H_dU_0(t)\right)\equiv 0$. It follows
that $M$ must also equal zero, which contradicts the supposition
that the minimizer is abnormal. Thus, all optimal controls except
$\varepsilon(t) \neq \tr \left(H_d \mu\right) / \tr\left(\mu \mu
\right)$ are normal. Conversely, it can be verified that
$\varepsilon=-\tr\left( H_d,\mu\right)/\tr\left( \mu,\mu\right)$
satisfies the maximum principle with $\lambda_0=0$ and $M$ chosen
such that $\tr\left( M \mu \right )= \tr \left(M,
\left[H_d,\mu\right]\right) = 0$. This implies that it is an
abnormal extremal, and since all extremals that are not of this form
are normal, this is the only abnormal extremal. Note that the only
abnormal extremal in this case is a constant control. For
higher-dimensional gate control problems using one control field, it
can be shown that there also exist constant controls that are
abnormal extremals, although it is not yet clear whether additional
abnormal extremals exist.

For the important problem of laser control of population transfer
examined in section \ref{statesoln}, where the control Hamiltonian
couples only neighboring energy levels (and the internal Hamiltonian
can be eliminated), the existence of abnormal extremals has been
studied for discrete quantum systems of arbitrary dimension; in
particular, for any time $t$ on the domain of a given solution to
any such state control problem, there exists an interval
$\left[t_1,t_2\right]$ around $t$ where the solution is not strictly
abnormal \cite{Boscain2007}. This property might be termed "weak
normality", by analogy to the phenomenon of weak resonance discussed
above.

The need to partition the time domain of the solutions in this way
arises for the same reason as in section \ref{statesoln}, i.e.
because the dynamics has singularities whenever a state vector
coefficient is zero. For example, if $\psi_1=\psi_2=0$, the control
$V_{1,2}$ has no effect on dynamics, i.e.,
$V_{1,2}\psi_2=V_{2,1}\psi_1=0$. In order to prove the property of
weak normality, one can define a subspace of $S^{n-1}$ on which the
corresponding control problem does not encounter singularities. An
auxiliary control problem can then be framed on this subspace, which
can be shown straightforwardly to possess no abnormal extremals.
Solutions to that control problem can be lifted to $S^{n-1}$ to
produce normal solutions to the original control problem.

We can accomplish these steps by first defining a partition $I
\bigcup J$ of $\{1,...,n\}$ satisfying the following condition:

\begin{equation}
j \in I \leftrightarrow \psi_j(t)=0 \quad \forall t \in
\left[t_1,t_2\right]
\end{equation}
\begin{equation}
j \in J  \leftrightarrow \psi_j(t) \neq 0 \quad \forall t \in
\left[t_1,t_2\right].
\end{equation}
In addition, it is necessary to subdivide $J$ into subspaces
connected by controls, because only in these subspaces are the
vector fields $F_{j,k}(\psi)=V_{j,k}\psi_k$ corresponding to the
controls identically nonzero.  Two indices $j,k$ of $J$ are
connected $(j \sim k)$ if there exists a sequence $j_1,...,j_s$ of
indices of $J$ such that $j=j_1,k=j_s$, and $\forall r < s$,
$V_{j_r,j_r+1}$ is a control, i.e. the two energy levels $E_j$ and
$E_k$ are connected if there exists path through state space where
successive states are coupled by control matrix elements. Denote by
$K_1,...,K_r$ the equivalence classes defined by $\sim$. Let
$m_1,...,m_r$ denote their respective cardinalities, and define
$M_0=0, ~ M_l=\sum_{k\leq l}m_k$. For convenience, we reorder the
indices in the partition such that $\forall l \leq r, \quad K_l =
\{M_{l-1}+1,...,M_l\}$ and $I = \{M_r+1,...,n\}.$ This simply shifts
the indices responsible for singularities to the upper end of the
spectrum.

The essential property of discrete quantum state control systems
that permits proof of weak normality is that it is possible to frame
this associated auxiliary control problem on an analytical
submanifold of the domain of the original problem, $S^{n-1}$.
Because $\sum_{j \in K_l}\mid \hat \psi_j(t)\mid^2$ is constant on
$\left[t_1,t_2\right]$, we can define the analytic submanifold on
which $\psi(t)$ evolves for $t \in \left[t_1,t_2\right]$ in terms of
the $\psi_j(t_1)$s as:

\begin{equation}
X=S^{m_1-1}(C_1)\times ... \times S^{m_r-1}(C_r) \times \prod_{j \in
I}\{\psi_j(t_1)\}
\end{equation}
where $C_l=\sqrt{\sum_{j\in K_l}\mid \psi_j(t_1) \mid^2}$. In other
words, because the subspaces labeled by $l$ are connected by matrix
elements of the control Hamiltonian, the corresponding components of
the state vector undergo unitary (state vector norm-preserving)
evolution on those subspaces. It follows that any extremal $\hat
\psi$ of the original control problem is also an extremal of the
auxiliary control problem on $X$, since $\hat \psi$ remains the same
if the controls $\hat V_{j,k}$, where at least one of the indices
$j,k$ is in $I$, are set to zero.

It is then straightforward to show \cite{Boscain2007} that the
cardinality of each subspace $\Delta_l$ generated by the control
elements $V_{j,k}$ with $j,k \in K_l$ is equal to the cardinality of
the Hilbert sphere $S^{m_l-1}(C_l)$, and hence that the cardinality
of $\Delta = \oplus_l \Delta_l$ is equal to the cardinality of $J$.
As such, no singularities can exist on this domain and the auxiliary
control problem has no abnormal extremals. We refer the reader to
ref \cite{Boscain2007} for details and for a discussion of how the
normality of solutions to the auxiliary control problem is preserved
upon lifting to $S^{n-1}$. Note that the conditions for normality
are equivalent whether the problem is formulated on the space of
quantum states or unitary propagators. Extending these results to
the whole domain of the solution is as of yet an unsettled question,
but this work represents the first step in that direction.

Note that the proofs of these analytical results pertaining to the
sparseness of abnormal extremals for discrete quantum control
problems make use of the compactness of the discrete unitary group
of dynamical propagators. Since this property is not shared by the
noncompact classical or continuous variable quantum dynamical
propagators,  the more general circumstance of abundant abnormal
extremals most likely holds for those cases.

Although critical points of the input-state map are unlikely to be
encountered directly during the search for optimal controls, unitary
propagators within a certain distance of these critical points will
be associated with nearly singular matrices $G$, thus possibly
compromising global tracking efficiency. Current numerical work
focuses on identifying the radius in $\U(N)$ within which such
ill-conditioned matrices occur, for various families of homologous
Hamiltonians.

\section{Open quantum systems}\label{open}

Quantum systems of practical interest in chemistry or physics are
always exposed to some kind of environment, which can render the
dynamics nonunitary and irreversible. Intuitively, environmentally
induced irreversible quantum dynamics would seem to downgrade the
quality of the control outcome.  It is therefore imperative to
determine whether the favorable features of quantum control
landscapes, derived in the context of ideal closed systems, are
preserved in the presence of environmental decoherence. In this
section, we examine the effects of strong decoherence on the
critical topology of open quantum system control problems.

The composite of the system and environment obeys the Schrodinger
equation:

\begin{eqnarray}
i\hbar \frac{\dd \rho_{total}}{\dd t} =
\left[H_{total},\rho_{total}\right]
\end{eqnarray}
The composite Hilbert space is $\hil = \hil_S \otimes \hil_E$, where
$\hil_S$ and $\hil_E$ are the Hilbert spaces of the system and
environment, respectively, and the initial state of the total system
is $\rho_{tot}(0) = \rho_S \otimes \rho_E$. The problem of
maximizing the expectation value of an observable of the system can
be expressed in terms of the Kraus dynamical propagators $K_{mn}$ of
the quantum system in the presence of the environment (see Appendix
\ref{appkraus}):

\begin{eqnarray}
J(K)&=&\tr\big(\sum_{m,n=1}^{\lambda} K_{mn}\rho
K_{mn}^{\dag}\Theta\big)\\
&=&\tr\{K(\rho(0)\otimes I_{\lambda})K^{\dag}(\Theta\otimes
I_{\lambda})\}
\end{eqnarray}
Using the terminology $K = F_E(U) = U(I_N \otimes \rho_E^{1/2})$,
where $U$ is a unitary propagator in $U(\lambda N)$, we can lift the
landscape topology problem onto the composite Hilbert space as

\begin{eqnarray}
J_{\rho_E}(U) &=& \tr\{F_{\rho_E}(U)(\rho(0)\otimes
I_{\lambda})F_{\rho_E}^{\dag}(U)\Theta'\}\\
&=&\tr(UPU^{\dag}\Theta')
\end{eqnarray}
where $P=\rho_S \otimes \rho_E$ and  $\theta'=\Theta \otimes
I_{\lambda}$. The composite system is assumed to be controllable
over $U(\lambda N)$. Under this assumption, Wu et al.
\cite{WuPech2006} showed that no suboptimal traps exist in the
control landscape for maximization of observable expectation values.
Moreover, the enhanced controllability attainable with open dynamics
actually broadens the range of attainable expectation values.

Because $K\left[\rho,N\right]$ is homeomorphic to the homogeneous
space of $U(\lambda N)$, a landscape mapping satisfying the
conditions of Theorem \ref{map} can be built from $U(\lambda N)$ to
$K\left[\rho,N\right]$. This theorem allows one to extract the
critical topology of the Kraus landscape in terms of the associated
unitary landscape. In particular, since the observable maximization
landscape for unitary evolution was demonstrated to have no
suboptimal traps (section II), we can immediately conclude that the
solution sets to open quantum system observable maximization
problems also have no (normal) traps, assuming the environment is
controllable. Although the latter condition may appear difficult to
achieve in practice, in most cases the interaction of a system with
its environment is dominated by local interactions, which may be
straightforward to control.

We note that the global observable and matrix tracking algorithms
described in section \ref{search} can be applied to open quantum
systems as well. In this case, a complete tomographic set of
observables $\left\{\Theta_i\right\}$ is composed of $N^4$ rather
than $N^2$ operators \cite{Hradil2003}, thus increasing the expense
of tracking, but not rendering it prohibitive.

\section{Conclusion and future challenges}

As we have seen, there are stark differences between the optimal
control landscapes for quantum and classical systems. In particular,
the geometric properties of the compact Lie group of
finite-dimensional quantum propagators endows the corresponding
control landscapes with remarkable properties that are considerably
simpler than those of classical systems\footnote{Formally, the
reduction in the dimensionality (hence complexity) of the search
space for discrete quantum controls originates in quantum
symmetries, as shown in section III. Future work may aim to frame
this statement within the context of Noether's theorem for optimal
control, which assigns so-called conserved currents to solutions to
the maximum principle, based on such symmetries.}. Although
quantization of a finite-dimensional classical system generally
produces an infinite-dimensional quantum system, the Hilbert spaces
of most quantum systems of practical interest that possess an
infinite number of levels can be effectively truncated to finite
dimensions. Thus, counterintuitively, locating optimal quantum
controls becomes in many ways easier than locating corresponding
classical controls. Given the apparent favorable scaling of
landscape search complexity with Hilbert space dimension, even if
distant energy levels play a role in the dynamics, the effort
involved in locating controls may still be minimal.

This is most important for the practical feasibility of quantum
control simulations and experiments on large molecules. The scaling
of the expense of quantum dynamical simulation suggests that the
computational problems inherent in quantum chemistry -
high-precision electronic structure calculations become
prohibitively expensive for most systems of practical interest -
should be exacerbated for optimal control of such systems. However,
the simple features of quantum control landscapes described above
indicate that optimal control search need not add additional
complexity to these problems.

The simple topology and geometry of quantum control landscapes,
moreover, can be exploited to develop both numerical and
experimental search algorithms that may outperform local or adaptive
algorithms. The further development of global experimental
algorithms is of particular interest, as these would take advantage
of landscape structure without suffering from the exponentially
unfavorable scaling of the cost of quantum simulation with Hilbert
space dimension.

Although the landscapes for control of finite-dimensional quantum
systems are thus simpler than those for classical systems, the need
for statistical inference of quantum observable expectation values,
states, or gradients thereof adds additional overhead to the cost of
identifying optimal controls in an experimental context. This
overhead is exacerbated when applying global algorithms that attempt
to take advantage of information regarding the quantum state or
dynamical propagator at each step along the landscape search
trajectory. An especially noteworthy challenge, therefore, is the
characterization of how the emerging methodologies of quantum
statistical inference may be employed to further reduce the search
complexity of quantum control problems.

The other feature of quantum dynamics that might be considered
prohibitive to their effective control, namely quantum decoherence,
was shown to not have a significant effect on some of the most
important properties of optimal control landscapes, in particular
their critical topology. Future work should more thoroughly explore
how the geometry of the control landscapes and the effectiveness of
global search algorithms are affected by noise, the nonunitary
evolution of incoherent quantum dynamics, and measurement.

\appendix

\section{Mathematical appendices}

\subsection{Critical topology}

\subsubsection{Landscape mapping}\label{appmap}

\begin{theorem}\label{map}
Suppose the function $x=f(y)$ is locally surjective near some point
$y_0 \in Y$, i.e., the Jacobian has full rank:
$$\textmd{rank}~\frac{\dd f}{\dd y}\mid_{y=y_0} = \textmd{dim}~
X\mid_{x=f(y_0)}$$ in some local coordinate system. Then $y_0$ is
critical for $L \circ f$ in $Y$ if and only if $x_0=f(y_0)$ is
critical for $L$ in $X$, and they have identical numbers of positive
and negative Hessian eigenvalues at $y_0$ and $x_0$, respectively.
Moreover, if the inverse image $f^{-1}(x_0)$ of every critical point
$x_0$ is connected, then the connected components of their critical
manifolds are one-to-one between the two landscapes
\cite{WuPech2006}.
\end{theorem}

\subsubsection{Hessian quadratic form: observable
maximization}\label{apphess}

On the domain of unitary propagators, the Hessian quadratic form
(HQF) for observable maximization can be written
$$\hil_A(U) = \tr\left[-4A^2\rho(0)U\Theta U^{\dag} + 4A\rho(0)AU\Theta
U^{\dag}\right]$$ at a unitary matrix $U$, expanded along an
arbitrary direction $A$ in the Lie algebra of $U(N)$. Consider a
particular solution $U_l$ that generates orderings $\lambda_j
\rightarrow \pi_l(\lambda_j)$ of the eigenvalues of $\Theta$, where
the array $\pi_l$ specifies an N-index permutation mapping. If we
define the matrix elements of $A$ as $A_{ij} =
\alpha_{ij}+i\beta_{ij}$, we obtain after some straightforward
calculations \cite{Mike2006a}
$$\hil_A(\hat U_l) =
-\sum_{j<k}(\alpha_{jk}^2+\beta_{jk}^2)\left[(\lambda_\pi(j)-\lambda_\pi(k))(\epsilon_j-\epsilon_k)\right]$$
from which the counting results presented in section II can be
derived. The number of positive principal axis directions equals the
number of $(j,k)$ pairs for which
$(\lambda_j-\lambda_k)(\epsilon_j-\epsilon_k) \geq 0$, and the
number of negative principal axis directions equals the number of
$(j,k)$ pairs for which
$(\lambda_j-\lambda_k)(\epsilon_j-\epsilon_k) \leq 0$.

\subsection{Maximum principle and adjoint control
systems}\label{max}

\begin{theorem} (Pontryagin maximum principle) Consider the problem
of steering the control system

$$\dot x = f(x,u), \quad x \in M, \quad u \in \Omega \subset
R^k,$$  where $M$ is the state space of the system, from some
initial state $x(0)=x_0$ to some final state $x_1$ while minimizing
a cost of the form $\int_0^T f^0(x,u) \dd t$. The maximum principle
states that if the couple $\bar u(t),\bar x(t)$ is optimal, there
exists an absolutely continuous vector $\lambda(t) \in \R^n$ and a
constant $\lambda \leq 0$, such that the PMP-Hamiltonian function
$h(x(t),\lambda(t),u(t)) = \langle \lambda(t),f(x(t),u(t))\rangle +
\lambda_0 f^0(x(t),u(t))$ satisfies
$$h(\bar x(t),\lambda(t),\bar u(t))=\textmd{max}_u h(\bar
x(t),\lambda(t),u)$$ and $$\lambda_j(t)=-\frac{\partial h}{\partial
x_j}, \quad j \in 1,...,n.$$ Moreover, denoting the tangent space to
the manifold $M$ at state $x(t')$ by $T_{x(t')}M$, we have $\langle
\lambda(0), T_{x(0)}M\rangle = \langle \lambda(T), T_{x(T)}M\rangle$
(transversality condition) \cite{Jurdjevic1997}. If the final time
$T$ is fixed, $h(\bar x(t),\lambda(t),\bar u(t))$ is constant,
whereas if $T$ is allowed to vary, $h(\bar x(t),\lambda(t),\bar
u(t))=0$.

If the control objective is to minimize the final time $T$ instead
of a cost of the form above, the optimal trajectory on
$\left[0,T\right]$ is associated with the Hamiltonian $-\lambda_0+
\langle \lambda(t),f(x(t),u(t))$. In this case, $\textmd{max}_u
h(\bar x(t),\lambda(t),u)=0$ in $\left[0,T\right]$, and we have the
additional condition that if $\lambda_0=0$, then $\lambda(t)\neq 0$
for any $t$ \cite{Jurdjevic1997}.

\end{theorem}

\begin{definition} (Normal, abnormal extremals) A trajectory $\bar x(t)$ satisfying the above condition is called an
extremal. If $\lambda_0 =0$, it is called an abnormal extremal; if
$\lambda_0 < 0$, it is called a normal extremal. If an extremal is
abnormal but not normal, it is called a strictly abnormal extremal.
\end{definition}

\begin{definition}\label{adj} (Adjoint control system) Consider the following control system $F$ on a Lie
group $G$:
$$\dot U = -\frac{i}{\hbar}\left[H_d+\sum_{j=1}^m u_j\mu_j\right]U.$$
Let $K$ denote the subgroup spanned by the control Hamiltonians
$\mu_j$, and denote the adjoint orbit of $-iH_d$ (the internal, or
drift Hamiltonian) under the action of the subgroup $K$ by
$Ad_K(-iH_d)$, i.e. $Ad_K(-iH_d)=\{k_1^{\dag}(-iH_d)k_1|k_1 \in
K\}.$ Then the adjoint control system of $F$ is defined as the
system $\dot P = \hil P, \quad \hil \in Ad_K(-iH_d),\quad P \in G,$
which evolves on the coset space $G/K$ \cite{KhaBro2001}.
\end{definition}

\begin{definition} (Infimizing time) For the control system $F$ above,
let  $\textmd{R}(I,t)$ denote the reachable set from the identity in
time $t$. Then $t^*(U_F)=\textmd{inf}\{t\geq 0|U_F \in
\textmd{R}(I,t)\} $ is called the infimizing time for producing the
propagator $U_F$.
\end{definition}

\begin{theorem} (Equivalence theorem) The infimizing time $t^*(U_F)$ for steering the system $$\dot U
= \left[ H_d + \sum_{j=1}^m u_iH_j\right]U$$ from $U(0)=I$ to $U_F$
is the same as the minimum coset time $L^*(KU_F)$ for steering the
adjoint system $$\dot P = \hil P, \quad \hil \in Ad_K(H_d)$$ from
$P(0)=I$ to $KU_F$ \cite{KhaBro2001}.
\end{theorem}

\begin{theorem} (Adjoint maximum principle)
For the above adjoint control system, denote the time-optimal
control law by $\bar \hil(t)$ and the corresponding optimal
trajectory by $\bar P(t)$. Define an adjoint auxiliary cost
function, $f(P) = \tr(\lambda^{\dag}\hil P), \quad P \lambda^{\dag}
\in p$. The corresponding adjoint PMP-Hamiltonian is
$h(P(t),\lambda(t),\hil(t))=\tr(\lambda^{\dag}(t)\hil(t)P(t))\equiv
\tr(N(t)\hil(t))$. The optimal adjoint control-trajectory pairs are
then the solutions to the Hamiltonian equations $\frac{\dd
\lambda(t)}{\dd t}=-\frac{\partial h}{\partial
P}=\hat\hil(t)\lambda(t)$. The adjoint maximum principle
\cite{KhaBro2002} demands that there exists a $N(t) \in p$
(directions in $G/K$ space) such that

$$\bar \hil(t) = \textmd{argmax}_{\hil}\tr(\hil N(t)), \quad \hil \in
Ad_K(-iH_d)$$
$$\frac{\dd \bar P(t)}{\dd t} = \bar \hil(t) \bar P (t)$$
$$\frac{\dd N(t)}{\dd t} = \left[\bar \hil(t),N(t)\right].$$
\end{theorem}

\subsection{Rotating wave approximation}\label{rwa}

The rotating wave approximation (RWA) consists of a unitary change
of coordinates (and controls) by which the internal (drift)
Hamiltonian can be eliminated in problems involving
atom-electromagnetic wave interactions, by virtue of the
electromagnetic radiation being nearly resonant, or where the
interaction Hamiltonian couples only neighboring states
\cite{Boscain2002}. We consider the latter case. Let
$\psi(t)=U(t)\psi'(t)$. Then the state vector in the rotated
coordinate system satisfies the Schrodinger equation:

$$i\frac{\dd \psi'(t)}{\dd t} = H'(t)\psi'(t)$$ where the Hamiltonian in
the rotated coordinate system is
$$H' = U^{-1}HU-iU^{-1}\frac{\dd U}{\dd t}.$$
In order to eliminate the internal Hamiltonian, we choose $U(t) =
\exp(-iDt)$; since $H=D+V(t)$,
$H'=iUDU^{\dag}+U(D+V(t))U^{\dag}=\exp(iDt)V(t)\exp(-iDt)$.
Redefining $\psi' \rightarrow \psi$ and $H:=-iH'$, we have
$$\frac{\dd \psi(t)}{\dd t} = H(t)\psi(t),$$
where $H$ is skew-Hermitian. The elements of this Hamiltonian are
either zero or are controls; as such, the drift is eliminated.
Assuming that the control Hamiltonian $V$ is off-diagonal (i.e.,
$V_{i,j}=0$ only if $i=j \pm 1$), the relation between the original
and "new" controls $H_{j,k}(t)$ is:
$$V_{j,k}(t)=H_{j,k}(t)\exp (i\left[(E_k-E_j)t+\pi/2\right]).$$
In the more general case where the control Hamiltonian is not
off-diagonal but the control fields are assumed to be roughly in
resonance with the system transition frequencies, the transformed
Hamiltonian takes on a similarly simple form under the approximation
that rapidly oscillating terms average to zero.

\subsection{Analytical solutions to state and gate control
problems}\label{appsoln}

\subsubsection{Low-dimensional gate control problems}

Consider the right-invariant control system described in Definition
4. Following section \ref{gatesoln}, let $G$ denote the special
unitary group $SU(N)$. Call the subalgebra generated by the controls
$\{\mu_1,...,\mu_m\}$ $l$, and the corresponding subgroup $K$. If we
decompose $G=p \oplus l$ such that $p$ is orthogonal to $l$, then
$p$ represents all possible directions to move in $G/K$ space.
Denote by $h \subset p$ a subspace of maximally commuting directions
or generators in $G/K$.

Specifically, in the case of two-qubit systems,
$G/K=SU(4)/SU(2)\otimes SU(2), \quad g= su(4)$, and $K = SU(2)
\otimes SU(2)$. In this case, it can be shown that the Lie algebras
$l, p,$ and $h$ are

\begin{widetext}
\begin{eqnarray*}
l &=& \textmd{span} \quad i\{I_x,I_y,I_z,S_x,S_y,S_z\} \\
p &=&\textmd{span} \quad
i\{I_xS_x,I_xS_y,I_xS_z,I_yS_x,I_yS_y,I_yS_z,I_zS_x,I_zS_y,I_zS_z\}\\
h&=& \textmd{span} \quad i\{I_xS_x, I_yS_y, I_zS_z\}.
\end{eqnarray*}
\end{widetext}
Decomposing the target unitary propagator as $U_F = k_2\exp(Y)k_1,$
(section \ref{gatesoln}), we have

\begin{multline*}
U_F = k_1
\exp\left[-i(\alpha_1I_xS_x+\alpha_2I_yS_y+\alpha_3I_zS_z)\right]k_2,\\
\quad k_1,k_2 \in SU(2) \otimes SU(2),
\end{multline*}
where the sub-Riemannian problem consists of generating $\exp(Y)$ in
the fastest possible way. If we define
$k_y^-=\exp(-i\pi/2I_y)\exp(-i\pi/2 S_y)$ and
$k_y^+=\exp(i\pi/2I_y)\exp(-i\pi/2S_y)$, we can verify that
$$k_y^{\pm}\exp(-iI_zS_z)(k_y^{\pm})^{-1} = \exp(\pm iI_xS_x)$$
and similarly for $k_x$, showing we can generate any element of the
Cartan subalgebra $h$.

In the case of three spins coupled by local interactions such that
$J_{12}=J_{23}, \quad J_{13}=0$, $G/K$ is a nonsymmetric space, but
is still a finite-dimensional Riemannian manifold. Khaneja and
coworkers \cite{KhaBro2002} considered the generation of unitary
propagators of the form

$$U=\exp(-i\theta I_{1\alpha}I_{2\beta}I_{3\gamma}), \quad
\alpha,\beta,\gamma \in \{x,y,z\},$$ which are hard to produce as
they involve trlinear terms in the effective Hamiltonian (trilinear
propagators). Applying the decomposition $U_F = k_2\exp(Y)k_1,$ it
is sufficient to produce

$$\exp(Y)= \exp(-i\theta I_{1z}I_{2z}I_{3z}), \quad \theta \in
\left[0,4\pi\right]$$ because all other propagators belonging to the
set $\exp(-i\theta I_{1\alpha}I_{2\beta}I_{3\gamma} | \alpha,\beta,
\gamma \in \{x,y,z\}\}$ of trilinear propagators can be produced
from $U_F$ in arbitrarily small time by selective hard pulses.

The corresponding adjoint control problem (Definition \ref{adj}) has
$\hil \in Ad_K(-i2\pi J(I_{1z}I{2z}+I_{2z}I_{3z}))$. For adjoint
control problems, an equivalent version of the Pontryagin maximum
principle exists (Appendix \ref{max}). It can be verified
\cite{KhaBro2002} that the following analytical time-dependent
control satisfies this principle:
\begin{multline*}
\bar \hil(t) = -i2\pi\J
\cdot\\
\cdot\big[(I_{1z}I_{2x}+I_{2x}I_{3z}\cos(\frac{\beta t}{T}) -
(I_{1z}I_{2y}+I_{2y}I_{3z})\sin(\frac{\beta t}{T})\big]
\end{multline*}
and steers the adjoint system from $P(0)=I$ to $P(T)\in KU_F$ in
minimal time. The minimum time $t^*(U_F)$ required to produce a
propagator of the form $U_F = \exp(-i\theta I_{1z}I_{2z}I_{3z}),
\quad \theta \in \left[0,4\pi\right]$ is then given by
$$t^*(U_F)=\frac{\sqrt{\kappa(4-\kappa)}}{2J}$$ where $\kappa =
\theta / 2\pi$.

%\subsubsection{Population transfer in 3-level systems}
\subsubsection{Low-dimensional state control problems}

Analytical solutions to problems of state-to-state population
transfer can be obtained for two- and three-level quantum systems,
for off-diagonal control Hamiltonians (i.e., $V_{jk} = 0$ if $j \neq
k \pm 1$), under the rotating wave approximation (Appendix
\ref{rwa}). We summarize the results for the optimal control of
population transfer in three-level systems using fluence as the
cost, with two controls that span the control Lie algebra
\cite{Boscain2002}, since this provides an example for how objective
function symmetry can endow integrability to quantum control systems
and more generally simplify the search for optimal controls.  Let us
consider the problem of transferring the population from pure state
$\mid 1 \rangle$ to pure state $\mid 3 \rangle$, with
$\mu_{1,12}=\mu_{2,21}=1, \quad \mu_{2,12}=\mu_{2,21}=1$. The
Schrodinger equation for the Hamiltonian in this case can be written

\begin{eqnarray*}
\dot c_1 &=& -i\varepsilon_1(t)c_2, \quad \\
\dot c_2 &=& -i(\varepsilon_1(t)c_1+\varepsilon_2(t)c_3),\\
\quad \dot c_3&=&-iu_2(t)c_2,
\end{eqnarray*}
where $c_i$ denote the coefficients of the wavefunction eigenstates
$x_i$. If we set $c_1=x_1+ix_2$, $c_2=x_4-ix_3$, $c_3=x_5+ix+6$, we
can write this concisely as $\dot x =
\varepsilon_1F_1+\varepsilon_2F_2$, where $x=(x_1,...,x_6)$ and
$F_1= (-x_3,-x_4,x_1,x_2,0,0)$ and $F_2=(0,0,x_5,x_6,-x_3,-x_4)$
denote the action of the control Hamiltonians $\mu_1$ and $\mu_2$ on
the state $x$. This is a problem on the 5-dimensional Hilbert
sphere, $S^5$. The initial condition for this problem is a point on
the circle $S_{in}^1\equiv \{\textbf{x}\in S^5 \mid
x_1^2+x_2^2=1\}$, whereas the target is a point on the circle
$S_{fin}^1 \equiv \{\textbf{x}\in S^5 \mid x_5^2+x_6^=1\}$. However,
the dimensionality of this problem can be reduced if we assume that
the controls are resonant (section \ref{statesoln}). In this case,
for each $\textbf{x}_0 \in S_{in}^1$, the orbit $O(\textbf{x}_0)$
(the reachable set of states) is a two-dimensional submanifold of
$S^5$, and hence the system is not fully controllable (i.e. not all
superpositions of states can be reached from arbitrary initial
conditions). Nonetheless, arbitrary eigenstate-eigenstate
transitions can be controlled. Let us define $\textbf{x}_0(\alpha)$
as the initial condition $x_1(0)=\cos(\alpha)$,
$x_2(0)=\sin(\alpha), \quad \alpha \in \left[0,2\pi\right]$. Then
this submanifold is the two-dimensional sphere defined by the
equation $x_1'^2+x_3'^2+x_5'^2=1$, where $\textbf{x}'=R\otimes I_3
\textbf{x}$. In other words, due to the isometry, all the points in
$S_{in}$ can be considered equivalently. Therefore, we can study the
optimal control problem on the orbit $O(\textbf{x}_0)$. Let us
consider the case where $\textbf{x}_0$ is defined by $x_1=1$; in
this case, $O(\textbf{x}_0)$ is the sphere defined by
$x_1^2+x_3^2+x_5^2=1$.
%which permits analytical solution despite the lack of controllability

%(, and hence the system is not fully controllable.)

We can then execute a change of variables to $y_1=x_1$,$y_2=x_3$,
$y_3=-x_5$, such that the control system can be rewritten as
$$\left(%
\begin{array}{c}
  \dot y_1  \\
  \dot y_2 \\
  \dot y_3 \\
\end{array}%
\right)= u_1F_1+u_2F_2;
\quad F_1=\left(%
\begin{array}{c}
  -y_2 \\
  y_1 \\
  0 \\
\end{array}%
\right),~~ F_2=\left(%
\begin{array}{c}
  0 \\
  -y_3 \\
  y_2 \\
\end{array}%
\right).$$ It is useful to frame the reduced problem in spherical
coordinates, where
\begin{eqnarray*}
y_1&=&\cos(\theta)\cos(\phi)\\
y_2&=&\sin(\theta)\\
y_3&=&\cos(\theta)\sin(\phi).
\end{eqnarray*}
The control system can then be written
$$\left(%
\begin{array}{c}
  \dot \theta  \\
  \dot \phi \\
\end{array}%
\right)= v_1G_1+v_2G_2,
$$
where $G_1=\partial_{\theta}, \quad
G_2=\tan(\theta)\partial_{\phi}$.

In spherical coordinates, the Hamiltonian associated with the
maximum principle is
\begin{eqnarray*}
h(\theta,\phi,P_{\theta},P_{\phi},v_1,v_2)&=&\langle P,v_1G_1+v_2G_2
\rangle + p_0(v_1^2+v_2^2) \\
&=& v_1P_{\theta}+v_2P_{\phi}\tan(\theta)+p_0(v_1^2+v_2^2).
\end{eqnarray*}
The maximum principle demands $\frac{\partial h}{\partial
v_1}=0,\quad \frac{\partial h}{\partial v_2}=0$; as such, $v_1 =
P_{\theta}, \quad v_2 = P_{\phi}\tan(\theta)$. The Hamiltonian
corresponding to these controls is $\hat h =
\frac{1}{2}(P_{\theta}^2+(\tan(\theta)P_{\phi})^2).$ The Hamiltonian
equations of motion following from the maximum principle are then:

$$\dot \theta = \frac{\partial \hat h}{\partial P_{\theta}} =
P_{\theta}, \quad \dot \phi = \frac{\partial \hat h}{\partial
P_{\phi}}=P_{\phi}\tan^2(\theta)$$
$$\dot P_{\theta}=\frac{\partial \hat h}{\partial \theta} =
-P_{\phi}^2\tan(\theta)(1+\tan^2(\theta)),\quad \dot
P_{\phi}=-\frac{\partial \hat h}{\partial\phi}=0.$$ This Hamiltonian
system is Liouville integrable, since there are two independent and
commuting constants of the motion $\hat h$ and $P_{\phi}=a$. The
solution for minimal fluence, with fixed transfer time $T$ can then
be shown to be \cite{Boscain2002}:

$$\min\left(\int_0^T(\varepsilon_1^2+\varepsilon_2^2)\dd t \right) =
\frac{3}{4}\pi^2\frac{1}{T}.$$

\subsection{Diffeomorphic homotopy on control
landscapes}\label{appmorph}

We sketch the derivation of the general diffeomorphic homotopy
procedure for Hamiltonian morphing and observable tracking
\cite{Rothman2006a, Rothman2006b, Rothman2005}. The condition for
remaining on a designated level set of an observable control
landscape is

$$\frac {\dd F(s)}{\dd s} = \frac{\dd \langle \Theta(s) \rangle_T}{\dd s}
= 0$$ In the following, we use the subscript $+$ to refer to a point
infinitesimally close to the system at point $s$.  The interaction
dynamical propagator in the interaction picture can then be written
$U_I(t,0)=U^{\dag}(t,0)U_+(t,0).$ To first order, the interaction
picture propagator at the final time $T$ is:

$$U_I(T,0)=I+\frac{\dd s}{i\hbar} \int_0^T \dd t \big( U^{\dag}(t,0)
\frac{\partial H(s,t)}{\partial s} U(t,0)\big)$$ If we Taylor expand
the Hamiltonian to first order at algorithmic time s, i.e.,

$$H_+(s,t) =
H(s+ds,t)=H(s,t)+ds\frac{\partial H(s,t)}{\partial s},$$the
expectation value of the observable of interest at the point $s+ds$
in Hamiltonian space can be expressed as:

\begin{multline} \langle \Theta(s+\dd s)\rangle_T = \langle \Theta(s) \rangle_T
~+\\
\frac{ds}{i\hbar}\langle \psi_0| \int_0^T
dt\big[\Theta(T),U^{\dag}(t,0)\frac{\partial H(s,t)}{\partial
s}U(t,0)\big]| \psi_0 \rangle.
\end{multline}
It can then be shown that the condition for remaining on the level
set can be written

\begin{widetext}
\begin{eqnarray} \label{dmorph} \frac{\dd F(s)}{\dd s} = \int_0^T \dd t \big(
a_0(s,t,T)\frac{\partial \varepsilon(s,t)}{\partial s} +
a_1(s,t,T)\varepsilon(s,t)+a_2(s,t,T)\big) = 0,
\end{eqnarray}
where
\begin{eqnarray*} a_0(s,t,T)&=&-\frac{1}{i\hbar}\langle \psi_0 |
\big[U^{\dag}(T,0)\Theta
U(T,0),U^{\dag}(t,0)\mu(s)U(t,0)\big]|\psi_0 \rangle\\
a_1(s,t,T)&=&-\frac{1}{i\hbar}\langle \psi_0 |
\big[U^{\dag}(T,0)\Theta U(T,0),U^{\dag}(t,0)\frac{\dd \mu(s)}{\dd
s}U(t,0)\big]|\psi_0\rangle\\
a_2(s,t,T)&=&-\frac{1}{i\hbar}\langle \psi_0 |
\big[U^{\dag}(T,0)\Theta U(T,0),U^{\dag}(t,0)\frac{\dd H_d(s)}{\dd
s}U(t,0)\big]|\psi_0 \rangle.
\end{eqnarray*}
\end{widetext}

A natural way to solve eqn (\ref{dmorph}) is to transform it into an
initial value problem for the laser field $\varepsilon(s,t)$.
Equation (\ref{dmorph}) can be reexpressed as the differential
equation
$$a_0(s,t,T)\frac{\partial \varepsilon(s,t)}{\partial s} +
a_1(s,t,T)\varepsilon(s,t)+a_2(s,t,T)=f(s,t)$$ where $f(s,t)$ is an
arbitrary function satisfying the constraint
$$\int_0^T f(s,t) \dd t = 0 \quad \forall s.$$ In general the coefficient $a_0(s,t,T)$ may vanish at some values of
s and t. Since there are \textit{a priori} no restrictions on
$a_0(s,t,T)$, we cannot preclude the possibility of singular
behavior with the class of D-MORPH controls admitted by equation
(\ref{dmorph}). Singular behavior is unattractive because it implies
the possible existence of similar undesirable behavior in the
control field $\varepsilon$. Regardless of the behavior of $a_0$, a
nonsingular class of D-MORPH solutions may be generated. Defining
the integral

$$b(s,T) = -\int_0^T(a_1(s,t,T)\varepsilon(s,t)+a_2(s,t,T))\dd t,$$
we have $$\int_0^T a_0(s,t,T)\frac{\partial
\varepsilon(s,t)}{\partial s} \dd t = b(s,T),$$ which is a Fredholm
integral equation of the first kind. More generally, in the case
that following a track of objective function values $\langle
\Theta(s) \rangle$ is desired, instead of remaining on a level set,
we have

$$\int_0^T a_0(s,t,T)\frac{\partial \varepsilon(s,t)}{\partial s} \dd t =
b(s,T) + \frac{\dd \langle \Theta(s,t)\rangle}{\dd s}$$ Then, it can
be shown \cite{Rothman2005} that this equation can be transformed
into the equivalent (nonsingular) differential equation

\begin{equation}\label{finalmorph}
\frac{\partial \varepsilon(s,t)}{\partial s} =  f(s,t)+ \frac
{\left(b(s,T)+\frac{\dd \langle \Theta(s,t) \rangle }{\dd s}-\gamma(s)\right)
a_0(s,t,T)}{\Gamma(s)}
\end{equation}
where $\Gamma(s) = \int_0^T \left[a_0(s,t,T)\right]^2 \dd t$ and
$\gamma(s) = \int_0^T a_0(s,t,T) f(s,t) \dd t$ is the projection of
the arbitrary function $f(s,t)$ onto $a_0(s,t,T)$ (the above expression can be multiplied
by an arbitrary shape coefficient $S(t)$).

The freedom to choose $f(s,t)$ corresponds to the multiplicity of
control field solutions on the level set, and arises naturally as a
consequence of the underspecified nature of the integral form of the
original D-MORPH equation. In particular, in the case of fluence
minimization, $f(s,t)=f_m(s,t)$ satisfies the condition
$$\int_0^T S(t)a_0(s,t,T)f_m(s,t)\dd t=0.$$ It can be shown \cite{Rothman2005} that with the choice
$$f(s,t) = -\frac{1}{\Delta s}\frac{\varepsilon(s,t)}{S(t)},$$ the
algorithm seeks to minimize the total field fluence at each step.
Analogous free functions that correspond to other auxiliary costs
(such as minimal time) in the Pontryagin maximum principle can be
constructed.

\subsection{Controllability on compact Lie groups}\label{appctrl}

\begin{definition} (Reachable sets and controllability) Consider a
control system $F$ defined on a manifold $M$. For each $T>0$, and
each $x_0$ in $M$, the set of points reachable from $x_0$ at time
$T$, denoted by $A(x_0,T)$, is equal to the set of terminal points
$x(T)$ of integral curves of $F$ that originate at $x_0$. The union
of $A(x_0,T)$, for $T \geq 0$, is called the reachable
set\cite{Jurdjevic1997} from $x_0$. The set of points reachable in T
or fewer units of time, defined as the union of $A(x_0,t), \quad t
\leq T$, is denoted $A(x_0,\leq T)$. A control system $F$ is
controllable if any point of M is reachable from any other point of
$M$, at any time $T>0$.
\end{definition}

\begin{definition} \label{invar} (Invariant control system on a Lie group)
A control system on a Lie group $G$, where $G$ is the Lie group
associated with a Lie algebra $h$, is defined by the equations
$$\dot U = AU(t) + \sum_{i=1}^m u_i(t)B_iU(t)$$ where $A$ and the
$B_i, \quad i=1,\cdots,m$ belong to $h$, $U(t)$ belongs to $G$, and
the $u_i(t)$ are scalar functions of time which play the role of the
external controls. The control system is said to be right-invariant
if the following condition holds:  If $U(t)$ is a solution
corresponding to the initial condition equal to the identity matrix,
the solution corresponding to the initial condition $F$ is given by
$U(t)F$.

\end{definition}

\subsection{Kraus superoperator formalism}\label{appkraus}

Consider a composite of system and environment whose Hamiltonian
$H_{total}$ consists of the Hamiltonians of the system, environment,
and their interaction. The total system evolution operator is
$U_{total}(t)$ on the total Hilbert space $\hil = \hil_S \otimes
\hil_E$, where $\hil_S$ and $\hil_E$ are the Hilbert spaces of the
system (of dimension $N$) and environment, respectively. The initial
state of the total system is $\rho_{tot}(0) = \rho_S \otimes
\rho_E$. We can obtain an expression for the system dynamics
$\rho_S(t)$ by tracing
$\rho_{total}(t)=U_{total}(t)\rho_{total}(0)U_{total}^{\dag}(t)$
over the environment \cite{WuPech2006}:

$$\rho_S(t) = \tr_E\{U_{total}(t)(\rho_S \otimes
\rho_E)U_{total}^{\dag}(t)\}$$

Define the $\lambda N$-dimensional matrix
$K(t)=\tr_E\{U_{total}(t)(I_N \rho_E^{1/2}).$ Divide $K$ into
$\lambda^2$ $N\times N$ matrices $K_{\alpha \beta}(t) =
|\beta\rangle\langle\alpha|)\},\quad
(\alpha,\beta=1,\cdots,\lambda)$, where $\alpha,\beta$ constitute an
arbitrary basis for $\hil_E$. These matrices form the Kraus
representation of the dynamical map as
$$\rho_S(t) =
\sum_{\alpha,\beta=1}^{\lambda}K_{\alpha\beta}(t)\rho(0)K_{\alpha
\beta}^{\dag}(t).$$

%\subsection{Symplectic representation of continuous variable quantum processes}\label{symp}
\subsection{Symplectic propagators}\label{symp}
Consider a continuous variable quantum system with a quadratic
Hamiltonian $H(t)$. $H(t)$ induces a Hamiltonian vector field, which
generates a one-parameter family of transformations $U(t)$ on the
Hilbert space $\hil$ that obeys the \sd equation
\begin{equation}\label{U-sd}
\frac{\partial {U}(t)}{\partial t}=- \frac{i}{\hbar}~H(t){U}(t),
\end{equation}
where the parameter is the time. The evolution propagator transforms
the quadrature vector of position and momentum operators $\hat
z=(\hat q_1,\cdots, \hat q_N;\hat p_1,\cdots,\hat p_N)^T$ linearly
through
$$U:~~\hat z_\alpha~\rightarrow ~U^\dagger(t) \hat z_\alpha U(t)=\sum_\beta S_{\alpha\beta}(t)\hat z_\beta,$$
where the $2N\times 2N$ matrix $S(t)$ is an element of the
symplectic group $\Sp(2N,\R)$ that satisfies $S^T J S=J$, with
$$J=\left(%
\begin{array}{cc}
   & I_N \\
  -I_N &  \\
\end{array}%
\right).$$ Thus, the matrix $S$ captures the Heisenberg equations of
motion for the operators $\hat z_i$, and the unitary propagator $U$
forms the metaplectic unitary representation of $S$ in $\Sp(2N,\R)$
\cite{ArvMuk1995}. Like the infinite-dimensional unitary group (but
unlike the finite-dimensional unitary group), $\Sp(2N,\R)$ is
noncompact.

Whereas a logical operation on $N$ discrete quantum bits (qubits) is
represented by a $2^N$-dimensional unitary matrix, the corresponding
operation on $N$ continuous quantum bits (qunits) can be represented
by a $2N$-dimensional symplectic matrix.

\section*{Acknowledgements}

The authors acknowledge support from DARPA.

%\bibliographystyle{abbrv}

%\bibliography{review}

\begin{thebibliography}{61}
\expandafter\ifx\csname natexlab\endcsname\relax\def\natexlab#1{#1}\fi
\expandafter\ifx\csname bibnamefont\endcsname\relax
  \def\bibnamefont#1{#1}\fi
\expandafter\ifx\csname bibfnamefont\endcsname\relax
  \def\bibfnamefont#1{#1}\fi
\expandafter\ifx\csname citenamefont\endcsname\relax
  \def\citenamefont#1{#1}\fi
\expandafter\ifx\csname url\endcsname\relax
  \def\url#1{\texttt{#1}}\fi
\expandafter\ifx\csname urlprefix\endcsname\relax\fi
\providecommand{\bibinfo}[2]{#2}
\providecommand{\eprint}[2][]{\url{#2}}

\bibitem[{\citenamefont{Demiralp and Rabitz}(1993)}]{Demiralp1993}
\bibinfo{author}{\bibfnamefont{M.}~\bibnamefont{Demiralp}} \bibnamefont{and}
  \bibinfo{author}{\bibfnamefont{H.~A.} \bibnamefont{Rabitz}},
  \bibinfo{journal}{Phys. Rev. A} \textbf{\bibinfo{volume}{47}},
  \bibinfo{pages}{809} (\bibinfo{year}{1993}).

\bibitem[{\citenamefont{Assion et~al.}(1998)\citenamefont{Assion, Baumert,
  Bergt, Brixner, and Kiefer}}]{Assion1998}
\bibinfo{author}{\bibfnamefont{A.}~\bibnamefont{Assion}},
  \bibinfo{author}{\bibfnamefont{T.}~\bibnamefont{Baumert}},
  \bibinfo{author}{\bibfnamefont{M.}~\bibnamefont{Bergt}},
  \bibinfo{author}{\bibfnamefont{T.}~\bibnamefont{Brixner}}, \bibnamefont{and}
  \bibinfo{author}{\bibfnamefont{B.}~\bibnamefont{Kiefer}},
  \bibinfo{journal}{Science} \textbf{\bibinfo{volume}{282}},
  \bibinfo{pages}{5390} (\bibinfo{year}{1998}).

\bibitem[{\citenamefont{Baumert et~al.}(1997)\citenamefont{Baumert, Brixner,
  Seyfried, Strehle, and Gerber}}]{Baumert1997}
\bibinfo{author}{\bibfnamefont{T.}~\bibnamefont{Baumert}},
  \bibinfo{author}{\bibfnamefont{T.}~\bibnamefont{Brixner}},
  \bibinfo{author}{\bibfnamefont{V.}~\bibnamefont{Seyfried}},
  \bibinfo{author}{\bibfnamefont{M.}~\bibnamefont{Strehle}}, \bibnamefont{and}
  \bibinfo{author}{\bibfnamefont{G.}~\bibnamefont{Gerber}},
  \bibinfo{journal}{Appl. Phys. B} \textbf{\bibinfo{volume}{65}},
  \bibinfo{pages}{779782} (\bibinfo{year}{1997}).

\bibitem[{\citenamefont{Herek}(2006)}]{Herek2006}
\bibinfo{author}{\bibfnamefont{J.~L.} \bibnamefont{Herek}},
  \bibinfo{journal}{J. Photochem. Photobiol.} \textbf{\bibinfo{volume}{180}},
  \bibinfo{pages}{225} (\bibinfo{year}{2006}).

\bibitem[{\citenamefont{Bartels et~al.}(2000)\citenamefont{Bartels, Backus,
  Zeek, Misoguti, Vdovin, Christov, Murnane, and Kapteyn}}]{Bartels2000}
\bibinfo{author}{\bibfnamefont{R.}~\bibnamefont{Bartels}},
  \bibinfo{author}{\bibfnamefont{S.}~\bibnamefont{Backus}},
  \bibinfo{author}{\bibfnamefont{E.}~\bibnamefont{Zeek}},
  \bibinfo{author}{\bibfnamefont{L.}~\bibnamefont{Misoguti}},
  \bibinfo{author}{\bibfnamefont{G.}~\bibnamefont{Vdovin}},
  \bibinfo{author}{\bibfnamefont{I.~P.} \bibnamefont{Christov}},
  \bibinfo{author}{\bibfnamefont{M.~M.} \bibnamefont{Murnane}},
  \bibnamefont{and} \bibinfo{author}{\bibfnamefont{H.~C.}
  \bibnamefont{Kapteyn}}, \bibinfo{journal}{Nature}
  \textbf{\bibinfo{volume}{406}}, \bibinfo{pages}{164} (\bibinfo{year}{2000}).

\bibitem[{\citenamefont{Palao and Kosloff}(2002)}]{PalKos2002}
\bibinfo{author}{\bibfnamefont{J.}~\bibnamefont{Palao}} \bibnamefont{and}
  \bibinfo{author}{\bibfnamefont{R.}~\bibnamefont{Kosloff}},
  \bibinfo{journal}{Phys. Rev. Lett.} \textbf{\bibinfo{volume}{89}},
  \bibinfo{pages}{188301} (\bibinfo{year}{2002}).

\bibitem[{\citenamefont{Grace et~al.}(2007)\citenamefont{Grace, Brif, Rabitz,
  Walmsley, Kosut, and Lidar}}]{Grace2006a}
\bibinfo{author}{\bibfnamefont{M.}~\bibnamefont{Grace}},
  \bibinfo{author}{\bibfnamefont{C.}~\bibnamefont{Brif}},
  \bibinfo{author}{\bibfnamefont{H.~A.} \bibnamefont{Rabitz}},
  \bibinfo{author}{\bibfnamefont{I.}~\bibnamefont{Walmsley}},
  \bibinfo{author}{\bibfnamefont{R.~L.} \bibnamefont{Kosut}}, \bibnamefont{and}
  \bibinfo{author}{\bibfnamefont{D.~A.} \bibnamefont{Lidar}},
  \bibinfo{journal}{In press}  (\bibinfo{year}{2007}), \eprint{eprint
  arXiv:quant-ph/0611189}.

\bibitem[{\citenamefont{Khaneja et~al.}(2001)\citenamefont{Khaneja, Brockett,
  and Glaser}}]{KhaBro2001}
\bibinfo{author}{\bibfnamefont{N.}~\bibnamefont{Khaneja}},
  \bibinfo{author}{\bibfnamefont{R.~W.} \bibnamefont{Brockett}},
  \bibnamefont{and} \bibinfo{author}{\bibfnamefont{S.~J.}
  \bibnamefont{Glaser}}, \bibinfo{journal}{Phys. Rev. A}
  \textbf{\bibinfo{volume}{63}}, \bibinfo{pages}{032308}
  (\bibinfo{year}{2001}).

\bibitem[{\citenamefont{Khaneja et~al.}(2002)\citenamefont{Khaneja, Glaser, and
  Brockett}}]{KhaBro2002}
\bibinfo{author}{\bibfnamefont{N.}~\bibnamefont{Khaneja}},
  \bibinfo{author}{\bibfnamefont{S.~J.} \bibnamefont{Glaser}},
  \bibnamefont{and} \bibinfo{author}{\bibfnamefont{R.~W.}
  \bibnamefont{Brockett}}, \bibinfo{journal}{Phys. Rev. A}
  \textbf{\bibinfo{volume}{65}}, \bibinfo{pages}{032301}
  (\bibinfo{year}{2002}).

\bibitem[{\citenamefont{Wu and Rabitz}(2007)}]{Wu2007}
\bibinfo{author}{\bibfnamefont{R.}~\bibnamefont{Wu}} \bibnamefont{and}
  \bibinfo{author}{\bibfnamefont{H.~A.} \bibnamefont{Rabitz}},
  \bibinfo{journal}{in preparation}  (\bibinfo{year}{2007}).

\bibitem[{\citenamefont{von Neumann}(1937{\natexlab{a}})}]{VonNeumann1937a}
\bibinfo{author}{\bibfnamefont{J.}~\bibnamefont{von Neumann}},
  \bibinfo{journal}{Tomsk. Univ. Rev.} \textbf{\bibinfo{volume}{1}},
  \bibinfo{pages}{286} (\bibinfo{year}{1937}{\natexlab{a}}).

\bibitem[{\citenamefont{von Neumann}(1937{\natexlab{b}})}]{VonNeumann1937b}
\bibinfo{author}{\bibfnamefont{J.}~\bibnamefont{von Neumann}},
  \bibinfo{journal}{Unpublished works, Institute for Advanced Study Archives}
  (\bibinfo{year}{1937}{\natexlab{b}}).

\bibitem[{\citenamefont{Rabitz et~al.}(2004)\citenamefont{Rabitz, Hsieh, and
  Rosenthal}}]{RabMik2004}
\bibinfo{author}{\bibfnamefont{H.~A.} \bibnamefont{Rabitz}},
  \bibinfo{author}{\bibfnamefont{M.~M.} \bibnamefont{Hsieh}}, \bibnamefont{and}
  \bibinfo{author}{\bibfnamefont{C.~M.} \bibnamefont{Rosenthal}},
  \bibinfo{journal}{Science} \textbf{\bibinfo{volume}{303}},
  \bibinfo{pages}{1998} (\bibinfo{year}{2004}).

\bibitem[{\citenamefont{Girardeau et~al.}(1998)\citenamefont{Girardeau,
  Schirmer, Leahy, and Koch}}]{Koch1998}
\bibinfo{author}{\bibfnamefont{M.~D.} \bibnamefont{Girardeau}},
  \bibinfo{author}{\bibfnamefont{S.~G.} \bibnamefont{Schirmer}},
  \bibinfo{author}{\bibfnamefont{J.~V.} \bibnamefont{Leahy}}, \bibnamefont{and}
  \bibinfo{author}{\bibfnamefont{R.~M.} \bibnamefont{Koch}},
  \bibinfo{journal}{Phys. Rev. A} \textbf{\bibinfo{volume}{58}},
  \bibinfo{pages}{2684} (\bibinfo{year}{1998}).

\bibitem[{\citenamefont{Ho and Rabitz}(2006)}]{HoRab2006a}
\bibinfo{author}{\bibfnamefont{T.~S.} \bibnamefont{Ho}} \bibnamefont{and}
  \bibinfo{author}{\bibfnamefont{H.~A.} \bibnamefont{Rabitz}},
  \bibinfo{journal}{J. Photochem. Photobiol. A} \textbf{\bibinfo{volume}{180}},
  \bibinfo{pages}{226} (\bibinfo{year}{2006}).

\bibitem[{\citenamefont{Wu et~al.}(2007{\natexlab{a}})\citenamefont{Wu, Hsieh,
  and Rabitz}}]{WuMike2007}
\bibinfo{author}{\bibfnamefont{R.}~\bibnamefont{Wu}},
  \bibinfo{author}{\bibfnamefont{M.~M.} \bibnamefont{Hsieh}}, \bibnamefont{and}
  \bibinfo{author}{\bibfnamefont{H.~A.} \bibnamefont{Rabitz}},
  \bibinfo{journal}{submitted}  (\bibinfo{year}{2007}{\natexlab{a}}).

\bibitem[{\citenamefont{Rabitz et~al.}(2006)\citenamefont{Rabitz, Hsieh, and
  Rosenthal}}]{Mike2006a}
\bibinfo{author}{\bibfnamefont{H.~A.} \bibnamefont{Rabitz}},
  \bibinfo{author}{\bibfnamefont{M.~M.} \bibnamefont{Hsieh}}, \bibnamefont{and}
  \bibinfo{author}{\bibfnamefont{C.~M.} \bibnamefont{Rosenthal}},
  \bibinfo{journal}{J. Chem. Phys.} \textbf{\bibinfo{volume}{51}},
  \bibinfo{pages}{204107} (\bibinfo{year}{2006}).

\bibitem[{\citenamefont{Shen et~al.}(2006)\citenamefont{Shen, Hsieh, and
  Rabitz}}]{Shen2006}
\bibinfo{author}{\bibfnamefont{Z.}~\bibnamefont{Shen}},
  \bibinfo{author}{\bibfnamefont{M.~M.} \bibnamefont{Hsieh}}, \bibnamefont{and}
  \bibinfo{author}{\bibfnamefont{H.~A.} \bibnamefont{Rabitz}},
  \bibinfo{journal}{J. Chem. Phys.} \textbf{\bibinfo{volume}{124}},
  \bibinfo{pages}{204106} (\bibinfo{year}{2006}).

\bibitem[{\citenamefont{Dirr et~al.}(2006)\citenamefont{Dirr, Helmke, Huper,
  Kleinsteuber, and Liu}}]{Helmke2006}
\bibinfo{author}{\bibfnamefont{G.}~\bibnamefont{Dirr}},
  \bibinfo{author}{\bibfnamefont{U.}~\bibnamefont{Helmke}},
  \bibinfo{author}{\bibfnamefont{K.}~\bibnamefont{Huper}},
  \bibinfo{author}{\bibfnamefont{M.}~\bibnamefont{Kleinsteuber}},
  \bibnamefont{and} \bibinfo{author}{\bibfnamefont{Y.}~\bibnamefont{Liu}},
  \bibinfo{journal}{J. Global Opt.} \textbf{\bibinfo{volume}{35}},
  \bibinfo{pages}{443} (\bibinfo{year}{2006}).

\bibitem[{\citenamefont{Glaser et~al.}(1998)\citenamefont{Glaser,
  Schulte-Herbruggen, Sieveking, Scheletzky, Nielsen, Sorensen, and
  Griesinger}}]{Glaser1998}
\bibinfo{author}{\bibfnamefont{S.~J.} \bibnamefont{Glaser}},
  \bibinfo{author}{\bibfnamefont{T.}~\bibnamefont{Schulte-Herbruggen}},
  \bibinfo{author}{\bibfnamefont{M.}~\bibnamefont{Sieveking}},
  \bibinfo{author}{\bibfnamefont{O.}~\bibnamefont{Scheletzky}},
  \bibinfo{author}{\bibfnamefont{N.~C.} \bibnamefont{Nielsen}},
  \bibinfo{author}{\bibfnamefont{O.~W.} \bibnamefont{Sorensen}},
  \bibnamefont{and}
  \bibinfo{author}{\bibfnamefont{C.}~\bibnamefont{Griesinger}},
  \bibinfo{journal}{Science} \textbf{\bibinfo{volume}{280}},
  \bibinfo{pages}{421} (\bibinfo{year}{1998}).

\bibitem[{\citenamefont{Rabitz et~al.}(2005)\citenamefont{Rabitz, Hsieh, and
  Rosenthal}}]{RabMik2005}
\bibinfo{author}{\bibfnamefont{H.~A.} \bibnamefont{Rabitz}},
  \bibinfo{author}{\bibfnamefont{M.~M.} \bibnamefont{Hsieh}}, \bibnamefont{and}
  \bibinfo{author}{\bibfnamefont{C.~M.} \bibnamefont{Rosenthal}},
  \bibinfo{journal}{Phys. Rev. A} \textbf{\bibinfo{volume}{72}},
  \bibinfo{pages}{052337} (\bibinfo{year}{2005}).

\bibitem[{\citenamefont{Ho and Rabitz}(2007)}]{HoRab2006b}
\bibinfo{author}{\bibfnamefont{T.~S.} \bibnamefont{Ho}} \bibnamefont{and}
  \bibinfo{author}{\bibfnamefont{H.~A.} \bibnamefont{Rabitz}},
  \bibinfo{journal}{To be submitted}  (\bibinfo{year}{2007}).

\bibitem[{\citenamefont{Frankel}(1962)}]{Frankel1965}
\bibinfo{author}{\bibfnamefont{T.}~\bibnamefont{Frankel}}, in
  \emph{\bibinfo{booktitle}{Differential and Combinatorial Topology: a
  symposium in honor of Marston Morse}}, edited by
  \bibinfo{editor}{\bibfnamefont{S.~S.} \bibnamefont{Cairns}}
  (\bibinfo{publisher}{Princeton University Press},
  \bibinfo{address}{Princeton, NJ}, \bibinfo{year}{1962}), pp.
  \bibinfo{pages}{37--53}.

\bibitem[{\citenamefont{Wu et~al.}(2007)\citenamefont{Wu, Chakrabarti, and
  Rabitz}}]{WuRaj2007}
\bibinfo{author}{\bibfnamefont{R.}~\bibnamefont{Wu}},
  \bibinfo{author}{\bibfnamefont{R.}~\bibnamefont{Chakrabarti}},
  \bibnamefont{and} \bibinfo{author}{\bibfnamefont{H.}~\bibnamefont{Rabitz}},
  \bibinfo{journal}{Submitted}  (\bibinfo{year}{2007}{\natexlab{c}}), \eprint
  {eprint arXiv:0708.2118 [quant-ph]}.

\bibitem[{\citenamefont{Lloyd and Braunstein}(1999)}]{LloBra1998}
\bibinfo{author}{\bibfnamefont{S.}~\bibnamefont{Lloyd}} \bibnamefont{and}
  \bibinfo{author}{\bibfnamefont{S.~L.} \bibnamefont{Braunstein}},
  \bibinfo{journal}{Phys. Rev. Lett.} \textbf{\bibinfo{volume}{82}},
  \bibinfo{pages}{1784} (\bibinfo{year}{1999}).

\bibitem[{\citenamefont{Wu et~al.}(2007{\natexlab{c}})\citenamefont{Wu,
  Chakrabarti, and Rabitz}}]{WuRaj2007b}
\bibinfo{author}{\bibfnamefont{R.}~\bibnamefont{Wu}},
  \bibinfo{author}{\bibfnamefont{R.}~\bibnamefont{Chakrabarti}},
  \bibnamefont{and} \bibinfo{author}{\bibfnamefont{H.~A.}
  \bibnamefont{Rabitz}}, \bibinfo{journal}{To be submitted}
  (\bibinfo{year}{2007}{\natexlab{c}}), \eprint{eprint arXiv:0708.3822
  [math-ph]}.

\bibitem[{\citenamefont{Jurdjevic}(1997)}]{Jurdjevic1997}
\bibinfo{author}{\bibfnamefont{V.}~\bibnamefont{Jurdjevic}},
  \emph{\bibinfo{title}{Geometric control theory}}
  (\bibinfo{publisher}{Cambridge University Press},
  \bibinfo{address}{Cambridge}, \bibinfo{year}{1997}).

\bibitem[{\citenamefont{Boscain et~al.}(2002)\citenamefont{Boscain, Charlot,
  Gauthier, Guerin, and Jauslin}}]{Boscain2002}
\bibinfo{author}{\bibfnamefont{U.}~\bibnamefont{Boscain}},
  \bibinfo{author}{\bibfnamefont{G.}~\bibnamefont{Charlot}},
  \bibinfo{author}{\bibfnamefont{J.~P.} \bibnamefont{Gauthier}},
  \bibinfo{author}{\bibfnamefont{S.}~\bibnamefont{Guerin}}, \bibnamefont{and}
  \bibinfo{author}{\bibfnamefont{H.~R.} \bibnamefont{Jauslin}},
  \bibinfo{journal}{J. Math. Phys.} \textbf{\bibinfo{volume}{43}},
  \bibinfo{pages}{2017} (\bibinfo{year}{2002}).

\bibitem[{\citenamefont{Boscain and Charlot}(2007)}]{Boscain2007}
\bibinfo{author}{\bibfnamefont{U.}~\bibnamefont{Boscain}} \bibnamefont{and}
  \bibinfo{author}{\bibfnamefont{G.}~\bibnamefont{Charlot}},
  \bibinfo{journal}{To be submitted}  (\bibinfo{year}{2007}).

\bibitem[{\citenamefont{D'Alessandro and Dahleh}(2001)}]{DAlessandro2001a}
\bibinfo{author}{\bibfnamefont{D.}~\bibnamefont{D'Alessandro}}
  \bibnamefont{and} \bibinfo{author}{\bibfnamefont{M.}~\bibnamefont{Dahleh}},
  \bibinfo{journal}{IEEE Trans. Autom. Control} \textbf{\bibinfo{volume}{46}},
  \bibinfo{pages}{866} (\bibinfo{year}{2001}).

\bibitem[{\citenamefont{Rothman et~al.}(2005)\citenamefont{Rothman, Ho, and
  Rabitz}}]{Rothman2005}
\bibinfo{author}{\bibfnamefont{A.}~\bibnamefont{Rothman}},
  \bibinfo{author}{\bibfnamefont{T.~S.} \bibnamefont{Ho}}, \bibnamefont{and}
  \bibinfo{author}{\bibfnamefont{H.~A.} \bibnamefont{Rabitz}},
  \bibinfo{journal}{Phys. Rev. A} \textbf{\bibinfo{volume}{72}},
  \bibinfo{pages}{023416} (\bibinfo{year}{2005}).

\bibitem[{\citenamefont{Rothman
  et~al.}(2006{\natexlab{a}})\citenamefont{Rothman, Ho, and
  Rabitz}}]{Rothman2006a}
\bibinfo{author}{\bibfnamefont{A.}~\bibnamefont{Rothman}},
  \bibinfo{author}{\bibfnamefont{T.~S.} \bibnamefont{Ho}}, \bibnamefont{and}
  \bibinfo{author}{\bibfnamefont{H.~A.} \bibnamefont{Rabitz}},
  \bibinfo{journal}{J. Chem. Phys.} \textbf{\bibinfo{volume}{123}},
  \bibinfo{pages}{134104} (\bibinfo{year}{2006}{\natexlab{a}}).

\bibitem[{\citenamefont{Rothman
  et~al.}(2006{\natexlab{b}})\citenamefont{Rothman, Ho, and
  Rabitz}}]{Rothman2006b}
\bibinfo{author}{\bibfnamefont{A.}~\bibnamefont{Rothman}},
  \bibinfo{author}{\bibfnamefont{T.~S.} \bibnamefont{Ho}}, \bibnamefont{and}
  \bibinfo{author}{\bibfnamefont{H.~A.} \bibnamefont{Rabitz}},
  \bibinfo{journal}{Phys. Rev. A} \textbf{\bibinfo{volume}{73}},
  \bibinfo{pages}{053401} (\bibinfo{year}{2006}{\natexlab{b}}).

\bibitem[{\citenamefont{Beltrani et~al.}(2007)\citenamefont{Beltrani, Dominy,
  and Rabitz}}]{Beltrani2007}
\bibinfo{author}{\bibfnamefont{V.}~\bibnamefont{Beltrani}},
  \bibinfo{author}{\bibfnamefont{J.}~\bibnamefont{Dominy}}, \bibnamefont{and}
  \bibinfo{author}{\bibfnamefont{H.~A.} \bibnamefont{Rabitz}},
  \bibinfo{journal}{Phys. Rev. A} \textbf{\bibinfo{volume}{74}},
  \bibinfo{pages}{043414} (\bibinfo{year}{2007}).

\bibitem[{\citenamefont{Roslund et~al.}(2006)\citenamefont{Roslund, Roth, and
  Rabitz}}]{Roslund2006}
\bibinfo{author}{\bibfnamefont{J.}~\bibnamefont{Roslund}},
  \bibinfo{author}{\bibfnamefont{M.}~\bibnamefont{Roth}}, \bibnamefont{and}
  \bibinfo{author}{\bibfnamefont{H.~A.} \bibnamefont{Rabitz}},
  \bibinfo{journal}{Phys. Rev. A} \textbf{\bibinfo{volume}{74}},
  \bibinfo{pages}{043414} (\bibinfo{year}{2006}).

\bibitem[{\citenamefont{Roslund and Rabitz}(2007{\natexlab{a}})}]{Roslund2007}
\bibinfo{author}{\bibfnamefont{J.}~\bibnamefont{Roslund}} \bibnamefont{and}
  \bibinfo{author}{\bibfnamefont{H.~A.} \bibnamefont{Rabitz}},
  \bibinfo{journal}{To be submitted}  (\bibinfo{year}{2007}{\natexlab{a}}).

\bibitem[{\citenamefont{Cardoza et~al.}(2005)\citenamefont{Cardoza,
  Trallero-Herrero, Langhojer, Rabitz, and Weinacht}}]{Cardoza2005}
\bibinfo{author}{\bibfnamefont{D.}~\bibnamefont{Cardoza}},
  \bibinfo{author}{\bibfnamefont{C.}~\bibnamefont{Trallero-Herrero}},
  \bibinfo{author}{\bibfnamefont{F.}~\bibnamefont{Langhojer}},
  \bibinfo{author}{\bibfnamefont{H.~A.} \bibnamefont{Rabitz}},
  \bibnamefont{and} \bibinfo{author}{\bibfnamefont{T.}~\bibnamefont{Weinacht}},
  \bibinfo{journal}{J. Chem. Phys.} \textbf{\bibinfo{volume}{122}},
  \bibinfo{pages}{124306} (\bibinfo{year}{2005}).

\bibitem[{\citenamefont{Wells et~al.}(2005)\citenamefont{Wells, Betsch,
  Conover, DeWitt, Pinkham, and Jones}}]{Wells2005}
\bibinfo{author}{\bibfnamefont{E.}~\bibnamefont{Wells}},
  \bibinfo{author}{\bibfnamefont{K.~J.} \bibnamefont{Betsch}},
  \bibinfo{author}{\bibfnamefont{C.~W.~S.} \bibnamefont{Conover}},
  \bibinfo{author}{\bibfnamefont{M.~J.} \bibnamefont{DeWitt}},
  \bibinfo{author}{\bibfnamefont{D.}~\bibnamefont{Pinkham}}, \bibnamefont{and}
  \bibinfo{author}{\bibfnamefont{R.~R.} \bibnamefont{Jones}},
  \bibinfo{journal}{Phys. Rev. A} \textbf{\bibinfo{volume}{72}},
  \bibinfo{pages}{063406} (\bibinfo{year}{2005}).

\bibitem[{\citenamefont{Wollenhaupt et~al.}(2005)\citenamefont{Wollenhaupt,
  Prakelt, Sarpe-Tudoran, Liese, and Baumert}}]{Wollenhaupt2005}
\bibinfo{author}{\bibfnamefont{M.}~\bibnamefont{Wollenhaupt}},
  \bibinfo{author}{\bibfnamefont{A.}~\bibnamefont{Prakelt}},
  \bibinfo{author}{\bibfnamefont{C.}~\bibnamefont{Sarpe-Tudoran}},
  \bibinfo{author}{\bibfnamefont{D.}~\bibnamefont{Liese}}, \bibnamefont{and}
  \bibinfo{author}{\bibfnamefont{R.}~\bibnamefont{Baumert}},
  \bibinfo{journal}{J. Mod. Opt.} \textbf{\bibinfo{volume}{52}},
  \bibinfo{pages}{2187} (\bibinfo{year}{2005}).

\bibitem[{\citenamefont{Roslund and Rabitz}(2007{\natexlab{b}})}]{Roslund2007b}
\bibinfo{author}{\bibfnamefont{J.}~\bibnamefont{Roslund}} \bibnamefont{and}
  \bibinfo{author}{\bibfnamefont{H.~A.} \bibnamefont{Rabitz}},
  \bibinfo{journal}{To be submitted}  (\bibinfo{year}{2007}{\natexlab{b}}).

\bibitem[{\citenamefont{Malley and Hornstein}(1993)}]{Malley1993}
\bibinfo{author}{\bibfnamefont{J.}~\bibnamefont{Malley}} \bibnamefont{and}
  \bibinfo{author}{\bibfnamefont{J.}~\bibnamefont{Hornstein}},
  \bibinfo{journal}{Stat. Sci.} \textbf{\bibinfo{volume}{8}},
  \bibinfo{pages}{433} (\bibinfo{year}{1993}).

\bibitem[{\citenamefont{Chakrabarti
  et~al.}(2007{\natexlab{a}})\citenamefont{Chakrabarti, Wu, and
  Rabitz}}]{RajWu2007}
\bibinfo{author}{\bibfnamefont{R.}~\bibnamefont{Chakrabarti}},
  \bibinfo{author}{\bibfnamefont{R.}~\bibnamefont{Wu}}, \bibnamefont{and}
  \bibinfo{author}{\bibfnamefont{H.~A.} \bibnamefont{Rabitz}},
  \bibinfo{journal}{To be submitted}(\bibinfo{year}{2007}{\natexlab{c}}), \eprint{eprint arXiv:0708.3384 [quant-ph]}.


\bibitem[{\citenamefont{Ramakrishna et~al.}(1995)\citenamefont{Ramakrishna,
  Salapaka, Dahleh, and Rabitz}}]{Ramakrishna1995}
\bibinfo{author}{\bibfnamefont{V.}~\bibnamefont{Ramakrishna}},
  \bibinfo{author}{\bibfnamefont{M.~V.} \bibnamefont{Salapaka}},
  \bibinfo{author}{\bibfnamefont{M.}~\bibnamefont{Dahleh}}, \bibnamefont{and}
  \bibinfo{author}{\bibfnamefont{H.~A.} \bibnamefont{Rabitz}},
  \bibinfo{journal}{Phys. Rev. A} \textbf{\bibinfo{volume}{51}},
  \bibinfo{pages}{960} (\bibinfo{year}{1995}).

\bibitem[{\citenamefont{Jurdjevic and Sussmann}(1972)}]{JurdSuss1972}
\bibinfo{author}{\bibfnamefont{V.}~\bibnamefont{Jurdjevic}} \bibnamefont{and}
  \bibinfo{author}{\bibfnamefont{J.}~\bibnamefont{Sussmann}},
  \bibinfo{journal}{J. Diff. Equat.} \textbf{\bibinfo{volume}{12}},
  \bibinfo{pages}{313} (\bibinfo{year}{1972}).

\bibitem[{\citenamefont{D'Alessandro}(2000)}]{DAlessandro2000}
\bibinfo{author}{\bibfnamefont{D.}~\bibnamefont{D'Alessandro}},
  \bibinfo{journal}{Syst. Control Lett.} \textbf{\bibinfo{volume}{41}},
  \bibinfo{pages}{213} (\bibinfo{year}{2000}).

\bibitem[{\citenamefont{Wu et~al.}(2006)\citenamefont{Wu, Tarn, and
  Li}}]{WuTarn2006}
\bibinfo{author}{\bibfnamefont{R.}~\bibnamefont{Wu}},
  \bibinfo{author}{\bibfnamefont{T.~J.} \bibnamefont{Tarn}}, \bibnamefont{and}
  \bibinfo{author}{\bibfnamefont{C.~W.} \bibnamefont{Li}},
  \bibinfo{journal}{Phys. Rev. A} \textbf{\bibinfo{volume}{73}},
  \bibinfo{pages}{012719} (\bibinfo{year}{2006}).

\bibitem[{\citenamefont{Jezek et~al.}(2003)\citenamefont{Jezek, Fiurasek, and
  Hradil}}]{Hradil2003}
\bibinfo{author}{\bibfnamefont{M.}~\bibnamefont{Jezek}},
  \bibinfo{author}{\bibfnamefont{J.}~\bibnamefont{Fiurasek}}, \bibnamefont{and}
  \bibinfo{author}{\bibfnamefont{Z.}~\bibnamefont{Hradil}},
  \bibinfo{journal}{Phys. Rev. A} \textbf{\bibinfo{volume}{68}},
  \bibinfo{pages}{012305} (\bibinfo{year}{2003}).

\bibitem[{\citenamefont{Maday and Turinici}(2003)}]{Maday2003}
\bibinfo{author}{\bibfnamefont{G.}~\bibnamefont{Maday}} \bibnamefont{and}
  \bibinfo{author}{\bibfnamefont{G.}~\bibnamefont{Turinici}},
  \bibinfo{journal}{J. Chem. Phys.} \textbf{\bibinfo{volume}{118}},
  \bibinfo{pages}{8191} (\bibinfo{year}{2003}).

\bibitem[{\citenamefont{Balint-Kurti et~al.}(2005)\citenamefont{Balint-Kurti,
  Manby, Ren, Artamonov, Ho, and Rabitz}}]{Balint-Kurti2005}
\bibinfo{author}{\bibfnamefont{G.~G.} \bibnamefont{Balint-Kurti}},
  \bibinfo{author}{\bibfnamefont{F.~R.} \bibnamefont{Manby}},
  \bibinfo{author}{\bibfnamefont{Q.}~\bibnamefont{Ren}},
  \bibinfo{author}{\bibfnamefont{M.}~\bibnamefont{Artamonov}},
  \bibinfo{author}{\bibfnamefont{T.~S.} \bibnamefont{Ho}}, \bibnamefont{and}
  \bibinfo{author}{\bibfnamefont{H.~A.} \bibnamefont{Rabitz}},
  \bibinfo{journal}{J. Chem. Phys.} \textbf{\bibinfo{volume}{122}},
  \bibinfo{pages}{084110} (\bibinfo{year}{2005}).

\bibitem[{\citenamefont{Amstrup et~al.}(1995)\citenamefont{Amstrup, Toth,
  Szabo, Rabitz, and Loerincz}}]{Amstrup1995}
\bibinfo{author}{\bibfnamefont{B.}~\bibnamefont{Amstrup}},
  \bibinfo{author}{\bibfnamefont{G.~J.} \bibnamefont{Toth}},
  \bibinfo{author}{\bibfnamefont{G.}~\bibnamefont{Szabo}},
  \bibinfo{author}{\bibfnamefont{H.~A.} \bibnamefont{Rabitz}},
  \bibnamefont{and} \bibinfo{author}{\bibfnamefont{A.}~\bibnamefont{Loerincz}},
  \bibinfo{journal}{J. Phys. Chem.} \textbf{\bibinfo{volume}{99}},
  \bibinfo{pages}{5206} (\bibinfo{year}{1995}).

\bibitem[{\citenamefont{Brixner et~al.}(2003)\citenamefont{Brixner, Damrauer,
  Kiefer, and Gerber}}]{Brixner2003}
\bibinfo{author}{\bibfnamefont{T.}~\bibnamefont{Brixner}},
  \bibinfo{author}{\bibfnamefont{M.~H.} \bibnamefont{Damrauer}},
  \bibinfo{author}{\bibfnamefont{B.}~\bibnamefont{Kiefer}}, \bibnamefont{and}
  \bibinfo{author}{\bibfnamefont{G.}~\bibnamefont{Gerber}},
  \bibinfo{journal}{J. Chem. Phys} \textbf{\bibinfo{volume}{118}},
  \bibinfo{pages}{3692} (\bibinfo{year}{2003}).

\bibitem[{\citenamefont{Levis et~al.}(2001)\citenamefont{Levis, Menkir, and
  Rabitz}}]{Levis2001}
\bibinfo{author}{\bibfnamefont{R.}~\bibnamefont{Levis}},
  \bibinfo{author}{\bibfnamefont{G.}~\bibnamefont{Menkir}}, \bibnamefont{and}
  \bibinfo{author}{\bibfnamefont{H.~A.} \bibnamefont{Rabitz}},
  \bibinfo{journal}{Science} \textbf{\bibinfo{volume}{292}},
  \bibinfo{pages}{709} (\bibinfo{year}{2001}).

\bibitem[{\citenamefont{Chakrabarti
  et~al.}(2007{\natexlab{b}})\citenamefont{Chakrabarti, Wu, and
  Rabitz}}]{RajWu2007b}
\bibinfo{author}{\bibfnamefont{R.}~\bibnamefont{Chakrabarti}},
  \bibinfo{author}{\bibfnamefont{R.}~\bibnamefont{Wu}}, \bibnamefont{and}
  \bibinfo{author}{\bibfnamefont{H.~A.} \bibnamefont{Rabitz}},
  \bibinfo{journal}{To be submitted}  (\bibinfo{year}{2007}{\natexlab{b}}),
  \eprint{eprint arXiv:0708.3513 [quant-ph]}.

\bibitem[{\citenamefont{Siegelmann et~al.}(1999)\citenamefont{Siegelmann, Hur,
  and Fishman}}]{Fishman1999}
\bibinfo{author}{\bibfnamefont{H.~T.} \bibnamefont{Siegelmann}},
  \bibinfo{author}{\bibfnamefont{A.~B.} \bibnamefont{Hur}}, \bibnamefont{and}
  \bibinfo{author}{\bibfnamefont{S.}~\bibnamefont{Fishman}},
  \bibinfo{journal}{Phys. Rev. Lett.} \textbf{\bibinfo{volume}{83}},
  \bibinfo{pages}{1463} (\bibinfo{year}{1999}).

\bibitem[{\citenamefont{Riviello et~al.}(2007)\citenamefont{Riviello, Moore,
  and Rabitz}}]{GregRab2007}
\bibinfo{author}{\bibfnamefont{G.~T.} \bibnamefont{Riviello}},
  \bibinfo{author}{\bibfnamefont{K.~A.} \bibnamefont{Moore}}, \bibnamefont{and}
  \bibinfo{author}{\bibfnamefont{H.~A.} \bibnamefont{Rabitz}},
  \bibinfo{journal}{In preparation}  (\bibinfo{year}{2007}).

\bibitem[{\citenamefont{Moore et~al.}(2007)\citenamefont{Moore, Hsieh, and
  Rabitz}}]{MooreRab2007}
\bibinfo{author}{\bibfnamefont{K.~A.} \bibnamefont{Moore}},
  \bibinfo{author}{\bibfnamefont{M.~M.} \bibnamefont{Hsieh}}, \bibnamefont{and}
  \bibinfo{author}{\bibfnamefont{H.~A.} \bibnamefont{Rabitz}},
  \bibinfo{journal}{To be submitted}  (\bibinfo{year}{2007}).

\bibitem[{\citenamefont{Dominy and Rabitz}(2007)}]{Dominy2007}
\bibinfo{author}{\bibfnamefont{J.}~\bibnamefont{Dominy}} \bibnamefont{and}
  \bibinfo{author}{\bibfnamefont{H.~A.} \bibnamefont{Rabitz}},
  \bibinfo{journal}{In preparation}  (\bibinfo{year}{2007}).

\bibitem[{\citenamefont{Rehacek et~al.}(2007)\citenamefont{Rehacek, Hradil,
  Knill, and Lvovsky}}]{Rehacek2007}
\bibinfo{author}{\bibfnamefont{J.}~\bibnamefont{Rehacek}},
  \bibinfo{author}{\bibfnamefont{Z.}~\bibnamefont{Hradil}},
  \bibinfo{author}{\bibfnamefont{E.}~\bibnamefont{Knill}}, \bibnamefont{and}
  \bibinfo{author}{\bibfnamefont{A.~I.} \bibnamefont{Lvovsky}},
  \bibinfo{journal}{Phys. Rev. A} \textbf{\bibinfo{volume}{75}},
  \bibinfo{pages}{042108} (\bibinfo{year}{2007}).

\bibitem[{\citenamefont{Agrachev and Sarychev}(1995)}]{Agrachev1995}
\bibinfo{author}{\bibfnamefont{A.~A.} \bibnamefont{Agrachev}} \bibnamefont{and}
  \bibinfo{author}{\bibfnamefont{A.~V.} \bibnamefont{Sarychev}},
  \bibinfo{journal}{Journal of Math. Syst. Est. Control}
  \textbf{\bibinfo{volume}{5}}, \bibinfo{pages}{1} (\bibinfo{year}{1995}).

\bibitem[{\citenamefont{Wu et~al.}(2007{\natexlab{d}})\citenamefont{Wu, Pechen,
  Rabitz, Hsieh, and Tsou}}]{WuPech2006}
\bibinfo{author}{\bibfnamefont{R.}~\bibnamefont{Wu}},
  \bibinfo{author}{\bibfnamefont{A.}~\bibnamefont{Pechen}},
  \bibinfo{author}{\bibfnamefont{H.~A.} \bibnamefont{Rabitz}},
  \bibinfo{author}{\bibfnamefont{M.~M.} \bibnamefont{Hsieh}}, \bibnamefont{and}
  \bibinfo{author}{\bibfnamefont{B.}~\bibnamefont{Tsou}},
  \bibinfo{journal}{Submitted}  (\bibinfo{year}{2007}{\natexlab{d}}),
  \eprint{eprint arXiv:0708.2119 [quant-ph]}.

\bibitem[{\citenamefont{Arvind et~al.}(1995)\citenamefont{Arvind, Dutta,
  Mukunda, and Simon}}]{ArvMuk1995}
\bibinfo{author}{\bibnamefont{Arvind}},
  \bibinfo{author}{\bibfnamefont{B.}~\bibnamefont{Dutta}},
  \bibinfo{author}{\bibfnamefont{N.}~\bibnamefont{Mukunda}}, \bibnamefont{and}
  \bibinfo{author}{\bibfnamefont{R.}~\bibnamefont{Simon}},
  \bibinfo{journal}{Phys. Rev. A} \textbf{\bibinfo{volume}{52}},
  \bibinfo{pages}{1609} (\bibinfo{year}{1995}).

\end{thebibliography}

\end{document}